\documentclass[11pt,a4paper,preprintnumbers,nofootinbib]{revtex4}
\usepackage[T1]{fontenc}
\usepackage[utf8]{inputenc}
\usepackage{mathtools}
\usepackage{amsmath}
\usepackage{amsthm}
\usepackage{amsmath}
\usepackage{amssymb}
\usepackage{amstext}
\usepackage[final]{graphicx}
\usepackage{float}
\usepackage{booktabs}
\usepackage{fixltx2e}
\usepackage{subfigure}
\usepackage{array}
\newcolumntype{C}[1]{>{\centering\arraybackslash$}p{#1}<{$}}

\usepackage[dvipsnames]{xcolor}
\usepackage{slashed}
\usepackage[section]{placeins}
\usepackage{tabularx}
\PassOptionsToPackage{hyphens}{url}
\usepackage{hyperref}
\usepackage{makecell}	
\allowdisplaybreaks
\usepackage[normalem]{ulem}

\newcommand{\MG}{\textsc{MadGraph5}\_aMC@NLO\ } 
\definecolor{lightgray}{gray}{0.91}

\begin{document}
\preprint{DO-TH 20/12}

\title{Multi-lepton signatures of vector-like leptons with flavor}
\author{Stefan~Bi{\ss}mann}
\author{Gudrun~Hiller}
\author{Clara~Hormigos-Feliu}
\affiliation{Fakult\"at Physik, TU Dortmund, Otto-Hahn-Str.4, D-44221 Dortmund, Germany}
\author{Daniel F.~Litim}
\affiliation{Department of Physics and Astronomy, University of Sussex, Brighton,
BN19QH, United Kingdom}

\begin{abstract}
 We investigate collider signatures of standard model extensions featuring  vector-like leptons  and a flavorful scalar sector. 
 Such a framework arises naturally within asymptotically safe model building, which tames the UV behavior of  the standard model towards the Planck scale and beyond.
 We focus on  values of Yukawa couplings and masses which allow to explain the present data on the muon and electron anomalous magnetic moments.
Using a CMS search based on $77.4 \, \rm{fb}^{-1}$ at the $\sqrt{s}=13$ TeV LHC we find that flavorful vector-like leptons are excluded for masses below around $300$~GeV if  they  are singlets under $SU(2)_L$, and around $800$~GeV  if they are doublets. 
 Exploiting the flavor-violating-like decays of the  scalars, we design novel null test observables based on opposite sign opposite flavor invariant masses.
 These multi-lepton distributions allow to  signal new physics  and to  extract mass hierarchies 
 in reach of near-future searches  at the LHC and the HL-LHC.
 
\end{abstract}

\maketitle

%======================================================================================================================================================
%

\section{Introduction}

The standard model (SM) is considered to be well established yet incomplete: It does not explain the puzzling structure of masses and mixings of elementary fermions.
It also displays meta-stability in the Higgs sector, and ultimately loses control towards (very) high energy as  the Higgs and hypercharge coupling run into Landau poles.
It is therefore commonly accepted that the SM has to be extended into a more complete one, with guidance  from both 
data and  top-down theory frontiers.

The concept of asymptotic safety 
\cite{Wilson:1971bg, Bailin:1974bq, Weinberg:1980gg}
opens up new directions  \cite{Litim:2014uca,Bond:2016dvk,Bond:2018oco,Bond:2017wut} to build models that remain both fundamental and predictive at  highest energies.
Concrete models which extend the SM into asymptotically safe ones include vector-like fermions as well as additional scalars \cite{Bond:2017wut, Kowalska:2017fzw}, which allow for phenomenological signatures that can be probed at colliders. 
A crucial difference to common  extensions of the SM
is that the scalars, which are singlets under the SM gauge group $SU(3)_C \times SU(2)_L \times U(1)_Y$, form a matrix in flavor space. This enhances the impact of Yukawa interactions
and allows for new, flavorful signatures once the SM and BSM flavor sectors are connected.

In this work, we study concrete such models with new vector-like leptons (VLLs) and flavor portal couplings which link  SM fermions  to the new matter particles by renormalizable Yukawa interactions.
Interestingly, while being asymptotically safe, or safe up to the Planck scale, such models
allow to explain  discrepancies between the SM and  current data of anomalous magnetic moments  (AMMs)  \cite{Hiller:2020fbu, Hiller:2019mou, Hiller:2019tvg}.
Presently, the electron and the muon AMMs deviate from the SM by $2.4~\sigma$ \cite{Hanneke:2008tm,  Parker:2018vye} and $3.5~\sigma$ \cite{Tanabashi:2018oca}, or 4.1 $\sigma$  \cite{Jegerlehner:2017lbd, Davier:2016iru}, respectively. Note also recent debates \cite{Borsanyi:2020mff,Aoyama:2020ynm,Crivellin:2020zul, Keshavarzi:2020bfy}.
In order to explain the current values of the AMMs, or values in a similar ballpark,  the VLLs can be as light as a few hundred GeV and thus can be probed  at the LHC.

Early searches for VLLs at LEP excluded heavy leptons lighter than $\sim 100~$GeV  \cite{Achard:2001qw}. At the LHC, ATLAS excluded VLLs transforming as singlets under $SU(2)_L$  in the range 114-176~GeV at 95~\% CL \cite{Aad:2015dha}. A recent CMS study based on $77.4 \, \rm{fb}^{-1}$  at 13~TeV searching for doublet VLLs coupling to third-generation leptons only \cite{Kumar:2015tna} excluded VLLs in the mass range 120-790~GeV at 95~\% CL \cite{Sirunyan:2019ofn}.
However, these VLL limits have been obtained within simplified models.
In this work we study collider  signatures of both the flavorful singlet and the doublet model in Refs.~\cite{Hiller:2020fbu, Hiller:2019mou, Hiller:2019tvg} 
in the multi-light lepton channel and
confront them to  the CMS search \cite{Sirunyan:2019ofn}. 
We further highlight how the specific features of the models such as lepton-flavor-violating-like decays suggest  new observables with null test potential. 
These are worked out for the LHC full Run 2 150 $\rm{fb}^{-1}$ data set and high luminosity (HL)-LHC with 3000  $\rm{fb}^{-1}$ at $\sqrt{s}=14$ TeV \cite{ApollinariG.:2017ojx}.

This paper is  organized as follows: 
In Sec.~\ref{sec:model} we present the BSM model frameworks and key parameters. In Sec.~\ref{sec:prod} we discuss the production and decay properties of the VLLs relevant for LHC phenomenology. We give  the settings used for the event generation in Sec.~\ref{sec:Simu}. We compare distributions to  the CMS measurements~\cite{Sirunyan:2019ofn} to obtain constraints on the models' parameter space. In Sec.~\ref{sec:Null-test} we construct new  observables that allow for null tests of the SM and work out
 projections for the full Run 2 data set. Perspectives for the higher luminosity scenario of  the HL-LHC are worked out  in Sec.~\ref{sec:Outlook}. 
 We discuss signatures  allowing for more general parameter regions  beyond the AMMs in Sec.~\ref{sec:beyondg2}. In Sec.~\ref{sec:conclusion} we summarize. A
comparison of  resonance heights before and after including detector effects for  Run 2 and the HL-LHC is provided in the Appendix.

\section{BSM framework \& setup}

\label{sec:model}

We start with the models of \cite{Hiller:2019mou, Hiller:2019tvg, Hiller:2020fbu}, which contain three generations of VLLs denoted by $\psi_{L,R}$. These are either 
colorless $SU(2)_L$ singlets with hypercharge $Y=-1$ (singlet model) or colorless $SU(2)_L$ doublets with $Y=-1/2$ (doublet model). The $SU(2)_L$-components in the latter read
$\psi_{L,R} = (\psi^0_{L,R}, \ \psi^-_{L,R})^T\,.$
 For the three generations of  left-handed
and right-handed SM leptons we use  $L=(\nu, \ \ell_L)^T$ and $E=\ell_R\,$, respectively,  and denote the Higgs doublet by $H$. 
All SM leptons and VLLs carry a lepton flavor index $i=1,2,3$, which is often suppressed to avoid clutter.
Both models also contain  complex scalars $S_{ij}$, with two flavor indices $i,j=1,2,3$,  and which are singlets under the SM gauge interactions.
In the interaction basis, the models'  BSM Yukawa sectors read
 \begin{equation} \label{Yukawa}
\begin{array}{l}
\!\!\mathcal{L}^{\text{singlet}}_{\text{Y}} = -\kappa \overline{L}_i H \psi_{Ri}  - \kappa'\overline{E}_i (S^{\dagger})_{ij}\psi_{Lj} - y\, \overline{\psi}_{Li} S_{ij} \psi_{Rj} + \mathrm{h.c.}\,, \\[1ex]
\!\!\!\!\mathcal{L}^{\text{doublet}}_{\text{Y}} = -\kappa \overline{E}_i {H}^{\dagger} \psi_{Li} - \kappa'\, \overline{L}_i S_{ij} \psi_{Rj} - y\, \overline{\psi}_{Li } S_{ij}  \psi_{Rj} + \mathrm{h.c.}\,,
\end{array}
\end{equation}
where the contraction of gauge indices is assumed. Here we followed \cite{Hiller:2020fbu} and identified $SU(3)$-flavor symmetries 
of the leptons with ones of the VLLs. This identification has important consequences for phenomenology: Each lepton flavor is conserved and leptons couple universally within
(\ref{Yukawa}), and  the BSM Yukawas $y, \kappa, \kappa^\prime$ become single couplings, instead of being  tensors. While
$y$ is key in variants of the asymptotically safe framework \cite{Bond:2017wut, Litim:2014uca},
in models like (\ref{Yukawa}) with mixed SM-BSM Yukawas its presence is not required to achieve a controlled UV-behavior  \cite{Hiller:2020fbu}.
As in addition the phenomenological implications of $y$  are less relevant we do not consider it in the numerical analysis.

After spontaneous symmetry breaking, the vector-like fermions and the leptons mix, see \cite{Hiller:2019mou, Hiller:2020fbu} for details. 
To be specific, we denote the lightest three mass eigenstates by leptons and the others by VLLs, and continue to use the notation as introduced above. 
$Z\to \ell\ell$ data \cite{Tanabashi:2018oca} constrains the mixing angles $\theta$  of left-handed (right-handed) leptons in  the singlet (doublet) model as  $\theta \simeq \kappa v_h/\sqrt{2}M_F < \mathcal{O}(10^{-2})$. Here,  $v_h \simeq 246 $ GeV is the Higgs vacuum expectation value (vev), and  we denote by $M_F$ the common mass of all
flavor and $SU(2)_L$-components of the VLLs. We learn that $\theta, \kappa \ll 1$, which allows for a small angle approximation.
At first order in $\kappa$ and $\kappa^\prime$, the interactions in the mass basis in the singlet model read
\begin{align}\label{Lint-singlet}
\begin{aligned}
\mathcal{L}_\text{int}^{\text{singlet}}=&
{\, -e\, \overline{\psi}\gamma^\mu\psi A_\mu}
{+\, \frac{g}{\cos\theta_w} \,
\overline{\psi}\gamma^\mu\psi Z_\mu}  
+ \Bigg(-\frac{\kappa}{\sqrt{2}}\, \bar\ell_L \psi_R\, h 
-\kappa^\prime \bar \ell_{R} S^\dagger \psi_L
\\ & 
+g_S\, \bar\ell_R S^\dagger \ell_L +
g_Z \, \overline{\ell}_L\gamma^\mu\psi_L\, Z_\mu + 
g_W\, \overline{\nu} \gamma^\mu\psi_L \, W_\mu^+  + \text{h.c.}\Bigg)\,,
\end{aligned}
\end{align}
where $A_\mu$ denotes the photon,  $h$ corresponds to the physical Higgs boson with $M_h = 125$ GeV and $e,g,\theta_w$ are the electromagnetic coupling, the $SU(2)_L$ coupling and the weak mixing angle, respectively. The remaining couplings fulfill
\begin{equation}\label{couplings-singlet}
\begin{aligned}
&g_S  = \frac{\kappa^\prime \kappa}{\sqrt{2}} \frac{v_h}{M_F}\,,
&& g_Z = {-\frac{\kappa g}{2\sqrt{2}\cos\theta_w} \frac{v_h}{ M_F}}\,,
&&&g_W = {\frac{\kappa g}{2} \frac{v_h}{M_F}}  \,.
\end{aligned}
\end{equation}

For the doublet model, we find 
\begin{align}\label{Lint-doublet}
\begin{aligned}
\mathcal{L}_\text{int}^{\text{doublet}}=&
{\, -e\, \overline{\psi^{-}}\gamma^\mu\psi^- A_\mu}
{ +\,
\frac{g}{2\cos\theta_w}\left[(2\sin^2\theta_w - 1) \overline{\psi^{-}}\gamma^\mu \psi^{-}  + \overline{\psi^0} \gamma^\mu\psi^0 \right]Z_\mu  }\\ & 
{ + \Bigg(\frac{g}{\sqrt{2}}\, \overline{\psi^{-}} \gamma^\mu\psi^0 W_\mu^-}
-\frac{\kappa}{\sqrt{2}}\, \bar\ell_R \psi_L^{-}\, h 
-\kappa^\prime \bar \ell_{L} S \psi_R^- 
-\kappa^\prime \bar \nu S \psi_R^0
+g_S\, \bar\ell_L S \,\ell_R  
\\ & +
g_Z \, \overline{\ell}_R\gamma^\mu\psi^-_R\, Z_\mu + 
g_W\, \overline{\ell}_R \gamma^\mu\psi_R^0 \, W_\mu^-  + \text{h.c.}\Bigg)\,,
\end{aligned}
\end{align}
with couplings
\begin{equation}\label{couplings-doublet}
\begin{aligned}
&g_S  = \frac{\kappa^\prime \kappa}{\sqrt{2}} \frac{v_h}{M_F}\,,
&& g_Z = {\frac{\kappa g}{2\sqrt{2}\cos\theta_w} \frac{v_h}{ M_F}}\,,
&&&g_W = {- \frac{\kappa g}{2} \frac{v_h}{M_F} } \,.
\end{aligned}
\end{equation}
The vertex $\overline{\nu} \gamma^\mu \psi_L^- \,W_\mu^+ $ arises only at higher order, see \cite{Hiller:2020fbu} for details, and can be safely neglected for the purpose of this analysis. 

The parameters we are concerned with  in the two models
are therefore $M_F, \kappa,  \kappa^\prime$ and the common mass of the BSM scalars, $M_S$.
Addressing the muon AMM anomaly $\Delta a_{\mu}  \equiv a_{\mu}^{\text{exp}} - a_{\mu}^{\text{SM}} = \ 268(63)(43)\cdot 10^{-11}$ \cite{Tanabashi:2018oca} at one loop eliminates $\kappa^\prime$ in terms of the BSM masses
 by \cite{Hiller:2020fbu, Hiller:2019mou}
\begin{align} \label{Amu}
\Delta a_{\mu} =\frac{\kappa'^2}{32\pi^2} \, \frac{m_{\mu}^2}{M_{F}^2} \,  f\left(\frac{M_S^2}{M_F^2}\right)\,, 
\end{align}
with  $f(t)  = (2 t^3+3 t^2-6 t^2 \ln t-6 t+1)/(t-1)^4$ positive for any $t$, and $f(0)=1$. 
For example, for  $M_S=500$~GeV and $M_F = \{100, 500, 1000 \}$~GeV one obtains $\kappa' \simeq \{3.6, 6.5, 10.4\}$, respectively.
The coupling $\kappa$ is needed to account for the electron AMM anomaly $\Delta a_e$  \cite{Hanneke:2008tm,  Parker:2018vye}, however, 
can be chosen with some freedom since parameters of the scalar sector also play a role here, see  \cite{Hiller:2020fbu, Hiller:2019mou} for details. 
For simplicity, we  fix $\kappa/\kappa^\prime = 10^{-2}$, consistent
with $Z$-decay constraints and both AMMs.

Let us briefly comment on the scalar sector of the BSM framework \cite{Litim:2015iea,Hiller:2020fbu, Hiller:2019mou}. The presence of $H$ and the flavorful scalar matrix field $S_{ij}$ allows for a substantial scalar potential with in total three quartic couplings plus a  Higgs portal one $\delta S^\dagger S H^\dagger H$. In addition to successful electroweak symmetry breaking 
 the diagonal entries of  $S$ can acquire a non-trivial vev, $v_s$. 
Interestingly, two different configurations exist: A universal ground state in which diagonal entries have the same vev, and one in which the vev points in a single flavor direction,
breaking universality spontaneously.
Both the portal $\delta$ and $v_s$ are instrumental in achieving the  chirally enhanced 1-loop contributions explaining  AMMs. However, as discussed in the next section, the impact of $\delta$, $v_s$ on the present collider study is negligible,  and we do not consider them in this work.

To summarize, in the following we perform a collider analysis in the two models, one with three flavors of singlet VLLs  (\ref{Lint-singlet}), and one with three flavors of doublet VLLs  (\ref{Lint-doublet}), featuring  nine flavored scalar singlets and with  parameters
\begin{align} \label{eq:para}
M_S, M_F \, , \kappa/\kappa^\prime=10^{-2}\, , \kappa^\prime=\kappa^\prime(M_S,M_F) \, .
\end{align}
Using \eqref{Amu} together with the muon AMM data to express $\kappa^\prime$ in terms of $M_S$ and $M_F$ renders the numerical predictions for
production and decay of the BSM sector in terms of the latter two masses.
We discuss more general settings  in Sec.~\ref{sec:beyondg2}

The reduction of the models' BSM parameter space (i.e.~BSM masses, Yukawas, and quartic couplings)  onto the set  (\ref{eq:para})  is sufficient to catch the leading degrees of freedom for a $pp$-collider study, and to validate the models up to the TeV energy range. To further demonstrate that all couplings reach the Planck scale without poles or instabilities requires a complete renomalization group analysis. This has been done previously for a wide range of BSM parameters using $M_F=2 M_S=1$ TeV  
\cite{Hiller:2020fbu,Hiller:2019mou}. In general, not every point in the BSM parameter space is guaranteed to be Planck safe. Still, since the widening of the mass range towards  $M_S, M_F$ within  0.1 to 1 TeV has only a minor effect on the RG running, we can find Planck safe trajectories within suitable ranges for the remaining BSM Yukawas and quartics, using the methods of   \cite{Hiller:2020fbu,Hiller:2019mou}. This completes the discussion of our setup.

\section{LHC production and decay \label{sec:prod} }

\begin{figure*}
	\centering
	\begin{minipage}{0.32\textwidth}
		\includegraphics[scale=1]{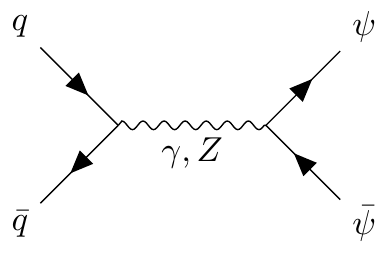}\\
		\centering\footnotesize{$a)$}
	\end{minipage}
	\begin{minipage}{0.32\textwidth}
		\includegraphics[scale=1]{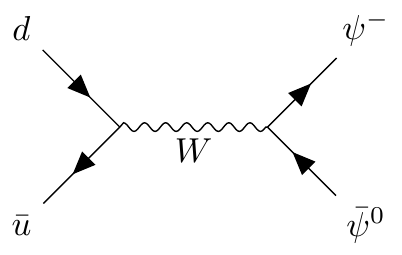}\\
		\centering\footnotesize{$b)$}
	\end{minipage}\\ \vspace{1em}
	\begin{minipage}{0.32\textwidth}
	\includegraphics[scale=1]{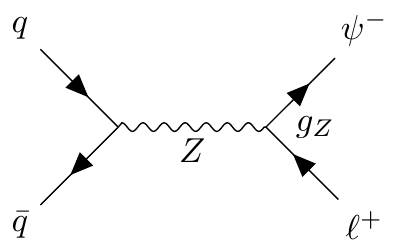}\\
	\centering\footnotesize{$c)$}
	\end{minipage}
	\begin{minipage}{0.32\textwidth}
	\includegraphics[scale=1]{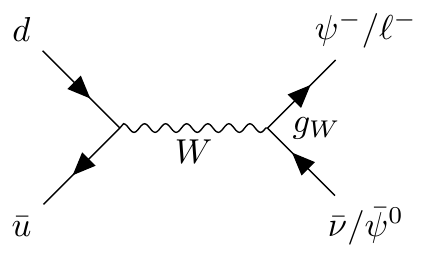}\\
	\centering\footnotesize{$d)$}
	\end{minipage}
 	\caption{Dominant pair (upper plots) and single (lower plots) production channels of vector-like leptons at $pp$ colliders. 
	Diagrams $c)$ and $d)$ involve the couplings $g_Z$ and $g_W$, which are  induced by  fermion mixing and  $\kappa$ (\ref{Lint-singlet}), (\ref{Lint-doublet}). In diagram $d)$, the final states $\psi^-\overline{\nu}$ ($\ell^-\overline{\psi}^0$) are only possible in the singlet (doublet) model.}
	\label{fig:production}
\end{figure*}

At the LHC, VLLs can be produced in pairs (upper plots) or singly (lower plots) in quark fusion through electroweak
 interactions shown in Fig.~\ref{fig:production}. 
Pair production occurs through $s$-channel photon or $Z$ (Fig.\ref{fig:production}a), and in the doublet model additionally through $s$-channel $W$ exchange (Fig.\ref{fig:production}b). 
Single production  is induced by the Yukawa portal coupling $\kappa$
through fermion mixing and $Z, W$-exchange (Fig.\ref{fig:production}c, d). %, respectively.
All three flavors are produced universally.
Additional contributions to VLL production arise through $s$-channel Higgs and BSM scalars $S_{ii}$ induced by Higgs-scalar mixing (not shown). 
Due to both quark-Yukawa and parton-luminosity suppression these contributions to matrix elements are suppressed by at least two orders of magnitude with respect to electroweak contributions and thus negligible. 
Further production channels through Yukawa interactions open up at lepton colliders, briefly discussed in  \cite{Hiller:2020fbu}.

\begin{figure}
	\centering
	\includegraphics[width=0.49\textwidth]{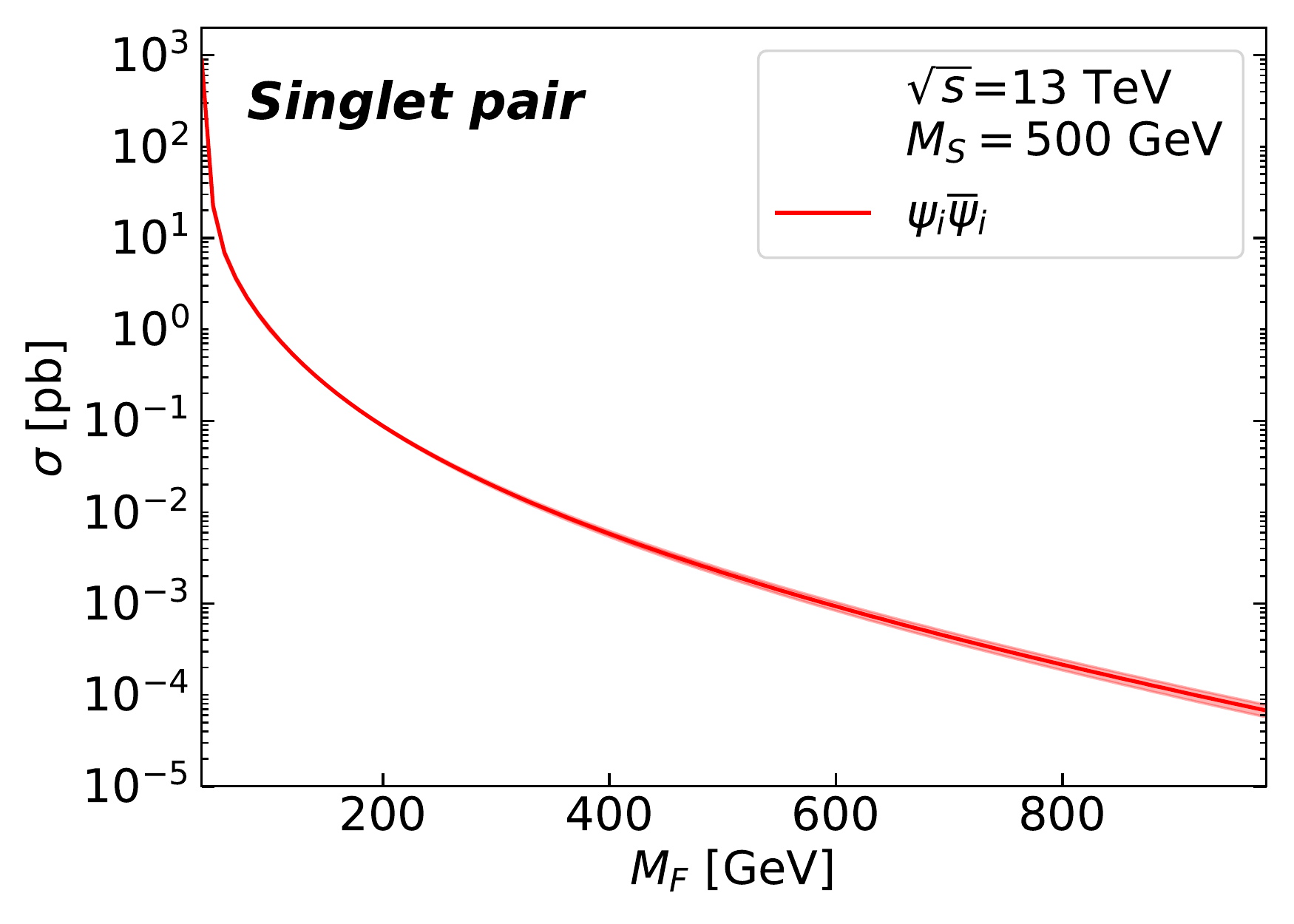}
	\includegraphics[width=0.49\textwidth]{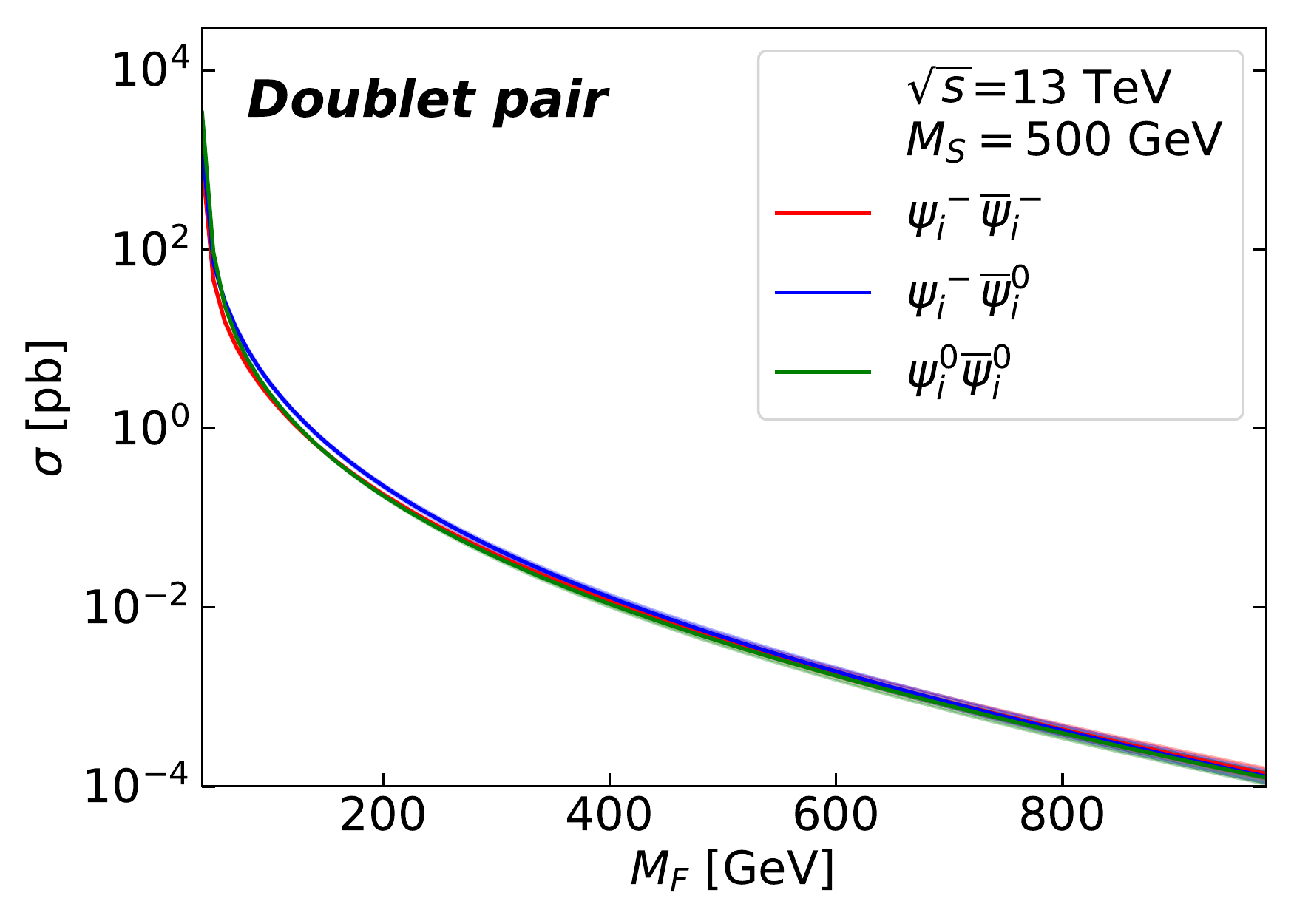} \\
	\includegraphics[width=0.49\textwidth]{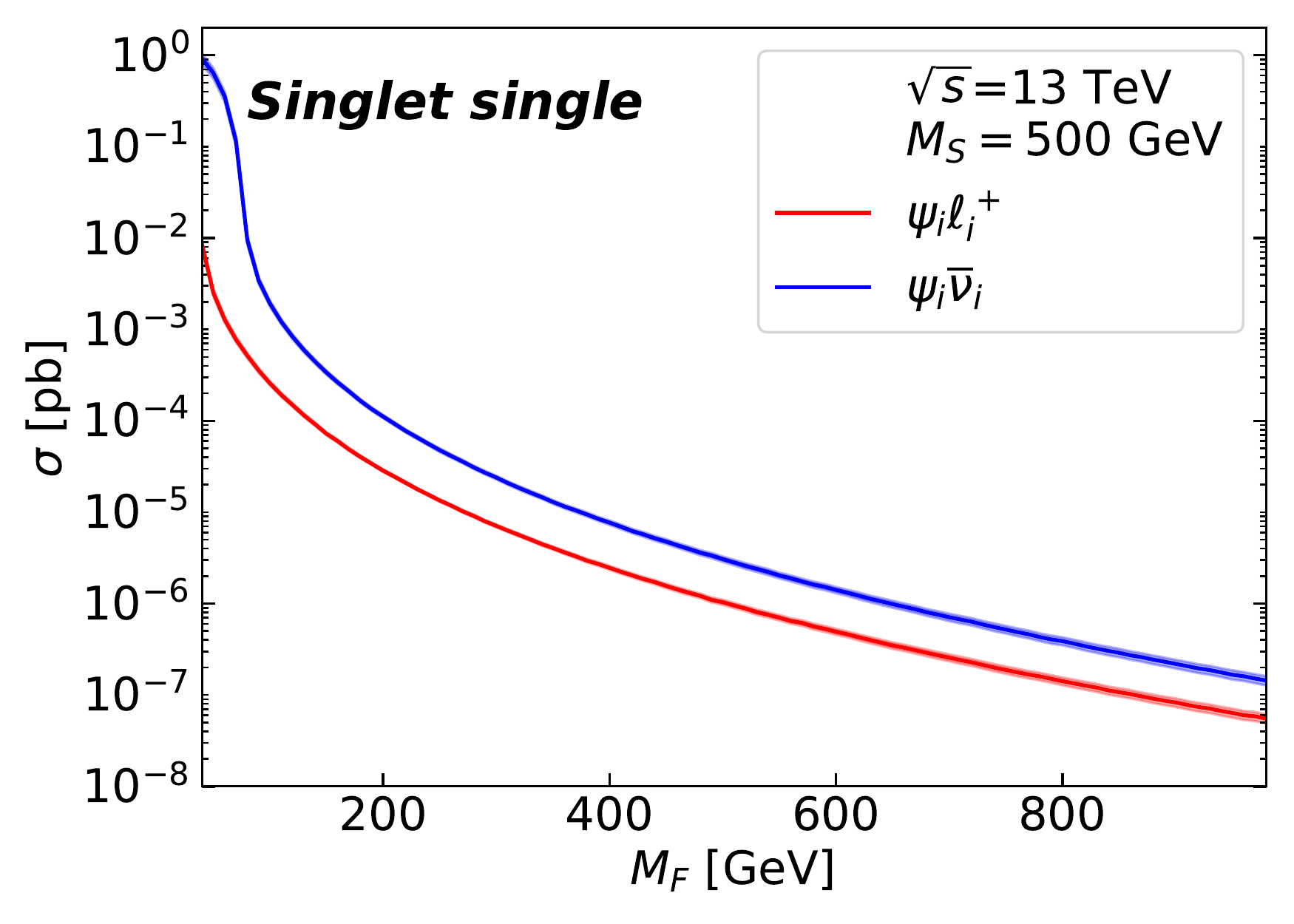}
	\includegraphics[width=0.49\textwidth]{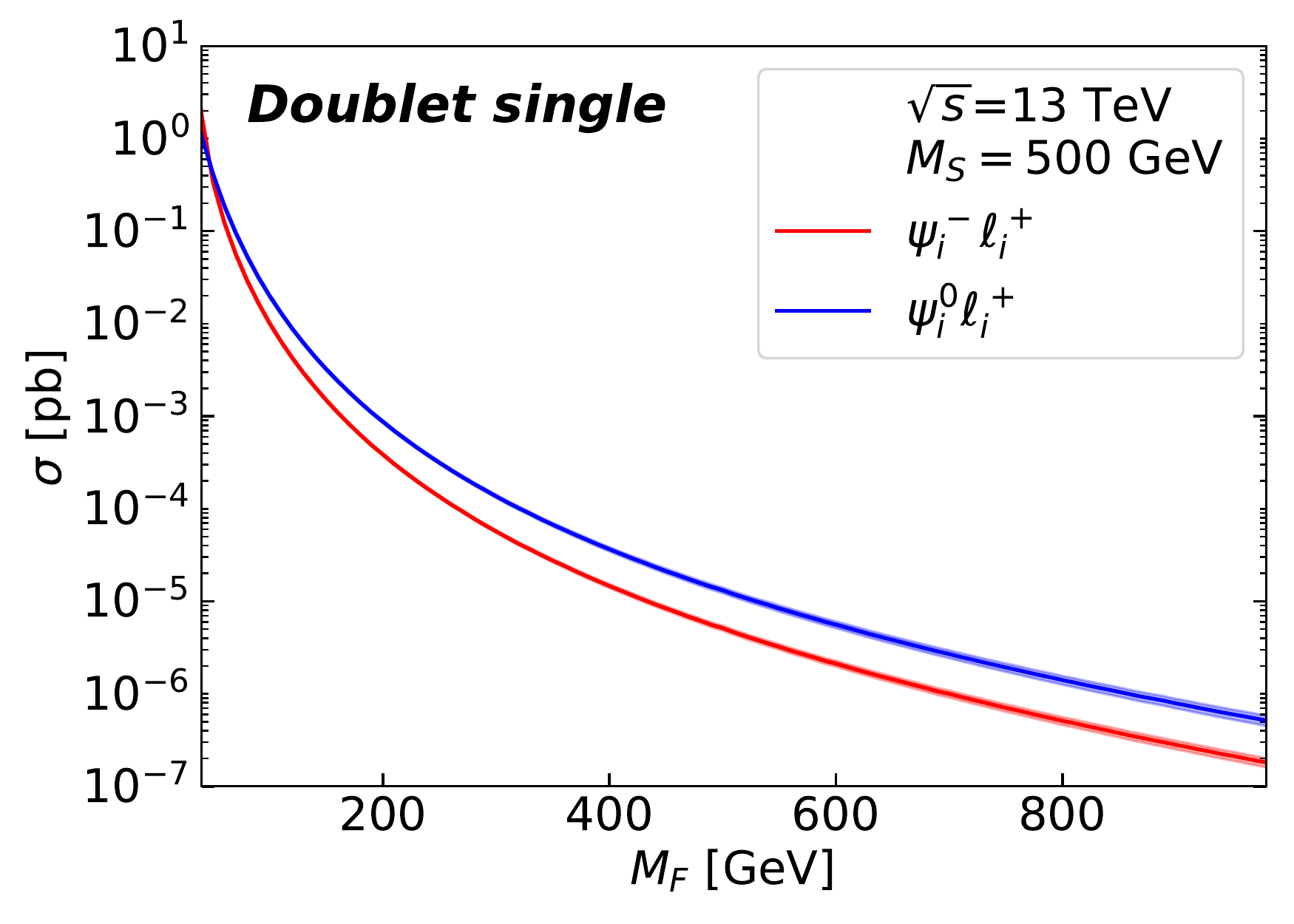}\\
	\caption{Cross sections for $\psi_i$ pair production (top) and single production (bottom) at $\sqrt{s} = 13~$ TeV for different vector-like lepton masses in the singlet model (left) and the doublet model (right) for $M_S=500$ GeV and the procedure described in \eqref{eq:para}.}
	\label{Fig:Cross_sec_LHC}
\end{figure}

In Fig.~\ref{Fig:Cross_sec_LHC} we show pair- and single-production cross sections for a single species  $\psi_i$ -- with lepton flavor index $i$ fixed --  at the LHC with $\sqrt{s}=13$ TeV as a function of $M_F$ for $M_S = 500$ GeV and the procedure in \eqref{eq:para}.
In both the singlet (left) and doublet model (right) the pair-production cross section is roughly three orders of magnitude larger than the single production cross section. 
This is due to the fact that single production is only induced by mixing between SM leptons and VLLs, while
 dominant pair production of VLLs  at the LHC occurs through electroweak gauge interactions.
$\kappa^\prime$, the larger of the BSM Yukawa couplings, is irrelevant also  for single  production, but  turns out to be important for BSM sector decays.

\begin{figure}
	\centering
	\includegraphics[width=0.49\textwidth]{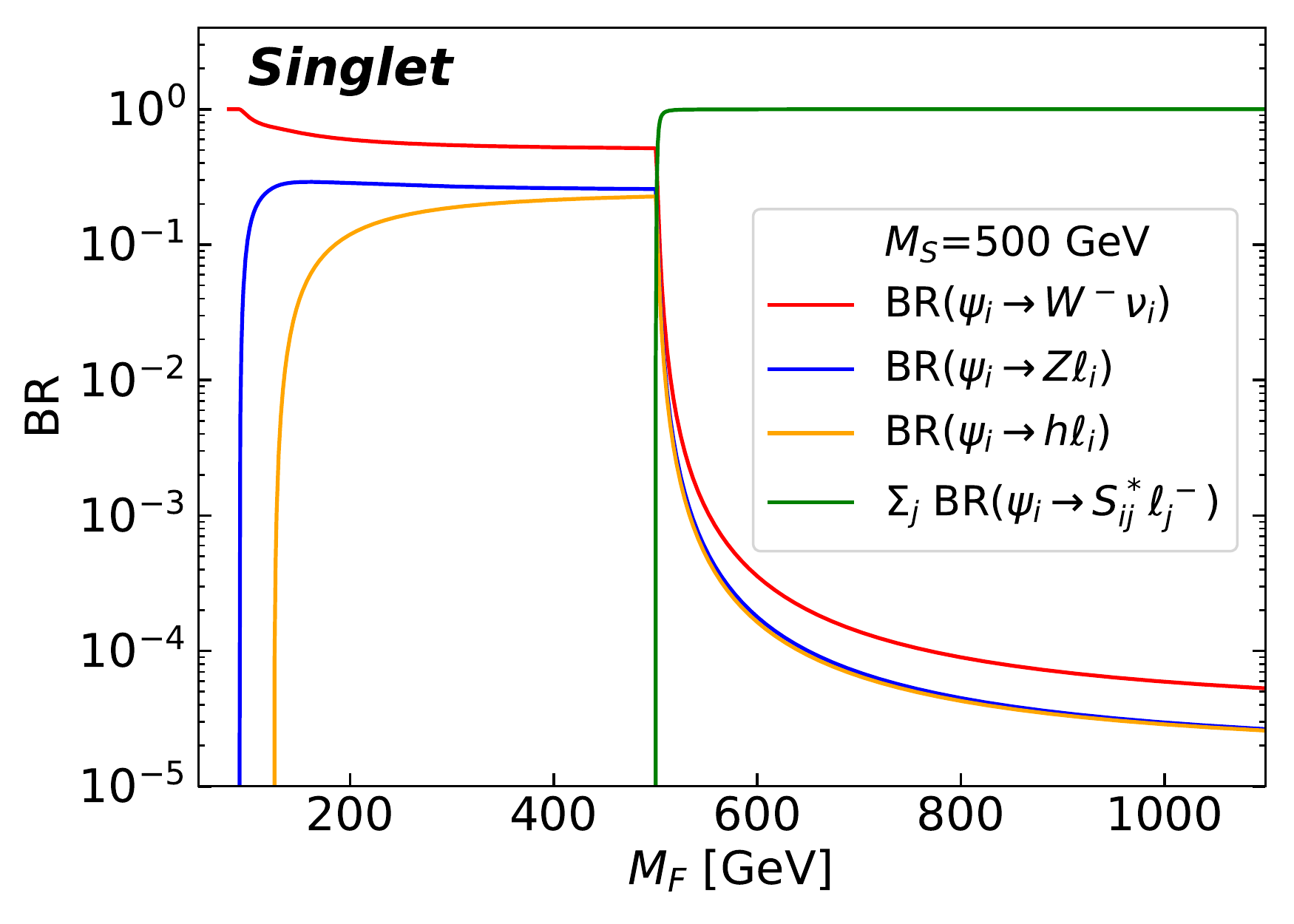}
	\includegraphics[width=0.49\textwidth]{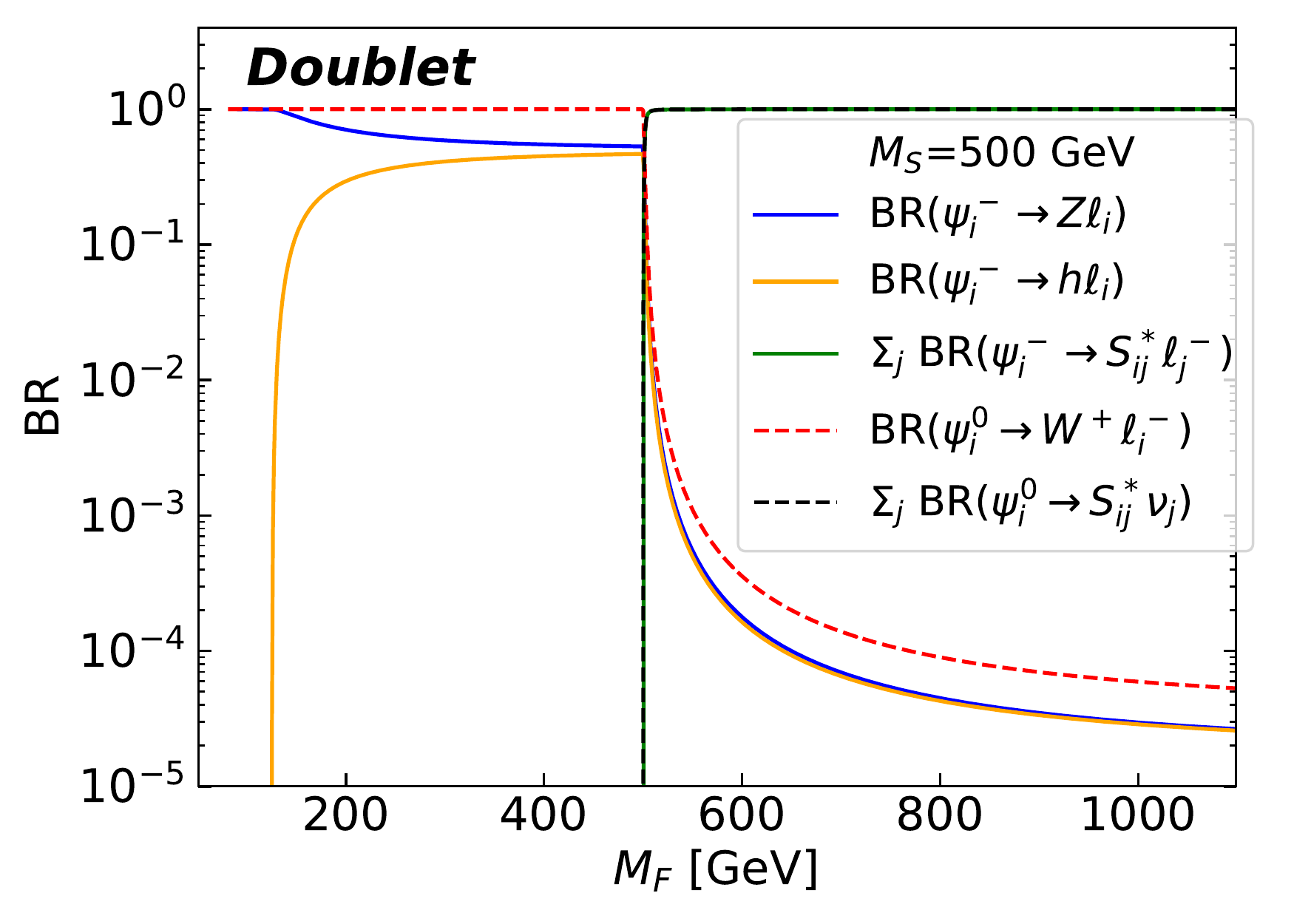}
	\caption{Branching ratios of the on-shell decays of the VLLs as a function of their mass in the singlet model (left) and the doublet model (right) for $M_S =500$~GeV
	and \eqref{eq:para}.
	Larger ratios $\kappa/\kappa'$ would enhance the  branching ratios of the electroweak decays.} 
	\label{Fig:BR_psi}
\end{figure}

In the singlet model, the decay rates of the possible decay channels of the VLLs are
\begin{equation}\label{rates-singlet}
\begin{aligned}
\Gamma(\psi_i \to h \ell_i ^-) & =\kappa^2 \frac{M_F}{64 \pi} (1 - r_h^2)^2  \,, \\
\Gamma(\psi_i  \to S^{\,*}_{ij}\, \ell^-_j) &= \kappa^{\prime\,2} \frac{M_F}{32 \pi} (1 - r_S^2)^2  \,,~~~~~( j ~\mbox{fixed}) \\
\Gamma(\psi_i \to W^- \nu_i  )&= g_W^2 \frac{M_F}{32 \pi} (1 - r_W^2)^2(2+1/r_W^2)\,,\\
\Gamma(\psi_i  \to Z \ell^-_i ) &= g_Z^2\,  \frac{M_F}{32 \pi} (1 - r_Z^2)^2(2+1/r_Z^2)\,,
\end{aligned}
\end{equation}
where $r_X = M_X/M_F$. For large  values of $\kappa'$ the decay $\psi_i  \to S^{\,*}_{ij}\, \ell^-_j$ dominates if kinematically allowed, as seen in Fig.~\ref{Fig:BR_psi} (left). Quantitatively, in the large-$M_F$ limit, the decays through the $S_{ij}$ dominate over Higgs-mediated decays (decays through weak bosons) for  $\kappa'\gtrsim \kappa/\sqrt{6}$ ($\kappa'\gtrsim \kappa / \sqrt{3}$ ).  In the doublet model, we obtain the decay rates
\begin{equation}\label{rates-doublet}
\begin{aligned}
\Gamma(\psi_i^- \to h \ell_i ^-) & =\kappa^2 \frac{M_F}{64 \pi} (1 - r_h^2)^2  \,, \\
\Gamma(\psi_i^-  \to S_{ji}\, \ell^-_j) &= \kappa^{\prime\,2} \frac{M_F}{32 \pi} (1 - r_S^2)^2   \,,~~~~~( j ~\mbox{fixed}) \\
\Gamma(\psi_i^0  \to S_{ji}\, \nu_j) &= \kappa^{\prime\,2} \frac{M_F}{32 \pi} (1 - r_S^2)^2  \,,~~~~~( j ~\mbox{fixed}) \\
\Gamma(\psi_i^-  \to Z \ell^-_i ) &= g_Z^2\,  \frac{M_F}{32 \pi} (1 - r_Z^2)^2(2+1/r_Z^2)\,,\\
\Gamma(\psi_i^0 \to W^+ \ell^-_i  )&= g_W^2 \frac{M_F}{32 \pi} (1 - r_W^2)^2(2+1/r_W^2)\,.\\
\end{aligned}
\end{equation}
The corresponding branching ratios  of the $\psi^-$ and the $\psi^0$ decays are shown in Fig.~\ref{Fig:BR_psi} (right). As in the singlet model, the decays to BSM scalars dominate for large $\kappa'$ if allowed by the mass hierarchy of the BSM sector. 
As already stated in the previous section we assume that the $\psi^-$ and $\psi^0$ are degenerate in mass;  we therefore neglect small isospin splitting 
induced by electromagnetic interaction
$\Delta m =  M_{\psi^{-1}} -M_{\psi^{0}}=g^2/(4 \pi)  \sin^2 \theta_W
M_Z/2    \simeq 0.4$ GeV, that also allows for rare  inter-multiplet decays $\psi^- \to \psi^0 W^{- \ast}$. 
The smallness of the  splitting prohibits that for instance searches in R-parity violating SUSY models into four light leptons \cite{Aaboud:2018zeb} apply to the VLL models.

The decays of VLLs to $S_{ij}$ plus lepton are a singular feature of the models, which distinguishes them from other theories with VLLs, such as \cite{Kumar:2015tna,Crivellin:2018qmi}. 
Moreover, the $S_{ij}$ can decay  through fermion mixing to lepton final states, in which they can be searched for.
Specifically, the singlet model  features the cascade decays
\begin{equation}\label{lfv-like-decay-singlet}
\psi_i \to S_{ij}^*\, \ell^-_j \to \ell_i^- \, \ell_j^+ \, \ell^-_j\,.
\end{equation}
 Similarly, in the doublet model the decays of the charged and neutral VLLs  proceed as
\begin{equation}\label{lfv-like-decay-doublet}
\begin{aligned}
&\psi_i^- \to S_{ji}\, \ell^-_j \to \ell_i^- \, \ell_j^+ \, \ell^-_j\,,
&&\psi_i^0 \to S_{ji}\, \nu_j \to \ell_i^- \, \ell_j^+ \, \nu_j\,.
\end{aligned}
\end{equation}
These processes  preserve flavor; however, the scalar decay yields a dilepton pair with different-flavor charged leptons for $i\neq j$,
 which looks  as if  lepton flavor has been violated and  cleanly signals new physics.
 While the scalars may also decay to dibosons through triangle loops, or to two VLLs through the coupling $y$ if $M_S< 2M_F$, see (\refeq{Yukawa}) and \cite{Hiller:2020fbu} for details, here we assume that these rates are negligible.

In this work we are  interested in final states with at least four light leptons (4L), where a light lepton is an electron or a muon, as in \cite{Sirunyan:2019ofn}. When the $\psi_i$ are pair-produced and decay through Eqs.~\eqref{lfv-like-decay-singlet} and \eqref{lfv-like-decay-doublet}, only certain flavor final states of each single decay can contribute to a 4L final state. These are given in Tab.~\ref{Tab:decays}.
Notice that the decay chains \eqref{lfv-like-decay-singlet}, \eqref{lfv-like-decay-doublet} allow to observe resonance structures from $S_{ij}$-decays in  clean different-flavor dilepton invariant mass distributions if the VLLs are sufficiently heavy, $M_F > M_S$.
We exploit this possibility  in Sec.~\ref{sec:Null-test}.

\begin{table}
	\centering
	\begin{tabular}{p{2.5cm}p{5cm}}\hline
		State	&	Decay modes \\\hline
		$\psi_1^{(-)}$ & $e^-e^+e^-, \ e^-\mu^+\mu^-, \ e^-\tau^+\tau^-$	\\
		$\psi_2^{(-)}$ & $\mu^-\mu^+\mu^-, \ \mu^-e^+e^-, \ \mu^-\tau^+\tau^-$	\\
		$\psi_3^{(-)}$ & $\tau^-e^+e^-, \ \tau^-\mu^+\mu^-$	\\
		$\psi_1^{0}$ & $e^-e^+\nu_e \,, \ e^-\mu^+\nu_\mu$	\\
		$\psi_2^{0}$ & $\mu^-\mu^+\nu_\mu  \,, \ \mu^-e^+\nu_e$
			\\ \hline
	\end{tabular}
	\caption{Decay modes of the VLLs through the $S$ scalars, \eqref{lfv-like-decay-singlet} and \eqref{lfv-like-decay-doublet}, giving  rise to a final state with four light leptons after $\bar \psi^{(-)}_i \psi^{(-)}_i$ or $\bar \psi^0_i \psi^-_i$ pair production. For the third generation $\psi_3^0$  no corresponding 4L final states arise.}
	\label{Tab:decays}
\end{table}

\section{Event simulation}\label{sec:Simu}

In this section  we describe the procedure used to generate a sample of events with 4L final states at the LHC.
We employ \textsc{FeynRules} \cite{Alloul:2013bka} to compute the Feynman rules at leading order (LO) for the models in Eqs.~\eqref{Lint-singlet} and \eqref{Lint-doublet}. The particles and Feynman rules are then implemented into UFO models \cite{Degrande:2011ua}.  
These UFO models are interfaced to the Monte Carlo generator \MG \cite{Alwall:2014hca} to compute production cross sections of single and pair production of VLLs at the LHC as well as distributions of final state particles at parton level. For each process we generate $5\times 10^4$ events. The decay of particles is handled with \textsc{MadSpin} \cite{Artoisenet:2012st}.
For the event generation the \textsc{NNPDF3.0} \cite{Ball:2013hta} PDF set is used. 
PDF and scale variation uncertainties are computed within \MG for each of the PDF sets. The scale variation uncertainties are computed by varying factorization and renormalization scales independently between $0.5\mu_0 \leq \mu_\textmd{F/R}\leq 2\mu_0$, where the scale $\mu_0$ is computed in \MG with different schemes. 
For the theory uncertainty we add PDF uncertainties, scale variation uncertainties and scheme variation uncertainties in quadrature.

For the event generation we adapt settings similar to the  CMS study \cite{Sirunyan:2019ofn}. We focus on the final states with at least four light leptons, that is, 
muons and electrons and require the missing transverse momentum, $p_T^\text{miss}$, to be smaller than $50$~GeV. 
This cut  serves to resemble the signal region considered by CMS  and to suppress contributions from neutrinos in the decay of the electroweak bosons.
Electrons and muons are required to have a minimal transverse momentum of $p_T^\ell\geq 20$~GeV. 
Similarly to the CMS analysis, we neglect all events with a light-lepton invariant mass, $m_{\ell\ell}$, smaller than $12~$GeV for all flavor and charge combinations. 
This cut serves to suppress resonances in the low-mass region.
 For the event generation we fix $\kappa=10^{-2}\kappa^\prime$. 
Masses of the new scalars and VLLs are varied between $M_S=300 - 1200$~GeV and $M_F = 100-1000$~GeV with $\kappa'$ computed according to Eq.~\eqref{Amu}. 
We also consider $ttZ$, triboson and $ZZ$ production as these processes contribute to the SM background for the distributions studied in \cite{Sirunyan:2019ofn}. 
The $ZZ$ production includes contributions from virtual photons via $p p \to \gamma^* \gamma^*,\gamma^* Z$. 
$ZZj$ final states are included via multijet merging in \textsc{PYTHIA8} \cite{Sjostrand:2014zea}.
We also take into account in the cross section gluon-fusion contributions $g g \to ZZ$, where the lowest order is induced at 1-loop.
SM background processes are computed at LO within \MG using the same set of PDF sets. Higher order corrections to SM production cross sections are taken from literature \cite{deFlorian:2016spz,Campanario:2008yg,Hankele:2007sb,Lazopoulos:2007ix,Binoth:2008kt,Cascioli:2014yka,Frixione:2015zaa, Caola:2015psa} and are taken into account by applying $k$ factors to the LO distributions.
To perform a simulation of the detector response we shower and hadronize the events with \textsc{PYTHIA8}  
and use \textsc{DELPHES3} \cite{deFavereau:2013fsa} for the fast detector simulation, yielding events at particle level. 
Jets are clustered with the anti-$k_t$ algorithm \cite{Cacciari:2008gp} with a radius parameter $R=0.5$ applying the \textsc{FastJet} package \cite{Cacciari:2011ma}. 
All criteria for the analysis are taken from the CMS default card for simplicity. 
In Tab.~\ref{Tab:Param} we summarize the values for the parameters and signal selection cuts used in the event generation and detector simulation.

\begin{table}
	\centering
	\begin{tabular}{p{5.5cm}p{3.5cm}p{3.5cm}}\hline
		Parameters	&	Signal selection	&	Reconstruction \\\hline
		$\alpha_s(M_Z) = 0.118$	&	$p_T^{\rm miss} < 50~$GeV	&	$\Delta M_Z=15~$GeV	\\
		$m_b = 4.7~$GeV &	$|\eta|\leq 2.5$	&	$\Delta M_S=5~$GeV	\\
		$M_Z = 91.188~$GeV	&	$R=0.5$	&	$\Delta M_H=5~$GeV	\\
		$M_h = 125~$GeV		&	$N_{\ell} \geq 4$	&	$\Delta M_F=100~$GeV	\\
		$M_W=80.379~$GeV	&	$p_T^{\rm jet} \geq 20~$GeV	&	-	\\
		$m_t = 172~$GeV		&	$p_T^{\ell} \geq 20~$GeV	&	-	\\ 
		-   &   $m_{\ell\ell}\geq12~$GeV    &   -\\
		\hline
	\end{tabular}
	\caption{Parameters used in the event generation, detector simulation and the reconstruction algorithm.}
	\label{Tab:Param}
\end{table}

\section{Constraints from  CMS data}\label{sec:results}

Here we  confront the  models \eqref{Lint-singlet}, \eqref{Lint-doublet} to the  CMS search \cite{Sirunyan:2019ofn} using the 4L final state, which is expected to be the channel most sensitive to contributions stemming from three generations of VLLs. 
In Sec.~\ref{sec:4L}, we study decay chains into 4L final states and their multiplicities. 
In  Sec.~\ref{sec:LT} we  compare the  distributions of the scalar sum of transverse momenta of the four light leptons $(e, \mu)$ with the largest transverse momenta, $L_T$, with CMS data to  obtain constraints on BSM masses.

\subsection{ 4L multiplicities   \label{sec:4L}}

4L final states stem from both single and pair production of VLLs. Due to the flavor structure of the BSM sector, the following decay chains include  4L final states
in the singlet model:
\begin{align}
\begin{aligned}\label{decaychain-singlet}
p p &\rightarrow \psi_i \bar \psi_i \rightarrow \ell_i^- \ell_i^+ \ell_j^+\ell_j^- \ell_k^+\ell_k^- \quad \text{for $i,j,k=1,2,3$} \,, {\ (20)}\\
p p &\rightarrow \psi_i \bar \psi_i \rightarrow \ell_i^- \ell_i^+ q_j \bar q_j \ell_k^+\ell_k^- \quad \text{for $i,k=1,2$}\,, {\ (15\times 4)}\\
p p &\rightarrow \psi_i \bar \psi_i \rightarrow \ell_i^- \ell_i^+ \ell_j^+\ell_j^- \nu_k \bar \nu_k  \quad \text{for $i,j=1,2$, $k=1,2,3$} \,,{\ (12)}\\
p p &\rightarrow \psi_i \bar \psi_i \rightarrow \nu_i \ell_i^+ \ell_j^+\ell_j^- \ell_k^- \bar \nu_k  \quad \text{for $i,j,k=1,2$}\,, {\  (8)}\\
p p &\rightarrow \psi_i \bar \psi_i \rightarrow \ell_i^- \bar\nu_i \ell_j^+\ell_j^- \ell_k^+ \nu_k  \quad \text{for $i,j,k=1,2$}\,,{\  (8) }\\
p p &\rightarrow \psi_i \ell_i^+ \rightarrow \ell_i^-\ell_j^+\ell_j^- \ell_i^+ \quad \text{for $i,j=1,2$}\,, {\ (4)}\\
p p &\rightarrow \bar\psi_i \ell_i^- \rightarrow \ell_i^+\ell_j^+\ell_j^- \ell_i^- \quad \text{for $i,j=1,2$}\,, {\ (4)} \\
\end{aligned}
\end{align}
 where $i\,,j\,,k$ are flavor indices and $q_i={u,d,c,s,b}$. We also indicate the values that the lepton flavor indices can take, and between  parentheses the number of 4L final states of each chain after summing over all indices. Note that, for the first decay chain in~\eqref{decaychain-singlet}, final states with four light leptons occur only when at most one of the three indices $i,j,k$ is equal to 3. For explicit expressions of flavors in the decays see Tab.~\ref{Tab:decays}.
 
  In the doublet model, the negatively charged state $\psi^-$ decays into 4L final states as in~\eqref{decaychain-singlet} with the exception of the
  decays with 8-fold multiplicity. These correspond to  $W$-mediated decays $\psi_i^- \rightarrow \nu_i \ell^-_j \bar \nu_j$, which 
  are subleading in the doublet model.
 Additionally, when the $\psi^0$ are produced, 4L final states arise through
\begin{align}
\begin{aligned}\label{decaychain-doublet}
p p &\rightarrow \psi_i^0 \overline{\psi}^0_i \rightarrow \nu_j \overline{\nu}_k \ell_j^+\ell_i^- \ell_i^+\ell_k^- \quad \text{for $i,j {,k }=1,2$}\,, {\ (8)}\\
p p &\rightarrow \psi_i^- \overline{\psi}^0_i \rightarrow \ell_i^- \ell_i^+  \overline{\nu}_j \ell_j^- \ell_k^+\ell_k^- \quad \text{for $i,j{,k }=1,2$}\,,{\ (8)}\\
p p &\rightarrow \psi_i^0 {\psi}^+_i \rightarrow \ell_i^- \ell_i^+ \ell_j^+\nu_j \ell_k^+\ell_k^- \quad \text{for $i,j{,k }=1,2$}\,,{\ (8)}\\
p p &\rightarrow \psi_i^- \overline{\psi}^0_i \rightarrow \ell_i^- \ell_i^+  \overline{q}_j q_j \ell_k^+\ell_k^- \quad \text{for $i{,k }=1,2$}\,,{\ (15\times 4)}\\
p p &\rightarrow \psi_i^0 {\psi}^+_i \rightarrow \ell_i^- \ell_i^+ \overline{q}_j q_j \ell_k^+\ell_k^- \quad \text{for $i{,k }=1,2$}\,.{\ (15\times 4)}\\
\end{aligned}
\end{align}
\begin{figure*}
	\centering
	\begin{minipage}{0.32\textwidth}
		\includegraphics[scale=0.7]{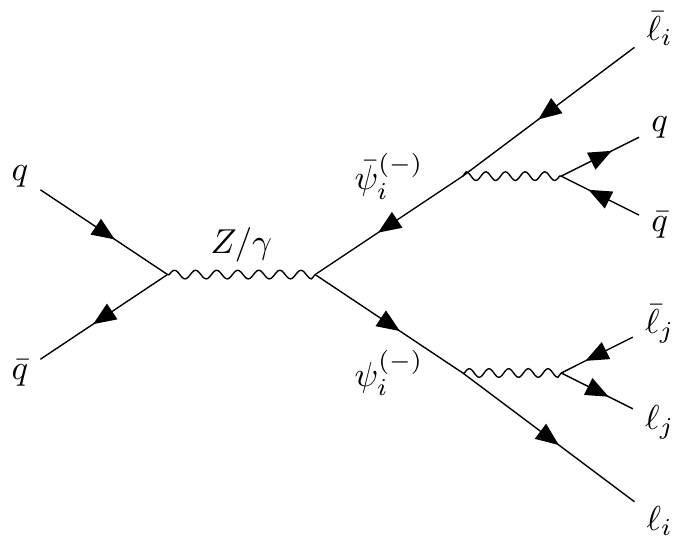}\\
		\centering\footnotesize{$a)$}
	\end{minipage}
	\begin{minipage}{0.32\textwidth}
		\includegraphics[scale=0.7]{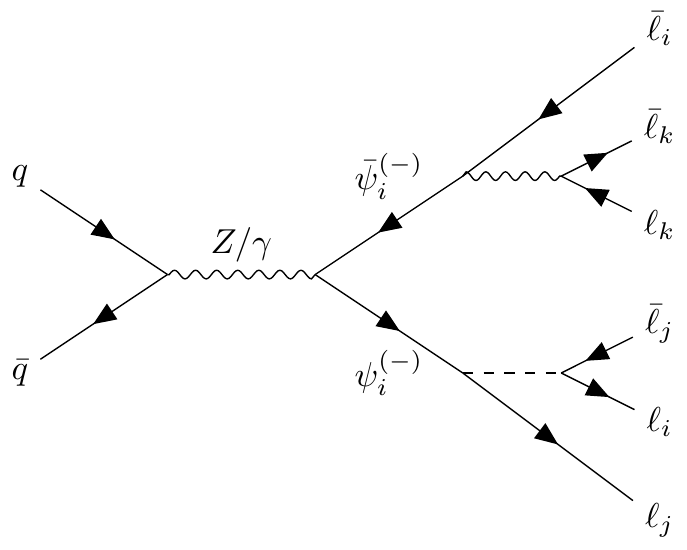}\\
		\centering\footnotesize{$b)$}
	\end{minipage}
	\begin{minipage}{0.32\textwidth}
		\includegraphics[scale=0.7]{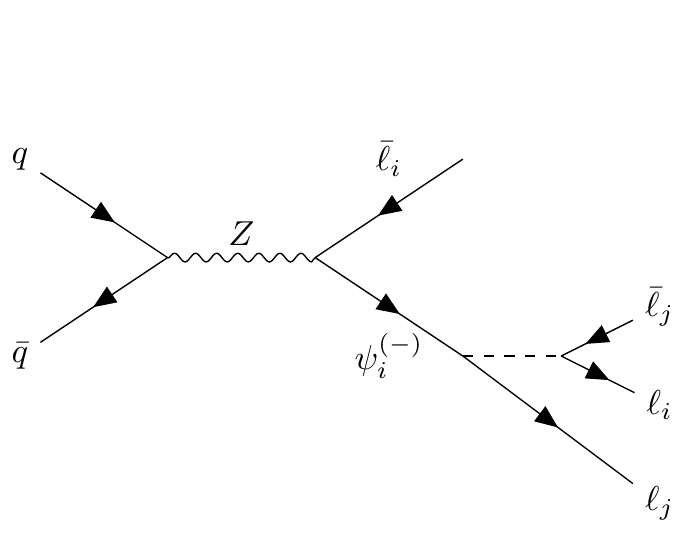}\\
		\centering\footnotesize{$c)$}
	\end{minipage}
	\caption{Examples of signal channel Feynman diagrams  with at least four light leptons in the final state  in the singlet \eqref{Lint-singlet} and doublet \eqref{Lint-doublet} model.
		Only first- and second-generation
		vector-like leptons can contribute via the diagram with jets in the final state (a) and the single production diagram (c). }
	\label{Fig:4ll_singlet}
\end{figure*}
The first decay chain in Eq.~\eqref{decaychain-singlet}, involving a six charged-lepton final state, is the only one where production of the third generation $\psi_3$ can give rise to a 4L final state. In all other cases, $\psi_3$ production  yields at most three light leptons in the final state, since a $\tau^+\tau^-$ pair is always produced due to flavor conservation. In Fig.~\ref{Fig:4ll_singlet} we give examples of Feynman diagrams for the different decay chains,  with jets (a) or without them (b), and from single production (c).

\begin{figure}
	\includegraphics[width=0.49\textwidth]{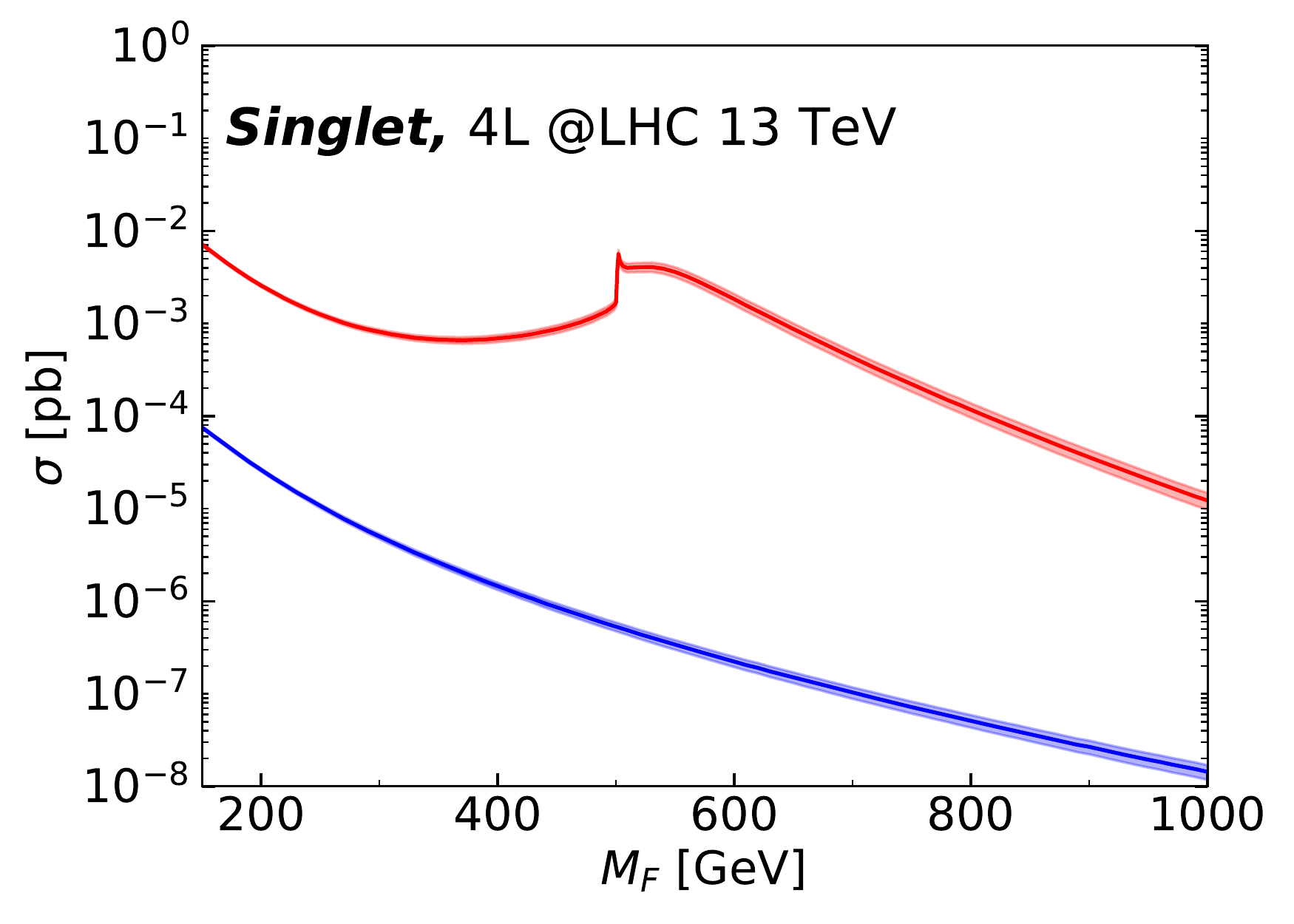}
	\includegraphics[width=0.49\textwidth]{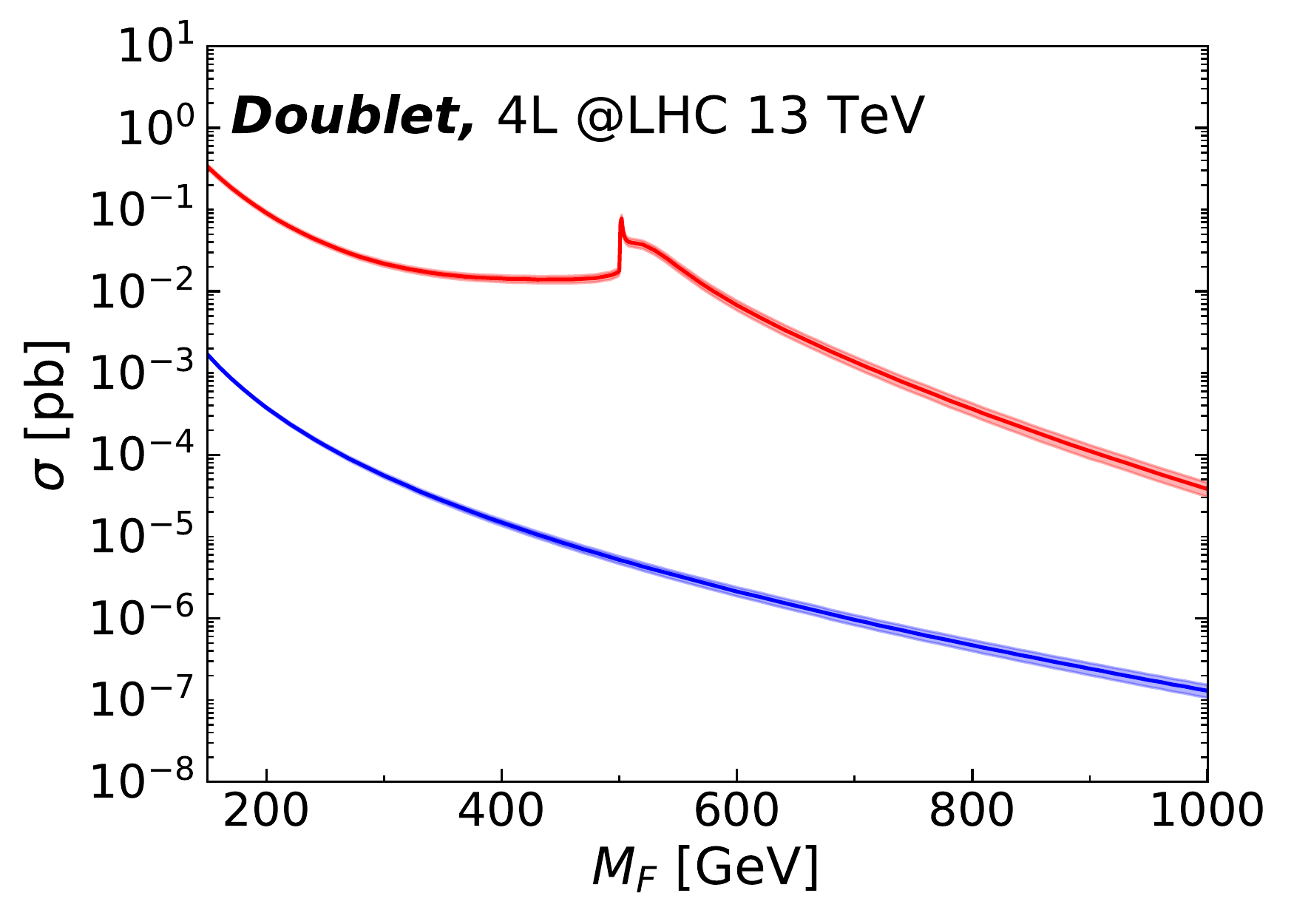}
	\caption{Cross section for BSM production of at least four light leptons at a $pp$ collider with $\sqrt{s} = 13$ TeV in the singlet (left) and the doublet (right) model 
	as a function of the VLL mass for  $M_S=500$~GeV. The red curves correspond to the VLL models~\eqref{Lint-singlet} and~\eqref{Lint-doublet}, while the blue curves correspond to third-generation VLL models as in \cite{Kumar:2015tna}. The band widths include uncertainties discussed in Sec.~\ref{sec:Simu}.}
	\label{Fig:4ll_production}
\end{figure}

In Fig.~\ref{Fig:4ll_production} we show the cross section at the $\sqrt{s} = 13$ TeV LHC for BSM production of at least four light leptons in terms of the VLL mass for the singlet model (left) and the doublet model (right) for $M_S=500$~GeV, together with cross sections of the models in  Ref.~\cite{Kumar:2015tna}. In general, our cross sections are larger by roughly two orders of magnitude.
This enhancement stems from the first and second generation of  VLLs, which present a richer multiplicity of decays into 4L final states than the  $\psi_{3}$. For $M_F < M_S$, the enhancement originates predominantly from the additional final states with two jets and four light leptons, while for $M_F > M_S$ cross sections increase further, up to a factor of approximately $10^4$. This  effect is caused by the VLLs  decaying mainly through on-shell production of scalars $S_{ij}$, see  Fig.~\ref{Fig:BR_psi}, and their subsequent decay  into light leptons.

\subsection{ $L_T$ distributions and CMS constraints \label{sec:LT}}

\begin{figure}
	\centering
	\includegraphics[width=0.49\textwidth]{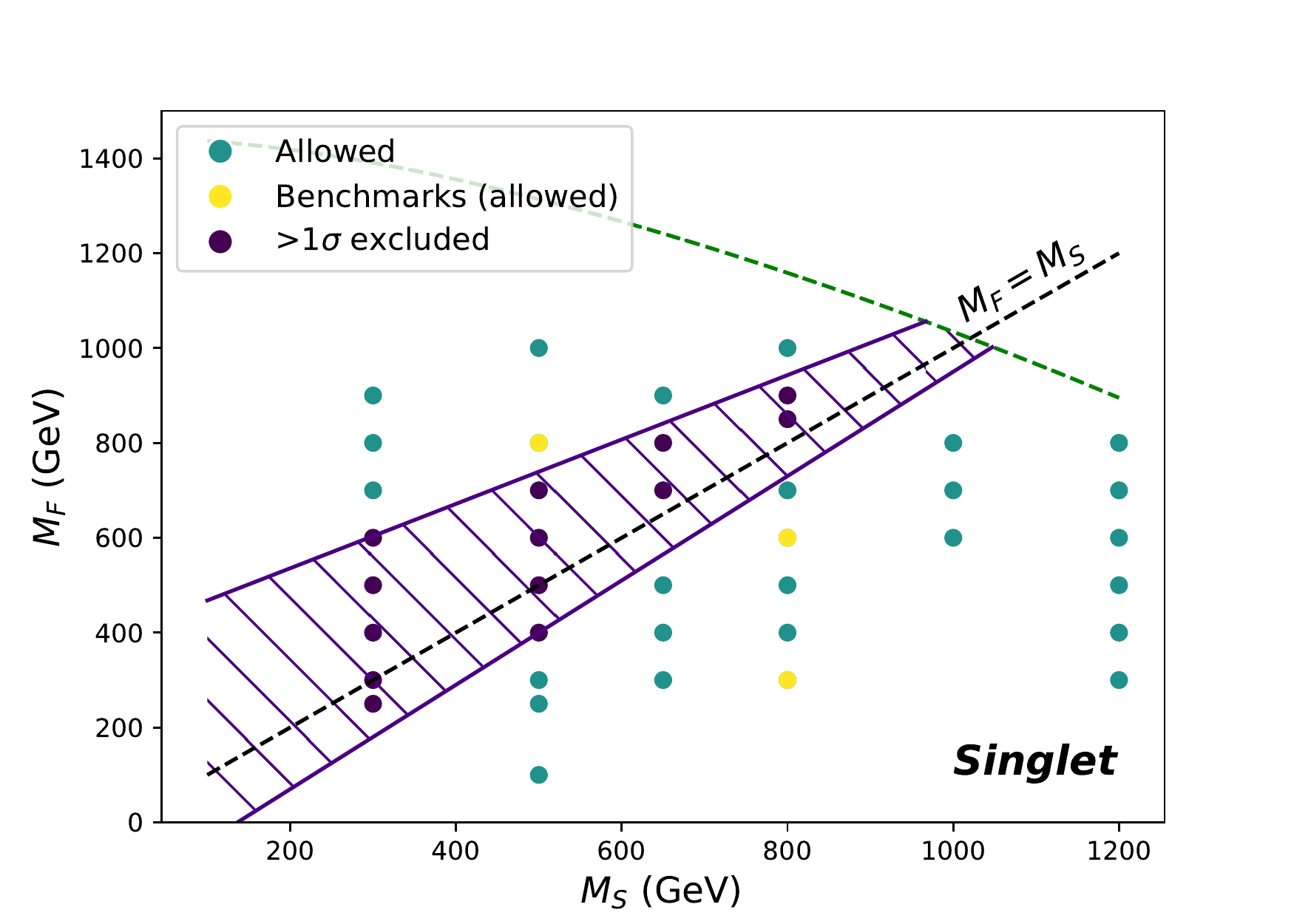}
	\includegraphics[width=0.49\textwidth]{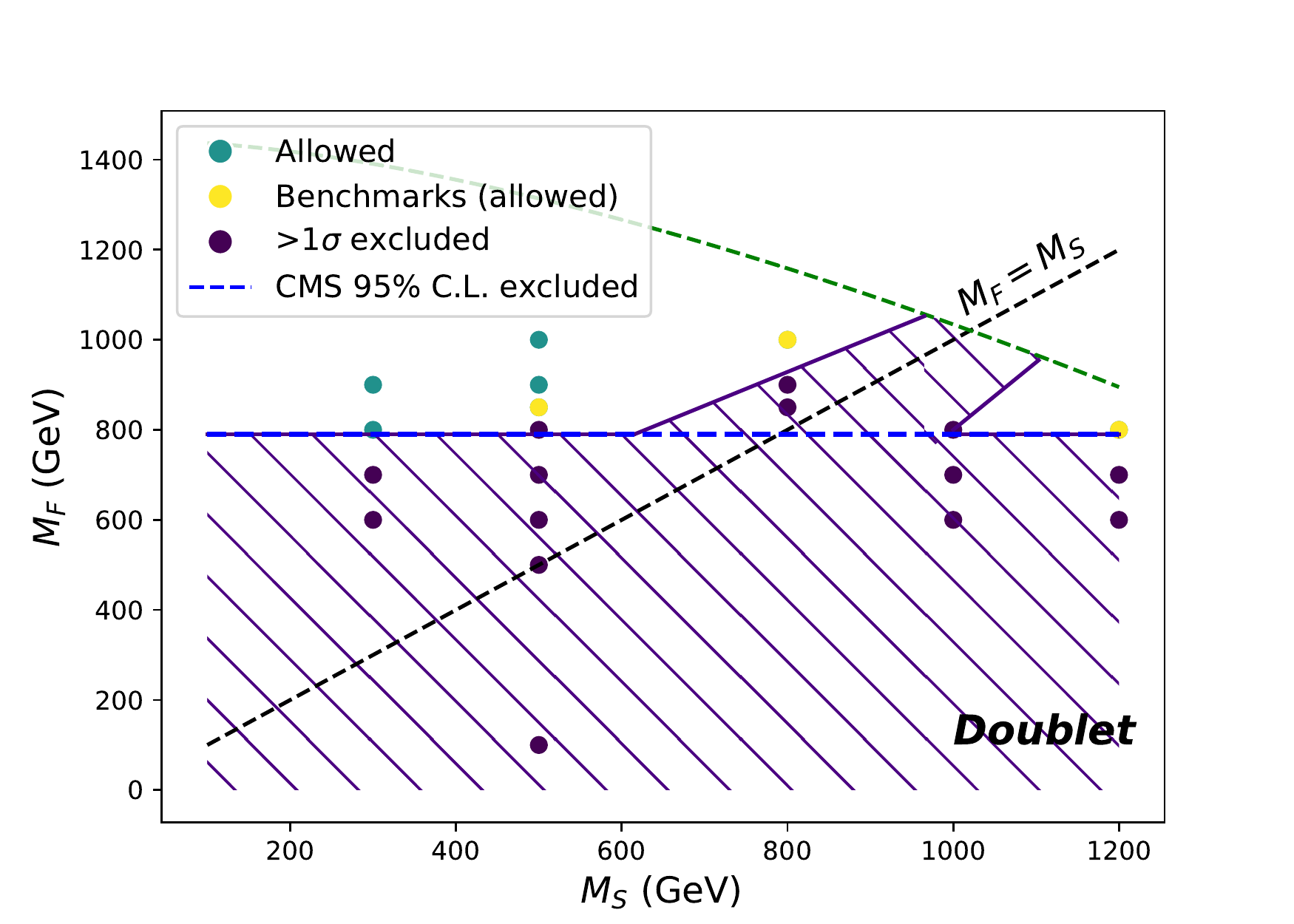}
	\caption{Allowed (green and yellow points) and excluded  (purple points) values of the VLL mass $M_F$ and the BSM scalar mass $M_S$ with $\kappa'$ fixed \eqref{eq:para}. For the points marked as allowed, all bins in the sum of transverse momenta $L_T$ of the 4L final sates fall within 1$\sigma$ of central values measured by CMS \cite{Sirunyan:2019ofn}. For benchmark points marked as yellow circles we show the $L_T$ distributions in Fig.~\ref{fig:L_T_delphes-benchmark}.  
	Above the green dashed curve $\kappa^\prime$ becomes  non-perturbative.}
		\label{fig:parameter-space}
\end{figure}

\begin{figure}
	\centering
    \includegraphics[width=0.49\textwidth]{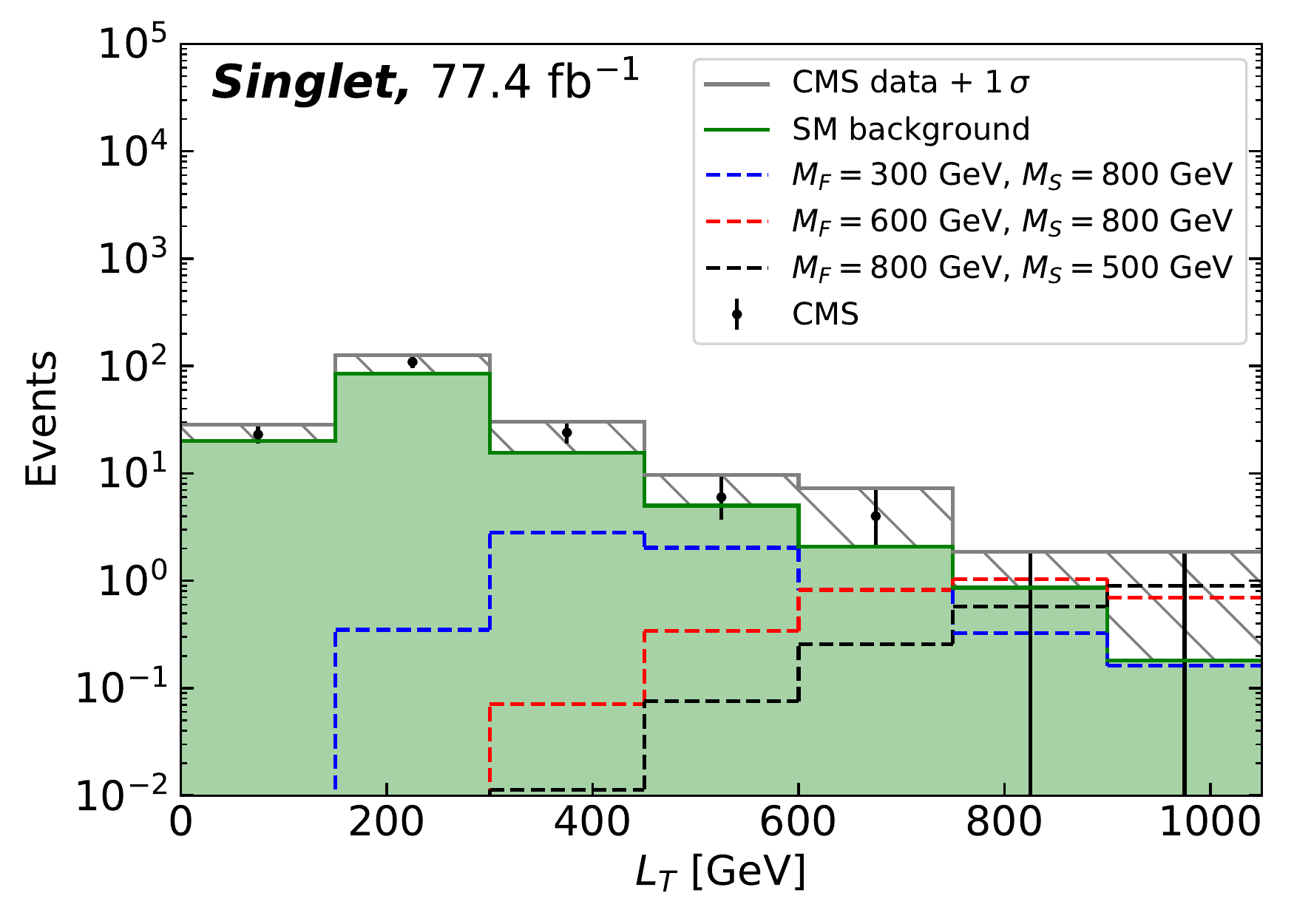}
	\includegraphics[width=0.49\textwidth]{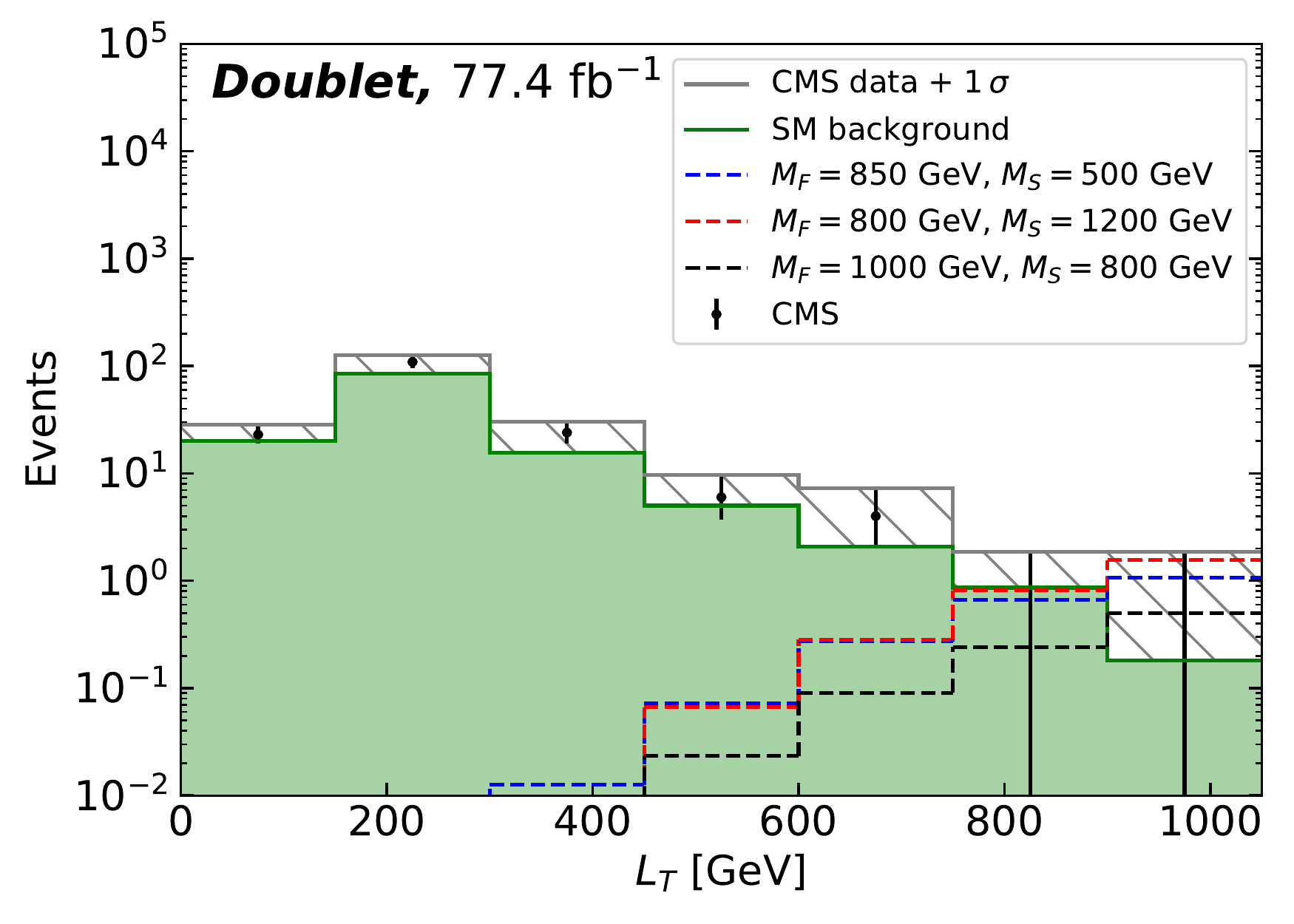}\\
	\caption{$L_T$ distributions in the singlet (left) and the doublet model (right) for SM background processes in our simulation (green shaded area) and for the different benchmark masses of vector-like fermions and new scalars  (yellow circles  in Fig.~\ref{fig:parameter-space}). The observables are shown for an integrated luminosity of $77.4$ fb${}^{-1}$ and subsequent detector simulation.  Also shown are CMS data \cite{Sirunyan:2019ofn} (black points), including the range covered up to $1 \sigma$ (hatched area), see text for details.}
	\label{fig:L_T_delphes-benchmark}
\end{figure}

CMS has searched for VLLs employing the scalar sum of the leading four light leptons' transverse momenta, $L_T$, finding no significant discrepancies with the background \cite{Sirunyan:2019ofn}. To work out  the implications of this analysis for the  models \eqref{Lint-singlet} and  \eqref{Lint-doublet} 
we compute the  $L_T$ distributions for different  values of $M_S$ and $M_F$ and fixed BSM Yukawas  \eqref{eq:para}.
After performing the detector simulation we compare the distributions to CMS data for 4L final states. 
We also compute  $L_T$ distributions for the dominant SM background processes of $ZZ$, triboson and $t\bar tZ$ production. 
We include the control region veto, two dilepton pairs with invariant masses $76~{\rm GeV}<m_{2\ell}<106~$GeV, and set the bin width to $150$ GeV as in \cite{Sirunyan:2019ofn}.

Since our simulation of the SM background is performed at LO and only a fast detector simulation is publicly available, differences with the one by CMS are expected.
In contrast to CMS, we can not perform a fit of the background distribution to a control region.
This prohibits a quantitative  reinterpretation of the  data at precision level, which could be obtained from an actual experimental analysis only.
Still, we find that our background simulation is in reasonable agreement with the shape and the bin content of the $L_T$ distribution.
In view of the differences between our SM prediction and the one from CMS, and to make progress, in the following we refer to benchmarks as 'excluded' if the BSM distribution overshoots the CMS data in at least one of the bins by more than one sigma.

Our findings are summarized in Fig.~\ref{fig:parameter-space}, showing which masses are compatible with data  (green and yellow circles) and which are not (purple circles)  for the singlet (left)  and the  doublet (right) model.  We scanned 40 points in the singlet and 20 in the doublet model, and expect these to indicate the main features
of the parameter space in the  $M_S,M_F$-plane. The purple hatched region is excluded, while the remainder is  still to be probed. Fig.~\ref{fig:parameter-space} also shows for which masses the coupling $\kappa^\prime$ required to accommodate the 
present $(g-2)_\mu$ anomaly becomes non-perturbative (above the green dashed curve).

We observe the following pattern: For both models the region $M_F\sim M_S$ is excluded.
This is a result of the 4L cross sections' enhancement around the on-shell $S$-production threshold, as can  be seen also  in Fig.~\ref{Fig:4ll_production}. 
In the singlet model large areas of parameter space outside of  the $M_F\sim M_S$ region remain unconstrained. 
Conversely, for the doublet model a significantly larger part of the parameter space is already probed due to the  larger 4L cross sections, see Fig.~\ref{Fig:4ll_production}. We find that values  of $M_F$ below $800$~GeV are excluded, 
consistent with the CMS 95\% C.L. limit of 790 GeV. 
Still, departing from the $M_F\sim M_S$ region we find areas in the doublet model parameter space that are  in agreement with the 4L data. 

We choose  three allowed benchmark points per model   to illustrate the analysis strategy described in the following sections. 
These points are marked as yellow circles in Fig.~\ref{fig:parameter-space}.
For the singlet model, one of the benchmarks features VLL masses as low as $300$~GeV and $M_S=800$~GeV, while another one presents 
$M_F = 600$~GeV for the same scalar mass. The third benchmark, with $M_F,M_S = 800,500$~GeV, displays the opposite mass hierarchy, 
allowing on-shell decays of the VLLs through the $S_{ij}$.
In the doublet model values below $M_F=800$~GeV are excluded regardless of $M_S$. We have chosen a benchmark which saturates this bound, with $M_F,M_S = 800,1200$~GeV.  The two remaining benchmarks present the inverse mass hierarchy, allowing for on-shell $\psi\to S\ell$ decays. The chosen parameters are $M_F,M_S = 850,500$~GeV and $M_F,M_S = 1000,800$~GeV, which in both cases lie at the frontier of the probed parameter space (see again Fig.~\ref{fig:parameter-space}).

In Fig.~\ref{fig:L_T_delphes-benchmark} we show the $L_T$ distributions for these benchmarks (long-dashed curves) to explicitly show that they pass  4L constraints,
that is, are within the CMS plus $1 \sigma$ range (hatched area).

\section{Optimized observables and Null Tests}
\label{sec:Null-test}

 In this section we design novel  observables which target specific flavor features of our models and can serve  as null tests of the SM. These optimized  observables consist of invariant mass distributions which aim at  reconstructing the masses of the new scalars and the VLLs. The latter are reconstructed through their decays to electroweak bosons, $h$ and $S$ plus lepton; final states with neutrinos are mostly removed  through cuts on the missing transverse momentum, see Tab.~\ref{Tab:Param}. 
 Thanks to the large values of $\kappa^\prime$, VLL decays to  $S_{ij}$ plus charged lepton are dominant when the $S_{ij}$ can be produced on-shell,
 see Fig.~\ref{Fig:BR_psi}, but remain significant also  for $M_F\lesssim M_S$.
Key modes to probe VLLs and scalars with flavor are the decays~\eqref{lfv-like-decay-singlet} and~\eqref{lfv-like-decay-doublet} of the negatively charged $\psi_i$ 
into six leptons
\begin{equation}
\begin{aligned}
	\psi_i \overline{\psi}_i &\rightarrow \ell^-_ j S_{ij}^* \ell^+_k S_{ik} \rightarrow \ell^-_j \ell^+_j\ell^-_i \ell^+_k \ell^-_k\ell^+_i \quad {\rm (singlet)\,,} \\
	\psi_i^- \psi_i^+ &\rightarrow \ell^-_ j S_{ji} \ell^+_k S_{ki}^* \rightarrow \ell^-_j \ell^+_j\ell^-_i \ell^+_k \ell^-_k\ell^+_i \quad {\rm (doublet)\,} \, , 
	\label{flavor_channel}
\end{aligned}
\end{equation}
which enable to construct   the $S_{ij}$ out of two leptons with opposite charge and same or different flavor.
Combining these two leptons with a third one carrying the same flavor and opposite charge as one of the leptons in the initial pair enables us to reconstruct the $\psi_i$. 
We reconstruct masses of the $Z$- and Higgs-boson as well as the masses of the new scalars $S_{ij}$ considering jets (for $Z$ and $h$ only) and charged leptons as the final states. These invariant masses computed from  two final-state particles are referred to as $m_{2\ell}$. The subset of invariant masses reconstructed from leptons with {\it different} flavor  are called $m_{2\ell}\_{\rm diff}$. Combining the reconstructed bosons with the remaining charged leptons gives the reconstructed masses of the VLLs, called $m_{3\ell}$ and $m_{3\ell}\_{\rm diff}$. 

For the $m_{i}\_{\rm diff}$ observables, in order to reconstruct both the $S_{ij}$ and the $\psi_i$ out of leptons with different flavors ($i\neq j$ and/or $i\neq k$ in Eq.~\eqref{flavor_channel}) we require to first find two pairs of different-flavor leptons with the same invariant mass within a small mass window $\Delta M_S$ (see Tab.~\ref{Tab:Param}), assuming a narrow width for the $S_{ij}$. This allows to search for $S$-mediated decays without any assumption on the scalar masses. When two candidates for the new scalars are found, the VLLs are reconstructed applying flavor-conservation conditions. 
The flavor and mass requirements sufficiently suppress SM background making the processes in Eq.~\eqref{flavor_channel} with $i\neq j$ and/or $i\neq k$ the 'golden channels' for our analysis.  

In Sec.~\ref{sec:2} we discuss the algorithm to construct the observables $m_{2\ell}$ and $m_{2\ell}\_{\rm diff}$, which could signal scalar resonances.
In Sec.~\ref{sec:3} we discuss how to obtain $m_{3\ell}$ and $m_{3\ell}\_{\rm diff}$ distributions, which could signal VLLs.
Projections for the full Run 2 data set are worked out in Sec.~\ref{sec:run2}.

\subsection{$m_{2\ell}$ and $m_{2\ell}\_{\rm diff}$   \label{sec:2}}
For each event with at least four light leptons, we compute all possible sets of two dilepton invariant masses from leptons of opposite charge, where each lepton contributes to only one of the invariant masses in the pair. This step includes $\tau$ leptons if present. If the event contains jets, we include all possible pairs of invariant masses where one of them is a dilepton invariant mass and the other is the dijet invariant mass. For each event, only one pair of invariant masses is added to the observable $m_{2\ell}$.  In order to be added, it must fulfill one of the following requirements:
\begin{itemize}
	\item[$a)$] Each invariant mass is equal either to $M_Z\pm \Delta M_Z$ or to $M_H \pm \Delta M_H$, according to the parameters in Tab.~\ref{Tab:Param}, and each dilepton pair contains two leptons of the same flavor. The states used to compute the masses are in this case $(\ell_i^+\ell_i^-)(\ell_j^+\ell_j^-)$, $(\tau^+\tau^-)(\ell_i^+\ell_i^-)$ or $(\ell_i^+\ell_i^-)(jj)$ with $i,j=1,2$. This condition reconstructs $Z$ and Higgs bosons. 
	\item[$b)$] The difference between both invariant masses is less than $\Delta M_S$  (see Tab.~\ref{Tab:Param}), while none of the other invariant mass pairs present a smaller difference, and each invariant mass is computed from same-flavored leptons. The states used to compute the masses are in this case $(\ell_i^+\ell_i^-)(\ell_j^+\ell_j^-)$ with $i,j=1,2$. This condition reconstructs two scalars, $S_{ii}$ and $S_{jj}$. 
	\item[$c)$] Both invariant masses differ by less than $\Delta M_S$  (see Tab.~\ref{Tab:Param}), while none of the other invariant mass pairs present a smaller difference, and at least one of the invariant masses contains two leptons of different flavor. The states used to compute the masses are in this case $(\ell_i^+\ell_j^-)(\ell_k^+\ell_i^-)$ with $i,j,k=1,2,3$ leading to a maximum of one $\tau^+$ and  one $\tau^-$. This condition reconstructs two scalars, $S_{ij}$ and $S_{ik}$.
\end{itemize}

We check for these conditions in the above order ($a\to b \to c$) and stop when one of the requirements is fulfilled. 
We define the observable $m_{2\ell}\_{\rm diff}$ as 
invariant mass pairs that only fulfill condition $c)$, where two particles of approximately equal invariant mass are found and at least one of them is computed from different-flavor leptons. All SM contributions to this observable are purely statistical, and therefore any significant excess away from SM resonances is an indication of new physics which can be explained by the VLL models of Eqs.~\eqref{Lint-singlet} or~\eqref{Lint-doublet}.

\subsection{$m_{3\ell}$ and $m_{3\ell}\_{\rm diff}$ \label{sec:3}}

The $m_{3\ell}$ and $m_{3\ell}\_{\rm diff}$ observables are designed to reconstruct the invariant masses of the VLLs via their three-body decays. For each pair of two-particle invariant masses added to $m_{2\ell}$, we look for the additional lepton which stems from the decay of each $\psi$. We add to $m_{3\ell}$ the pairs of three-particle invariant masses which fulfill one the following conditions:

\begin{itemize}
	\item[$i)$] For two-particle invariant masses which reconstruct to a $Z$ or Higgs (condition $a$ in the previous section) each two-particle invariant mass is paired with an additional lepton present in the final state. The resulting three-particle invariant masses are added to $m_{3\ell}$ if their difference is smaller than $\Delta M_F$, and no other combination presents a smaller difference.
	\item[$ii)$] For two-lepton invariant masses which reconstruct to $S_{ii}$ and $S_{jj}$ (condition $b$ in the previous section) each two-lepton invariant mass is paired with an additional lepton present in the final state which has the {\it same} flavor of the two leptons in the two-lepton invariant mass. The resulting three-lepton invariant masses are added to $m_{3\ell}$ if their difference is smaller than $\Delta M_F$, and no other combination presents a smaller difference.
	\item[$iii)$] For two-lepton invariant masses which reconstruct to $S_{ik}$ and $S_{kj}$ (condition $c$ in the previous section) if a two-lepton invariant mass contains two same-flavor leptons, it is paired with an additional lepton present in the final state which has the {\it same} flavor. If it contains two {\it different}-flavor leptons, it is paired with an additional lepton which has the same flavor but opposite charge of one of the leptons in the two-lepton invariant mass. For each event, we find at most one possible combination that fulfills this condition. The corresponding three-lepton invariant masses are added to $m_{3\ell}$.
\end{itemize}

In the last two conditions, flavor requirements are designed to reflect flavor conservation in the decays of the $S_{ij}$. 
We define the observable $m_{3\ell}\_{\rm diff}$ as %the sum of
invariant mass pairs that only fulfill condition $iii)$. As it turns out, the selection of the third leptons via flavor rules allows to populate $m_{3\ell}\_{\rm diff}$ even when the $\psi$'s do not have a narrow width, which happens when the $\psi$ undergoes frequent on-shell decays to $S$, 
{\it i.e.}, for  $M_F>M_S$ and $\kappa^\prime$ large.
$m_{3\ell}\_{\rm diff}$ is a clean null test of the SM.

\subsection{Benchmark distributions for Run 2 \label{sec:run2}}

 \begin{figure}
	\centering
	\includegraphics[width=0.49\textwidth]{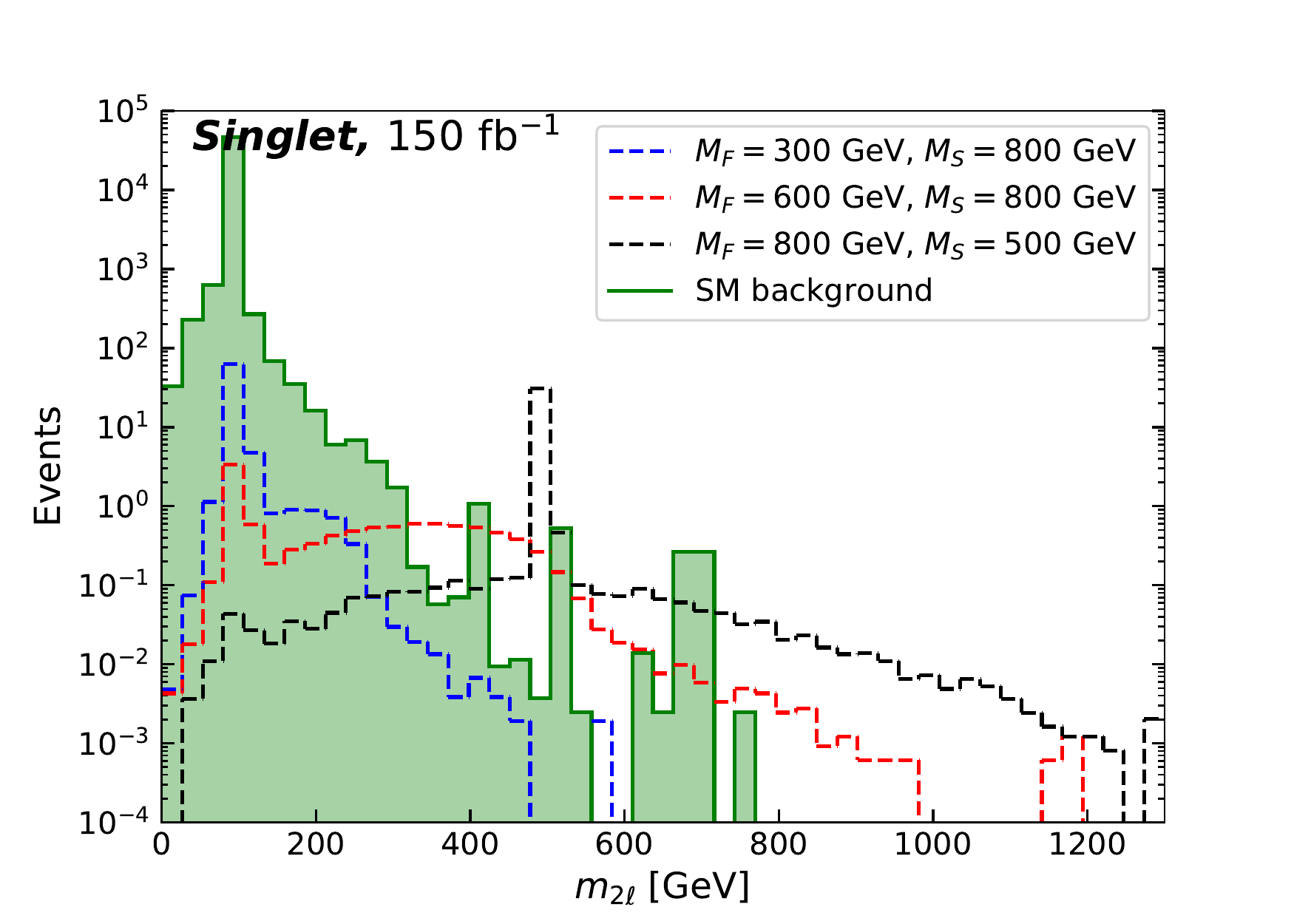}
	\includegraphics[width=0.49\textwidth]{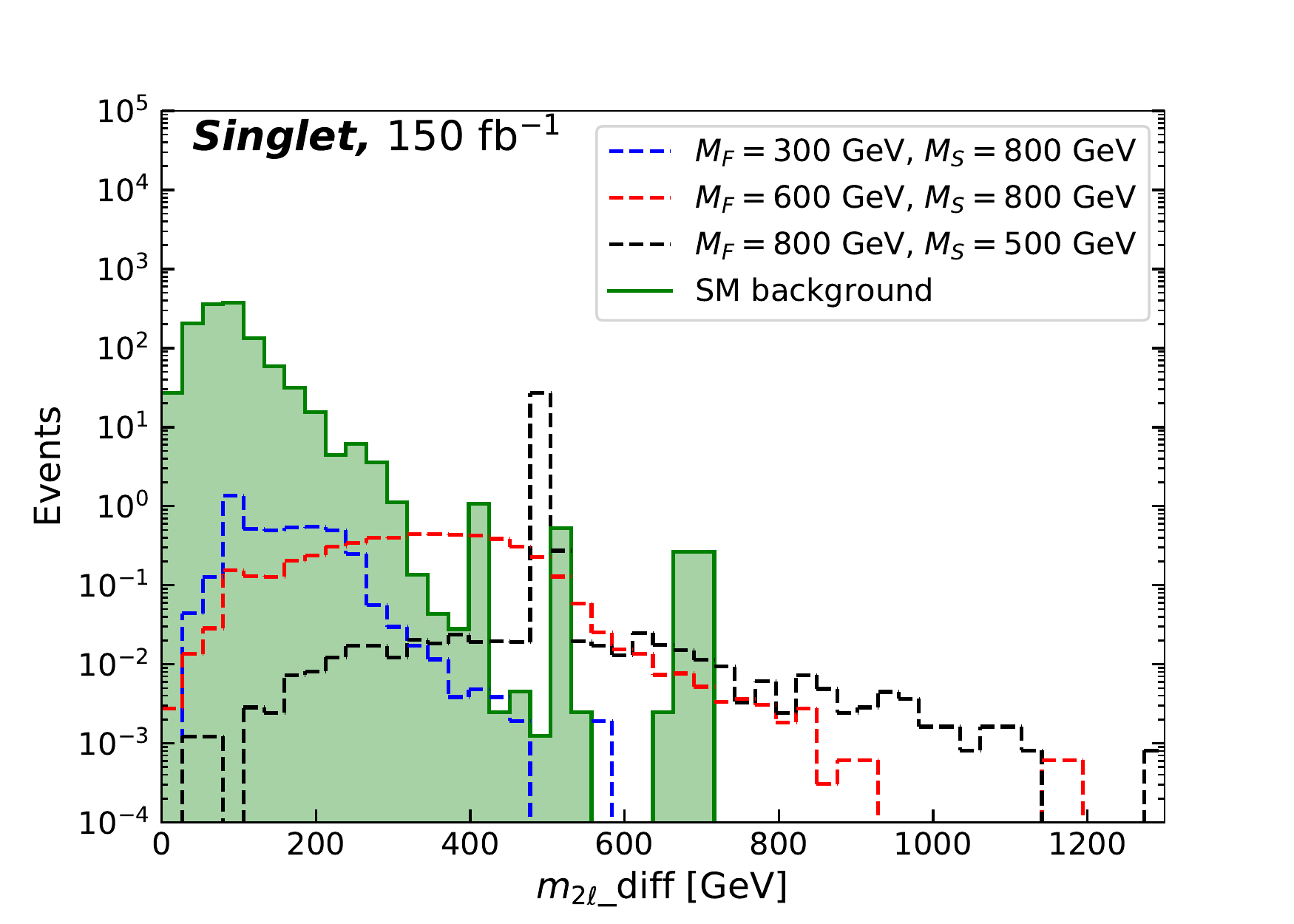}
	\includegraphics[width=0.49\textwidth]{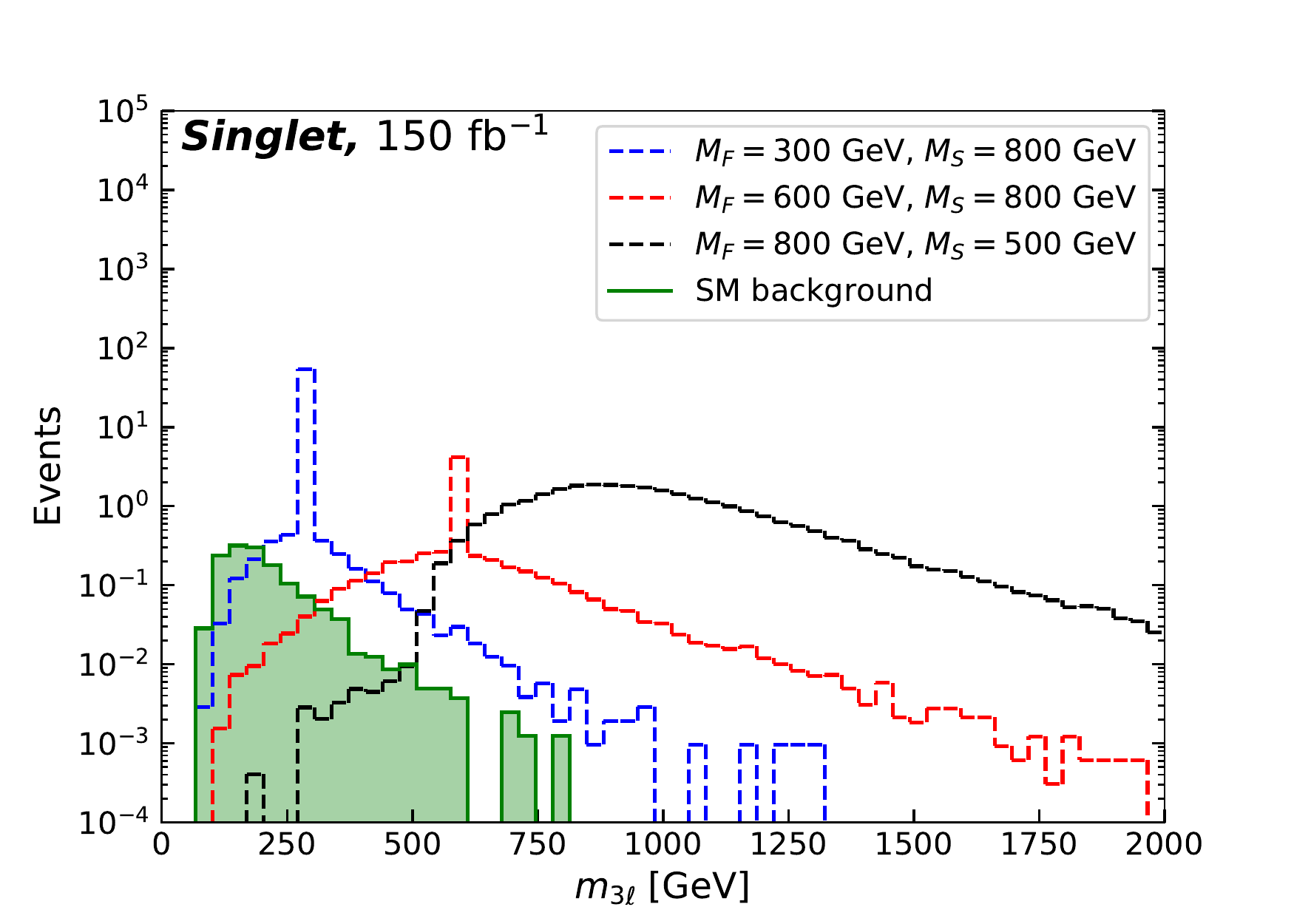}
	\includegraphics[width=0.49\textwidth]{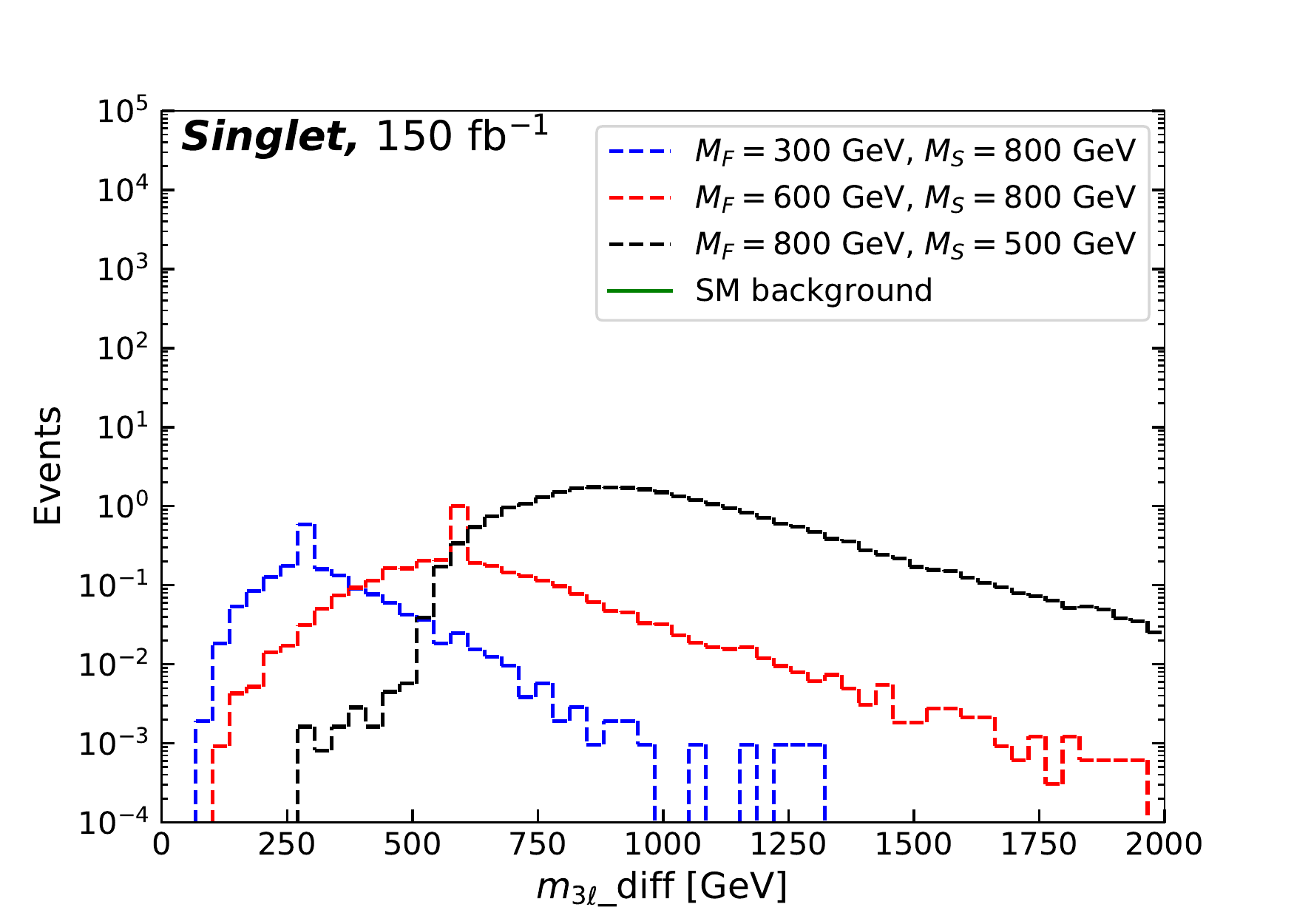}
	\caption{Di- and trilepton invariant mass distributions $m_{2\ell}$,  $m_{2\ell}\_{\rm diff}$, $m_{3\ell}$, and $m_{3\ell}\_{\rm diff}$ (see Sec.~\ref{sec:Null-test} for details) for the singlet model for different benchmark masses of the VLLs and the BSM scalars at a luminosity of $150$ fb${}^{-1}$ and $\sqrt{s}=13$~TeV. The coupling $\kappa'$ is fixed according to Eq.~\eqref{eq:para}.
	}
	\label{fig:singlet-bench}
\end{figure} 
 
\begin{figure}
	\centering
	\includegraphics[width=0.49\textwidth]{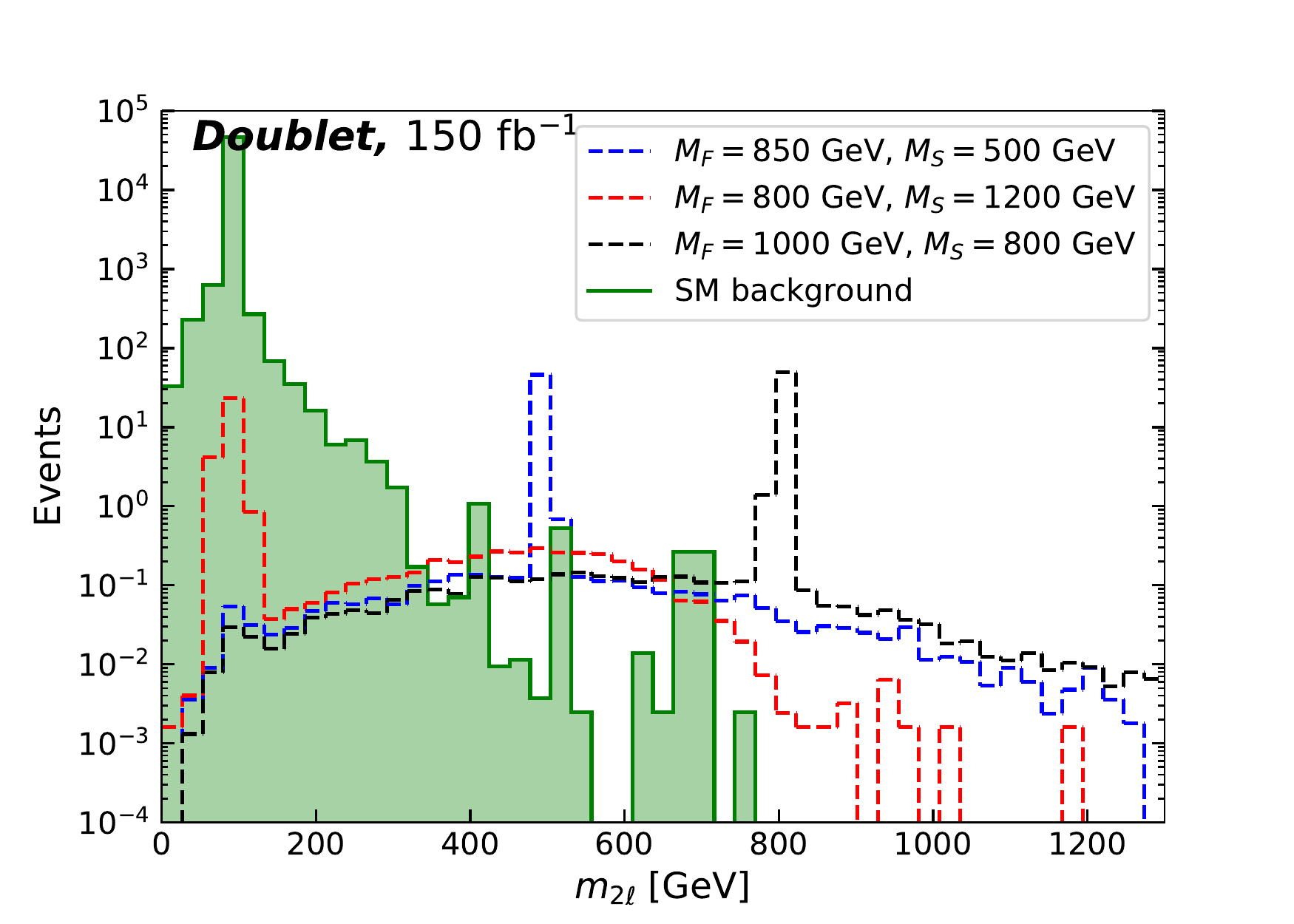}
	\includegraphics[width=0.49\textwidth]{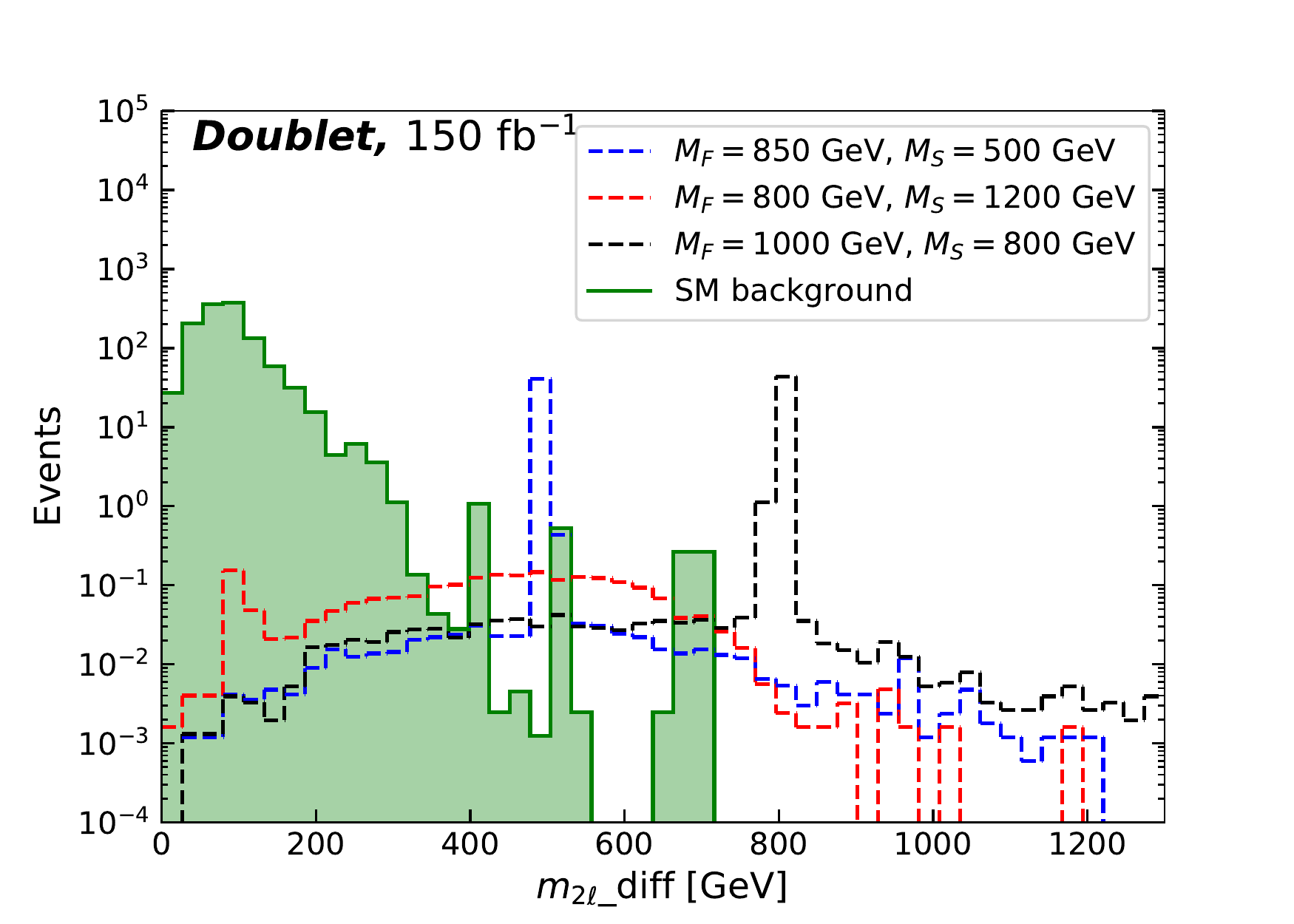}
	\includegraphics[width=0.49\textwidth]{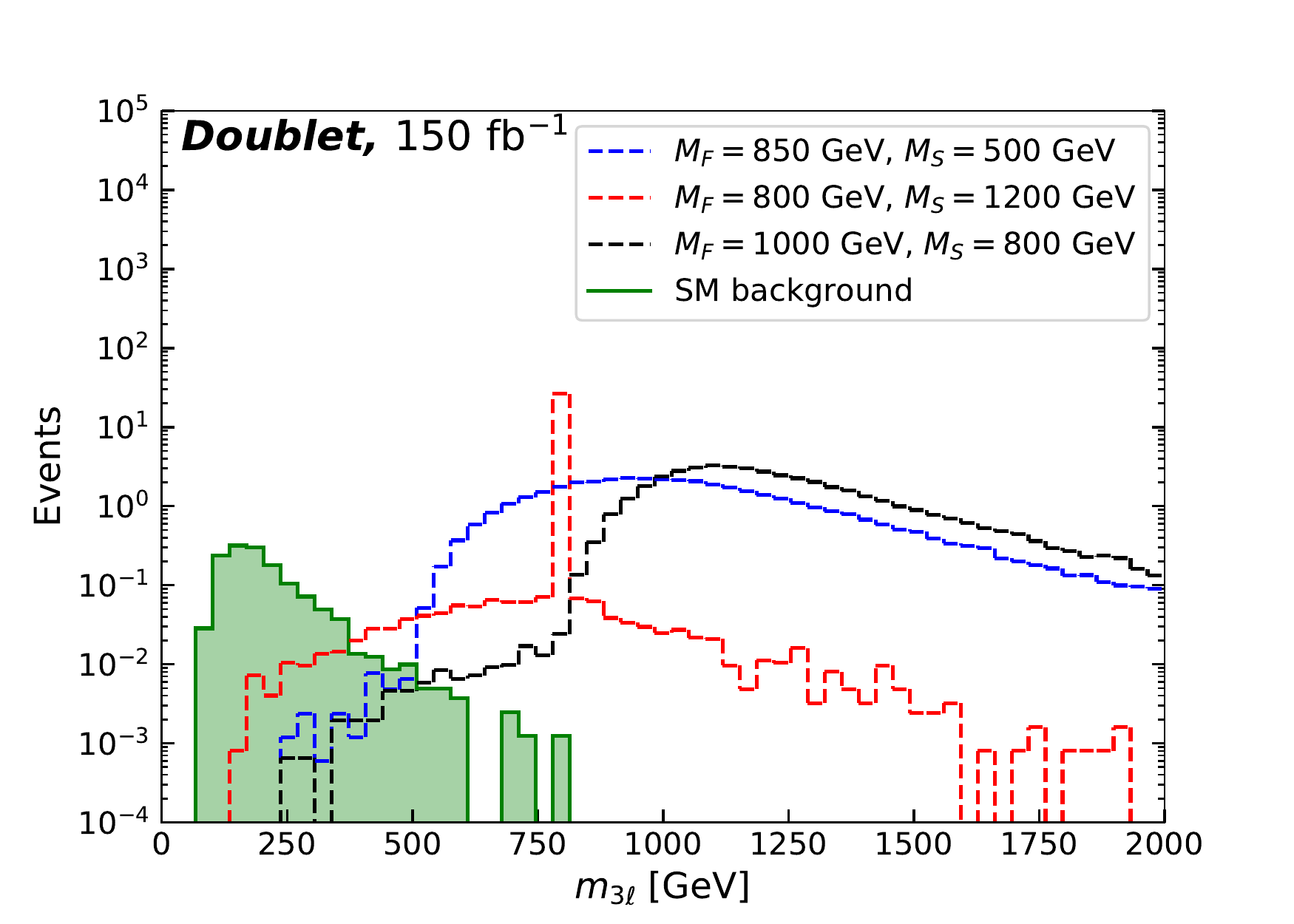}
	\includegraphics[width=0.49\textwidth]{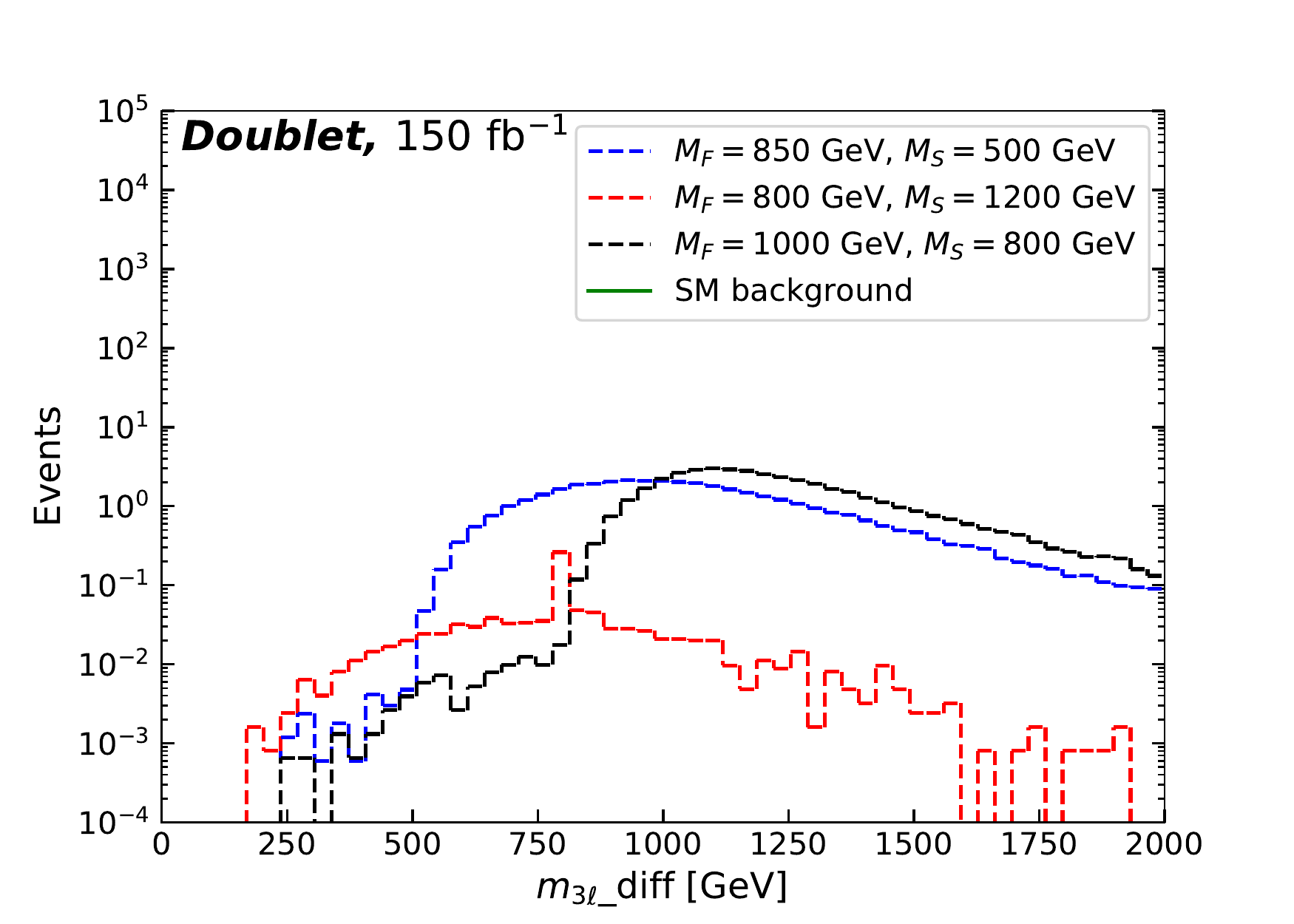}
	\caption{Di- and trilepton invariant mass distributions $m_{2\ell}$,  $m_{2\ell}\_{\rm diff}$, $m_{3\ell}$, and $m_{3\ell}\_{\rm diff}$ (see Sec.~\ref{sec:Null-test} for details) for the doublet model for different benchmark masses of the VLLs and the BSM scalars at a luminosity of $150$ fb${}^{-1}$ and $\sqrt{s}=13$~TeV. The coupling $\kappa'$ is fixed according to Eq.~\eqref{eq:para}.
	}
	\label{fig:doublet-bench}
\end{figure} 

We study the $m_{2\ell}$,  $m_{2\ell}\_{\rm diff}$, $m_{3\ell}$, and $m_{3\ell}\_{\rm diff}$  distributions for allowed benchmark values (yellow circles in Fig.~\ref{fig:parameter-space}) of the VLL mass $M_F$ and the BSM scalar mass $M_S$ for the full Run 2 data set. Results based on the algorithms described in Secs.~\ref{sec:2} and \ref{sec:3} are  shown in Fig.~\ref{fig:singlet-bench} and Fig.~\ref{fig:doublet-bench} for the singlet and doublet model, respectively.
The distributions are computed from $5\times 10^4$ generated events and rescaled to an integrated luminosity of $150~{\rm fb}^{-1}$, therefore small statistical fluctuations are present in the plots.

We learn the following generic features:

{\it i)} The results are qualitatively similar for the singlet and doublet models, with a larger cross section for the latter  yielding more populated distributions.
The "diff"-observables (plots to the right) are cleaner than their non-"diff" variants (plots to the left), {\it i.e.}, have more efficient SM background suppression.
For instance, note how $m_{2\ell}\_{\rm diff}$ reduces the SM background around the $Z$ mass in comparison to $m_{2\ell}$. 
The $m_{3\ell}, \ m_{3\ell}\_{\rm diff}$  (lower plots)  are cleaner than the $m_{2\ell}, \ m_{2\ell}\_{\rm diff}$ (upper plots) spectra.
The $m_{3\ell}\_{\rm diff}$ observable is  SM background free. 

{\it ii)} Resonance peaks from $S$-decays appear in the $m_{2\ell}, \ m_{2\ell}\_{\rm diff}$ spectra for the benchmarks with  $M_F > M_S$, that is, when on-shell production of the scalars 
takes place. Narrow resonance peaks from $\psi$-decays appear  in the $m_{3\ell}, \ m_{3\ell}\_{\rm diff}$ distributions for the other benchmarks, which present $M_F < M_S$. 
The latter condition eliminates the rapid on-shell decays to the scalars through the large $\kappa^\prime$ Yukawa.

{\it iii)} Distributions can signal BSM physics also  in the tails away from a  narrow resonance peak, or if none is present; see for instance the black curves in Fig.~\ref{fig:singlet-bench} and Fig.~\ref{fig:doublet-bench}.
All benchmarks display an excess above the SM in all distributions, with the exception of 
 light VLLs $M_F=300$ GeV and heavy-ish scalars $M_S=800$ GeV in the singlet model (blue curve), which are underneath the SM contributions in $m_{2\ell}, \ m_{2\ell}\_{\rm diff}$ spectra, but do show up in the $m_{3\ell}, \ m_{3\ell}\_{\rm diff}$ distributions.

\begin{figure}
	\centering 
	\includegraphics[width=0.49\textwidth]{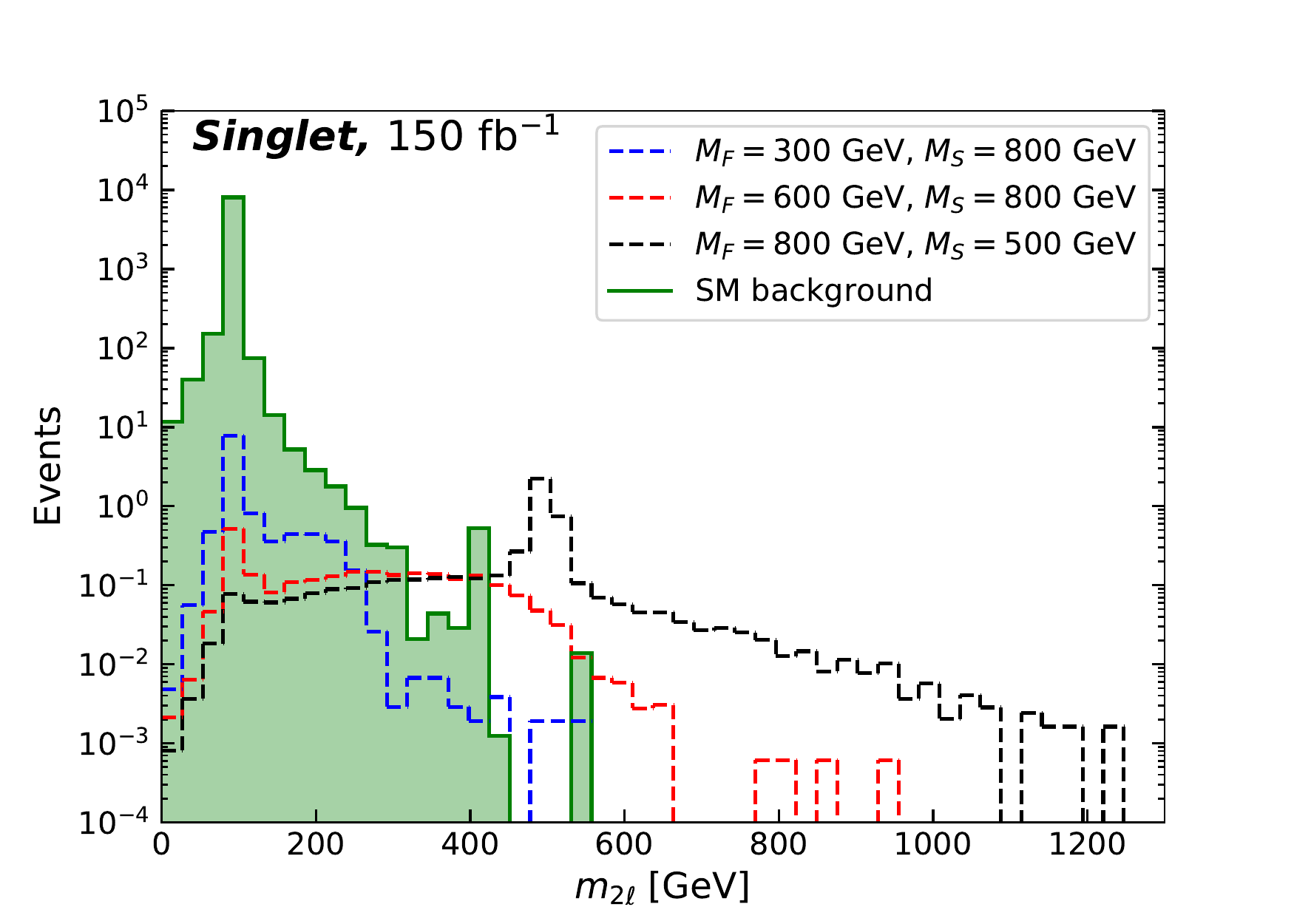}
	\includegraphics[width=0.49\textwidth]{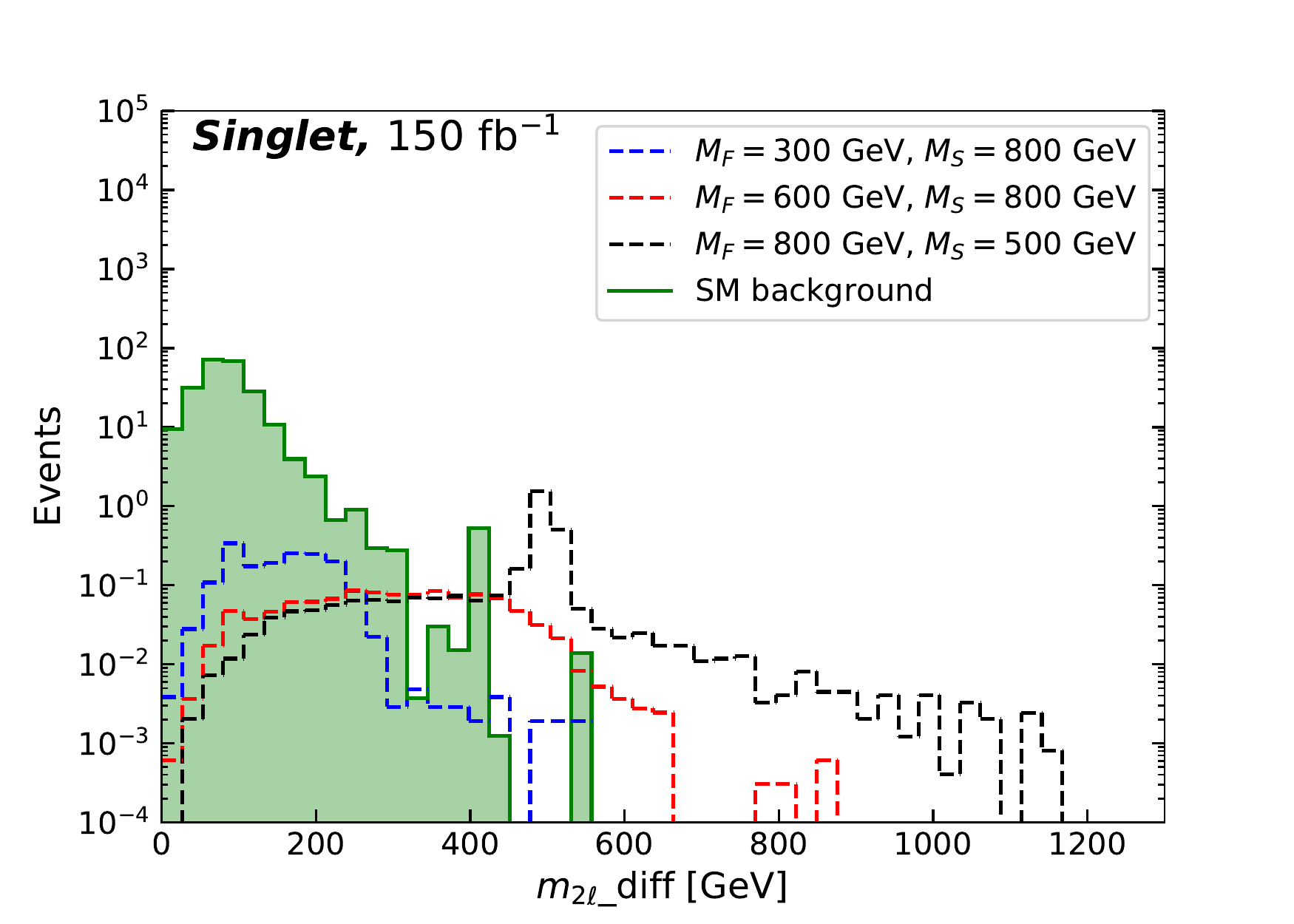}
	\includegraphics[width=0.49\textwidth]{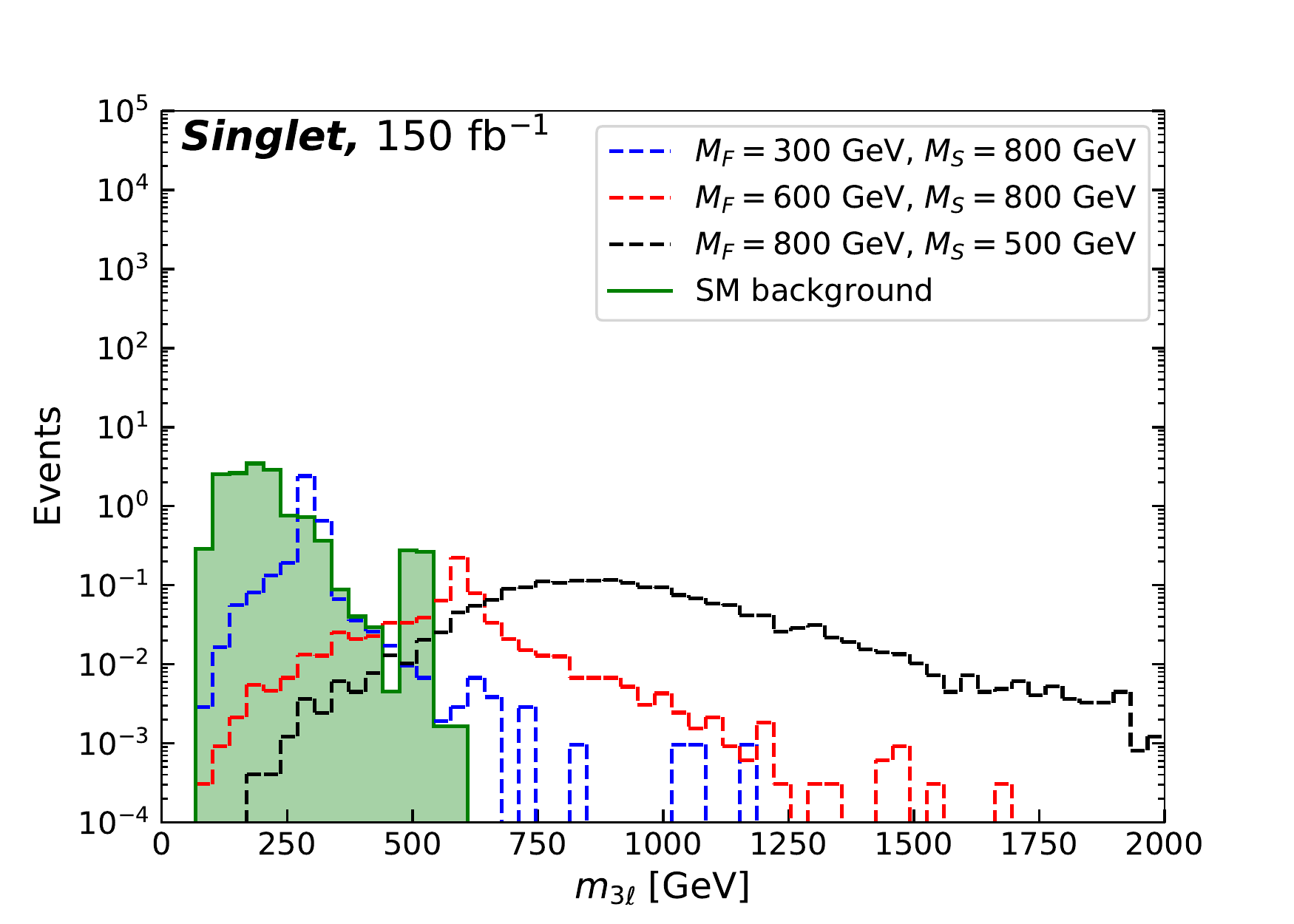}
	\includegraphics[width=0.49\textwidth]{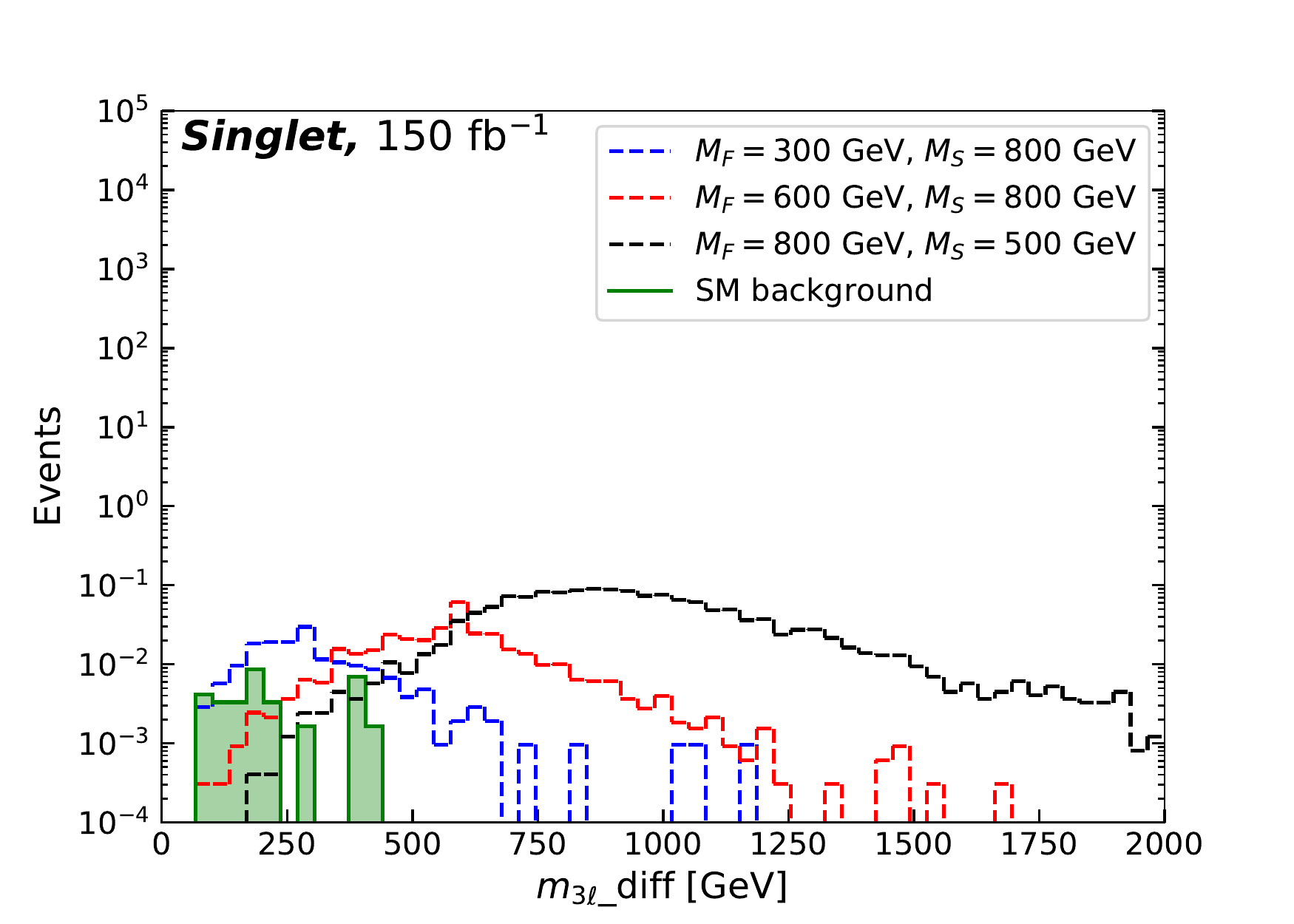}
	\caption{As in Fig.~\ref{fig:singlet-bench} after  detector simulation, see Sec.~\ref{sec:Null-test} for details.}
	\label{fig:singlet-bench-delphes}
\end{figure}

\begin{figure}
	\centering
	\includegraphics[width=0.49\textwidth]{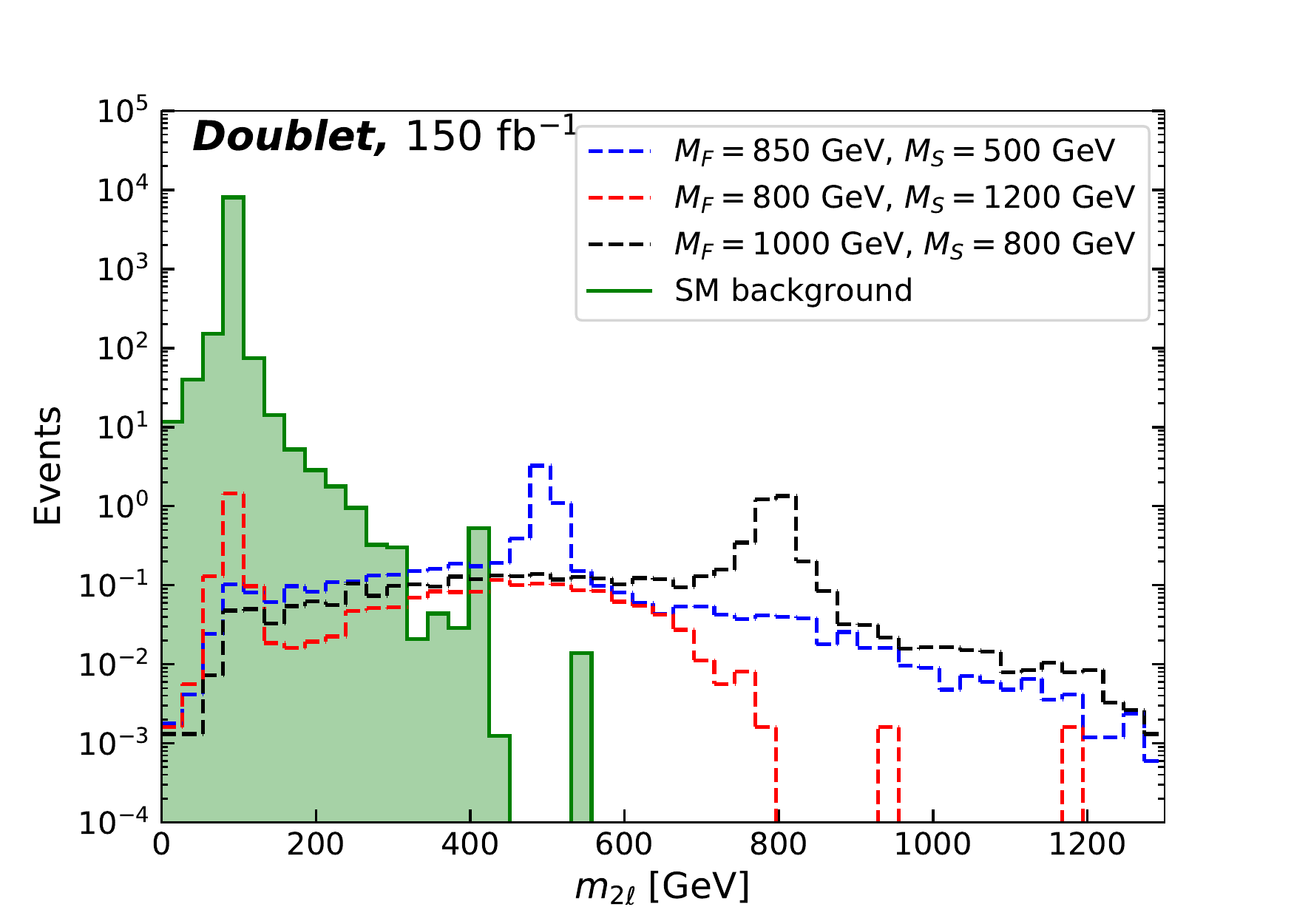}
	\includegraphics[width=0.49\textwidth]{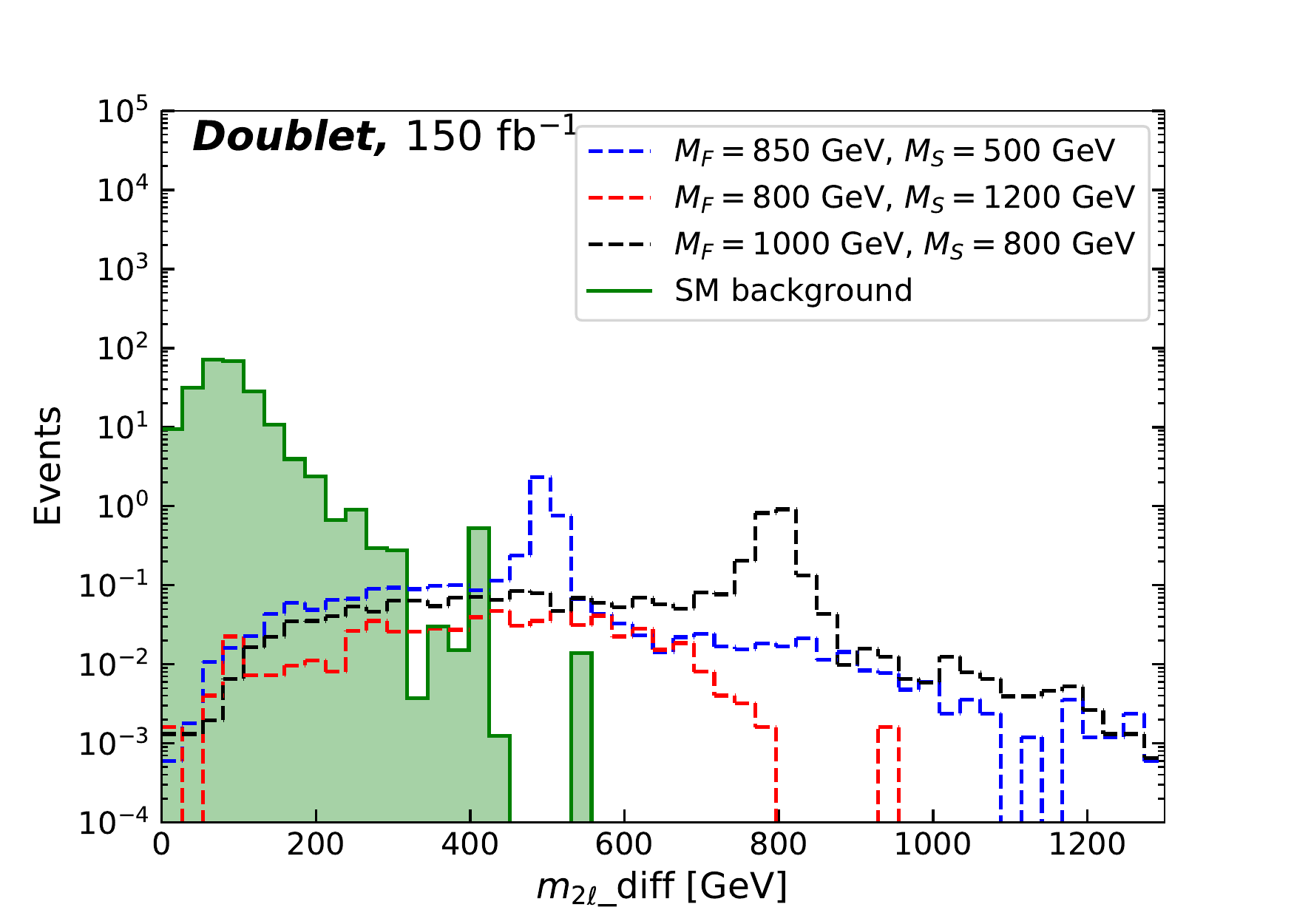}
	\includegraphics[width=0.49\textwidth]{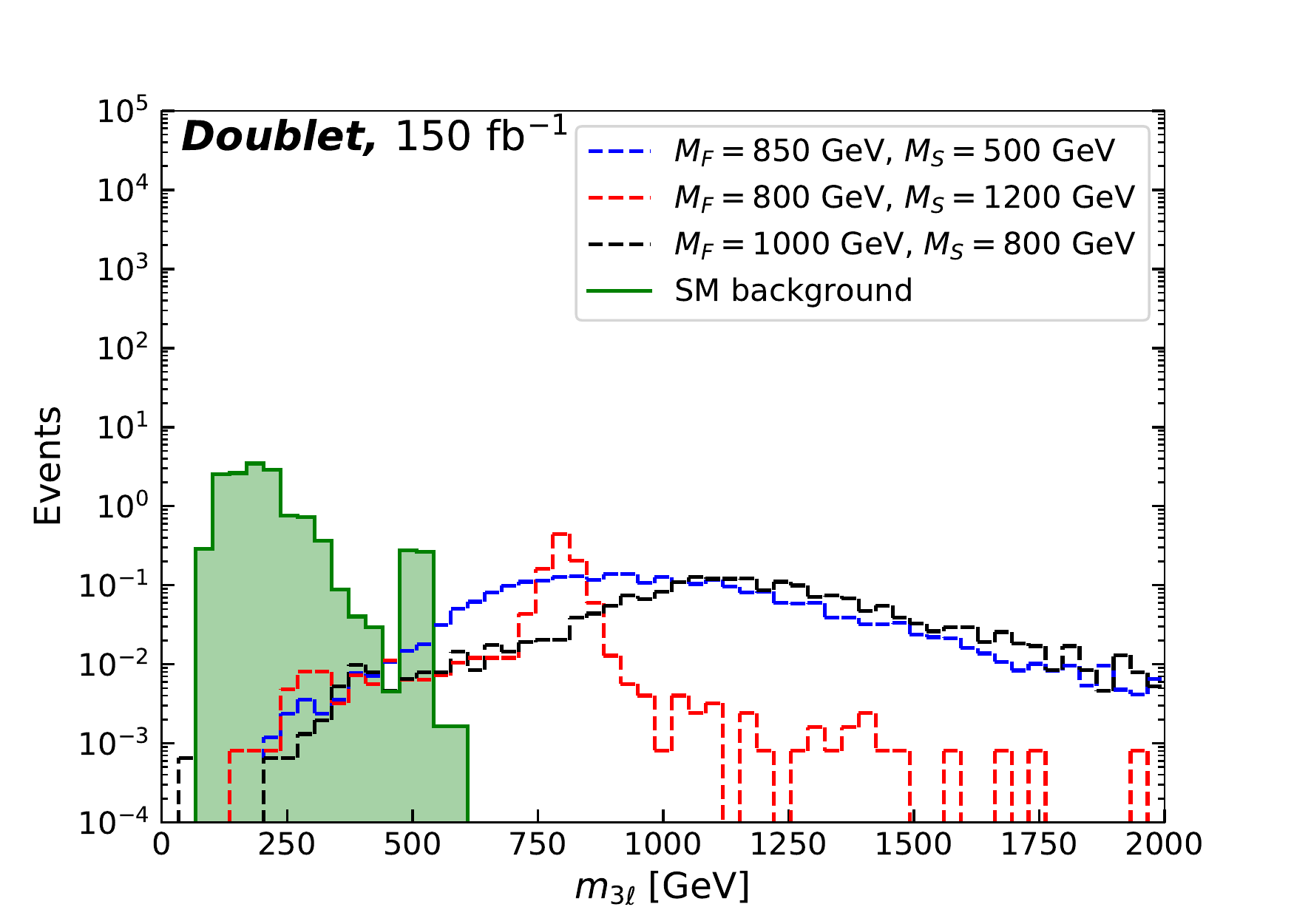}
	\includegraphics[width=0.49\textwidth]{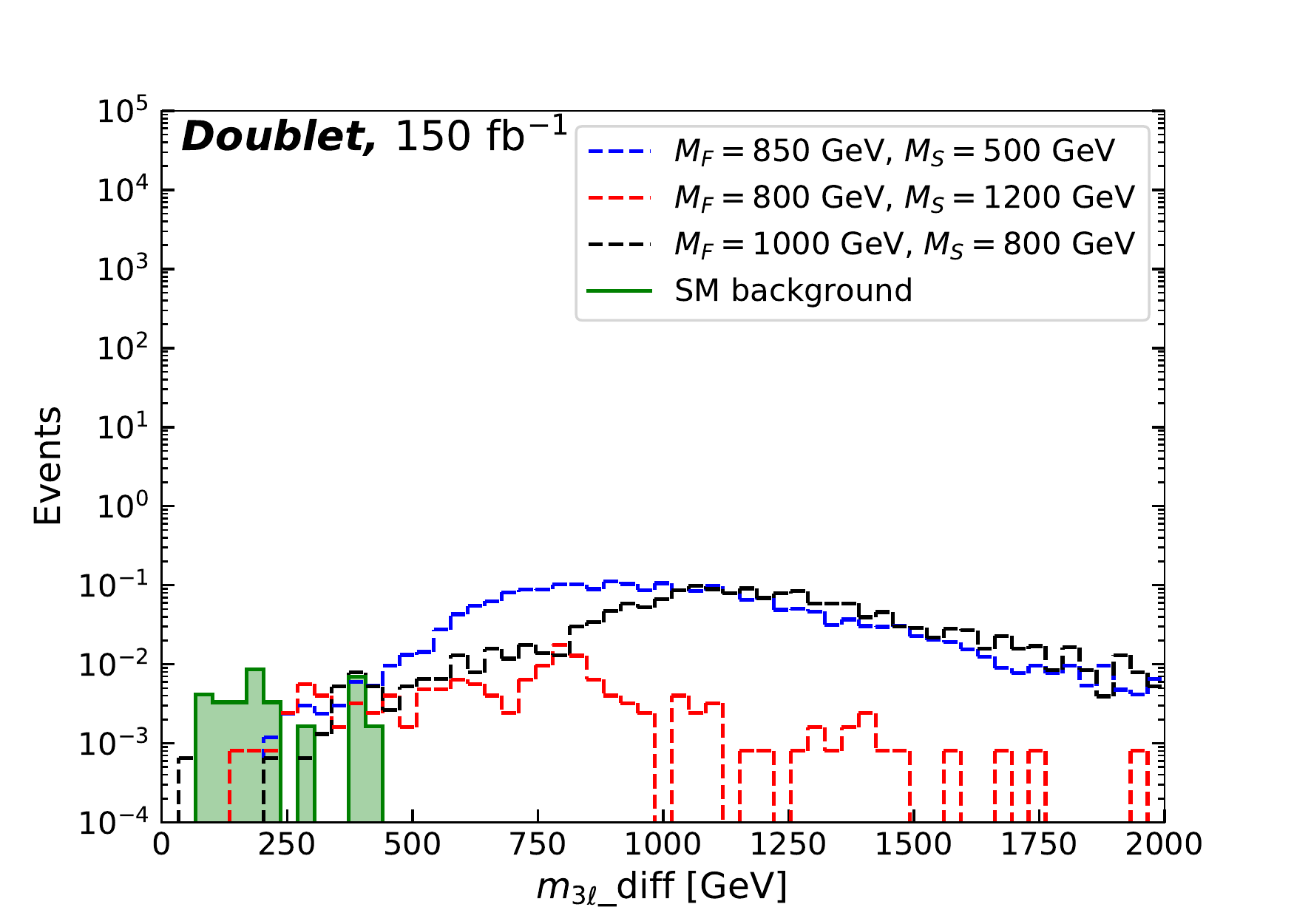}
	\caption{As in Fig.~\ref{fig:doublet-bench} after  detector simulation, see Sec.~\ref{sec:Null-test} for details.}
	\label{fig:doublet-bench-delphes}
\end{figure}

Including the effects of hadronization and finite detector resolution
we show in Figs.~\ref{fig:singlet-bench-delphes} and \ref{fig:doublet-bench-delphes} the observables for the singlet and doublet benchmark scenarios, respectively, after showering the events and applying a fast detector simulation.  As expected, we find that  peaks  become broader and event rates drop.
In the $m_{2\ell}, \ m_{2\ell}\_{\rm diff}$ distributions the number of events in the peaks is reduced by roughly one order of magnitude, leading to $\mathcal{O}(1)$ events in the peaks for on-shell $S$ production in the case $M_F=800$~GeV in the singlet and $M_F=850$ GeV, $1000$~GeV in the doublet model. 
In the case of the $m_{3\ell}, \ m_{3\ell}\_{\rm diff}$ observables, only in the singlet model and for small VLL masses $M_F=300$~GeV (blue curves) we find $\mathcal{O}(1)$ events in the peaks in the $m_{3\ell}$ distributions, while in all other scenarios and the $m_{3\ell}\_{\rm diff}$ distributions the number of signal events is below one.  
Scaling factors comparing the number of events in the peak  bins  before and after detector simulation  are given in Tab.~\ref{tab:scale-factors} in App.~\ref{app:delphes}, where we also discuss in more detail the effects of the detector simulation.

As we argued, the new observables have great sensitivity to flavorful BSM physics, and  would benefit from higher luminosity.
In the next section, we discuss perspectives for  the HL-LHC.

\section{Implications for the HL-LHC }\label{sec:Outlook}

\begin{figure}
	\centering
	\includegraphics[width=0.49\textwidth]{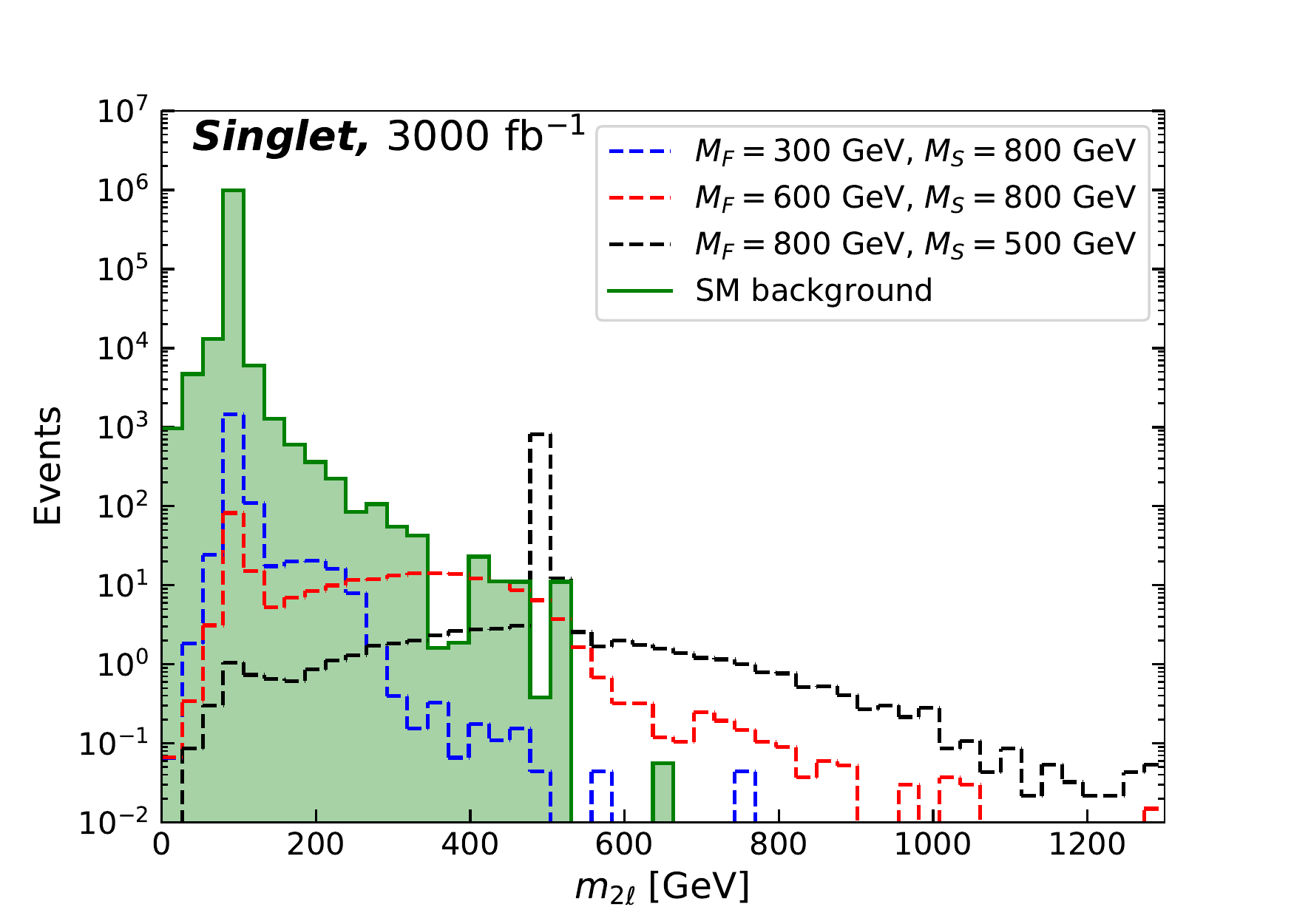}
	\includegraphics[width=0.49\textwidth]{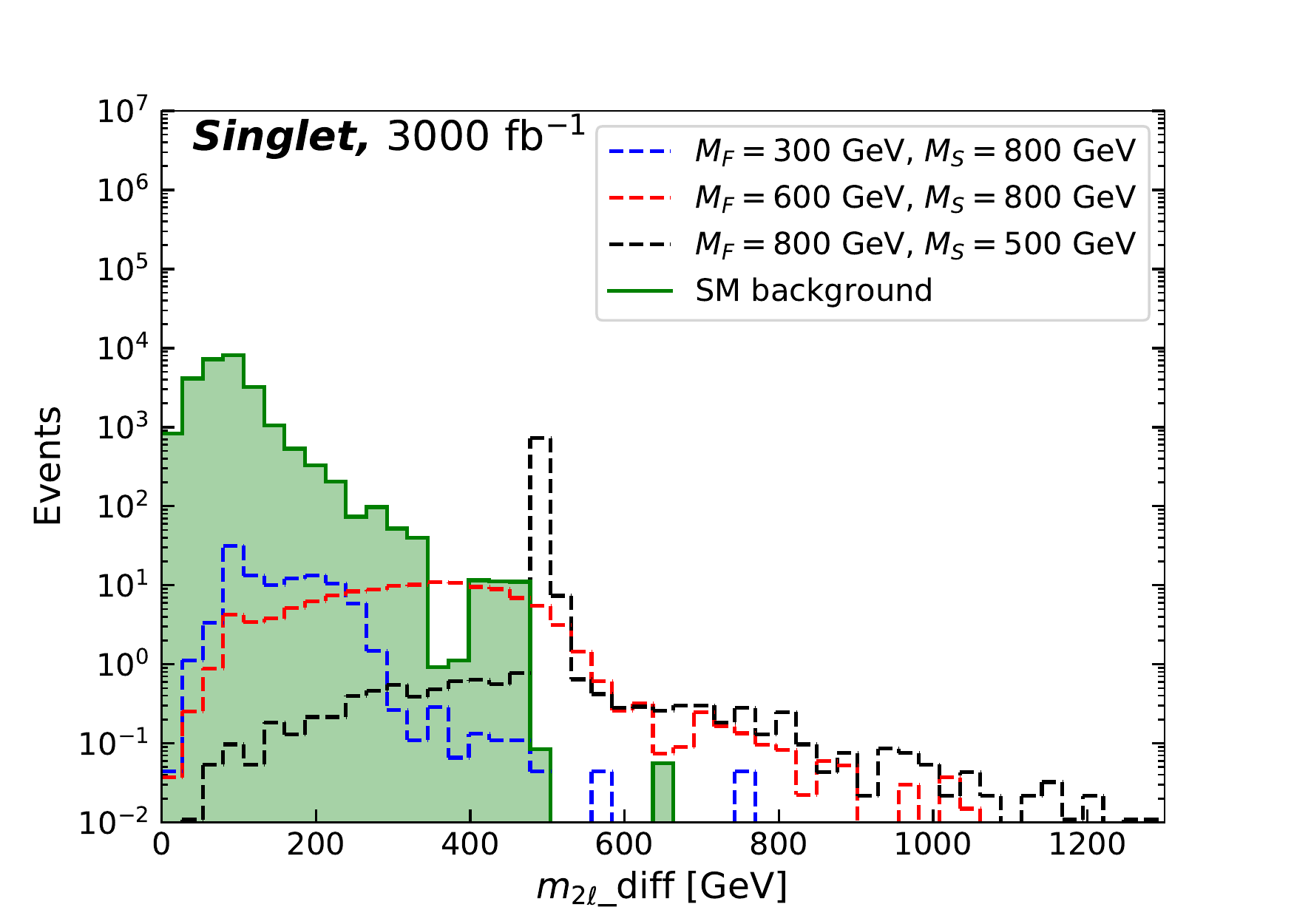}
	\includegraphics[width=0.49\textwidth]{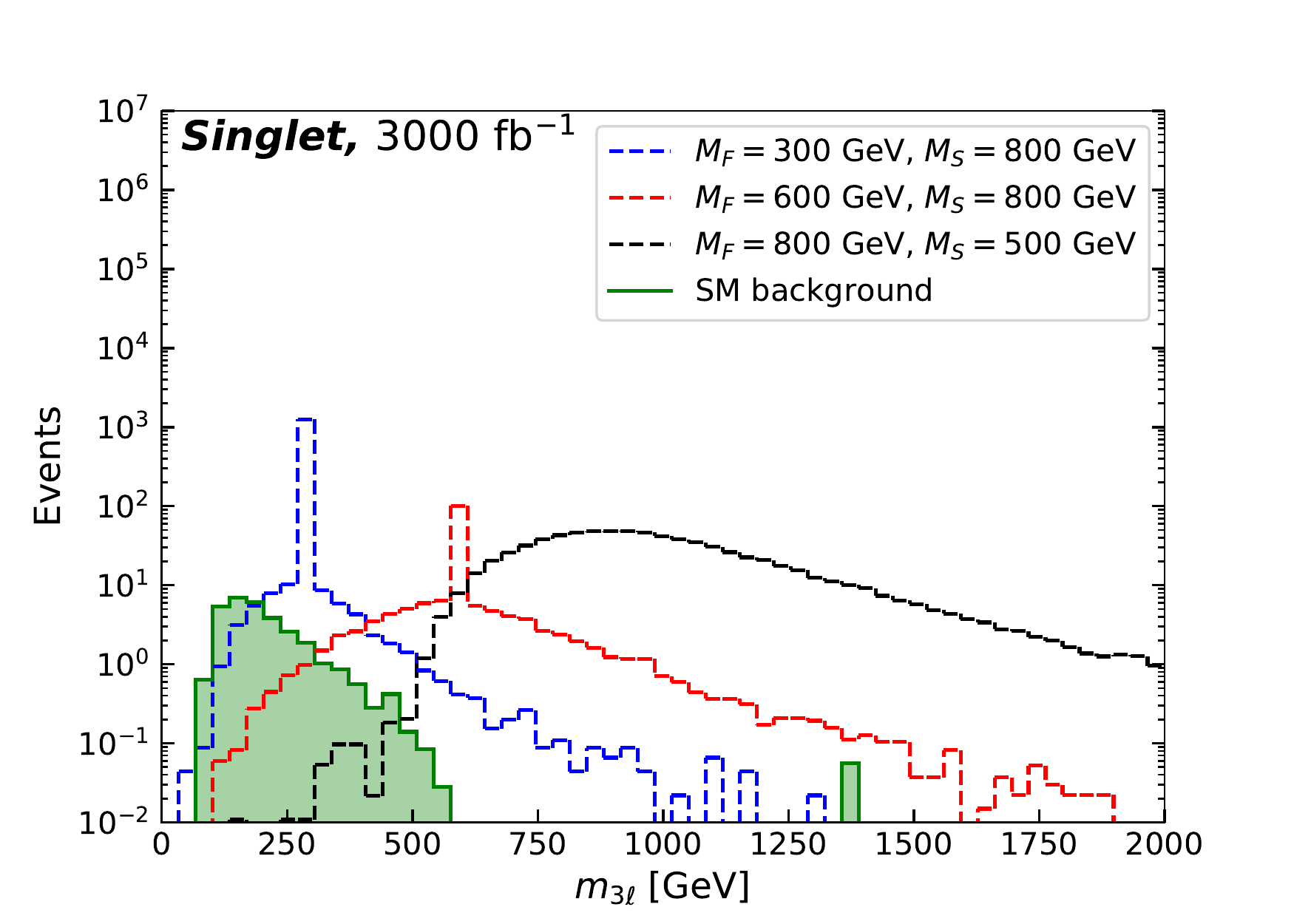}
	\includegraphics[width=0.49\textwidth]{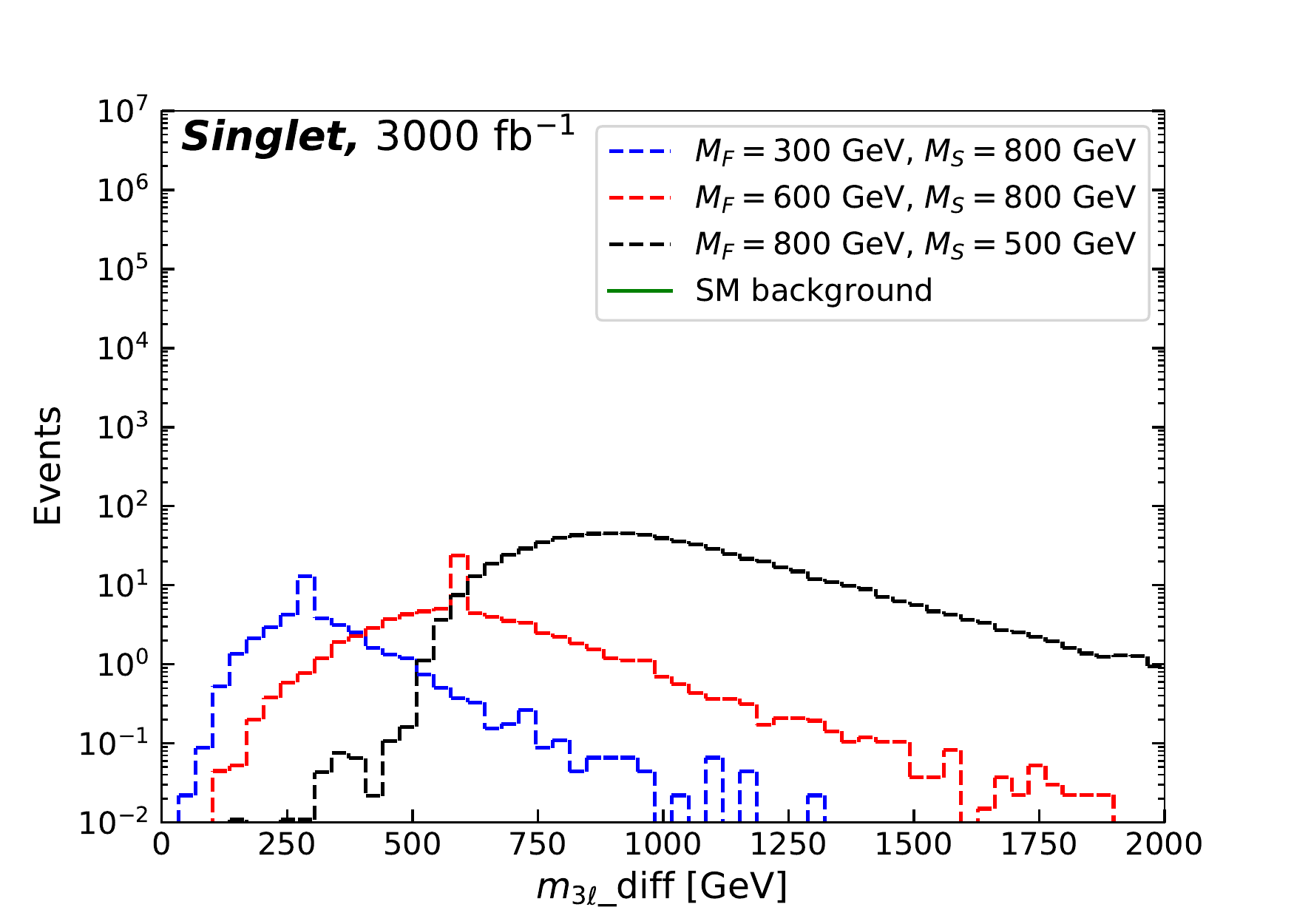}
	\caption{As in Fig.~\ref{fig:singlet-bench} but for  higher luminosity $3000$ fb${}^{-1}$ and $\sqrt{s}=14$~TeV.}
	\label{fig:singlet-bench-14tev}
\end{figure}

\begin{figure}
	\centering
	\includegraphics[width=0.49\textwidth]{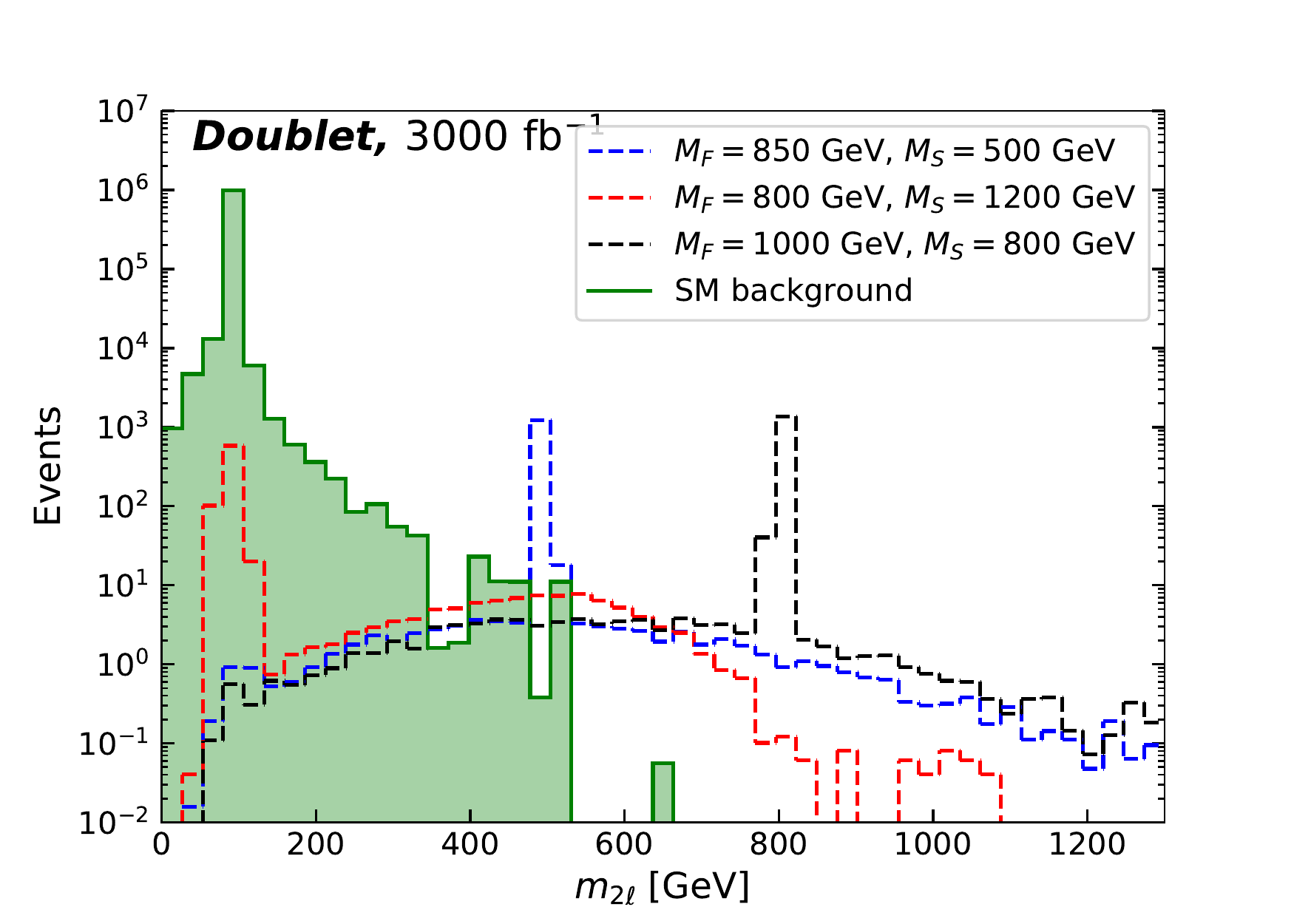}
	\includegraphics[width=0.49\textwidth]{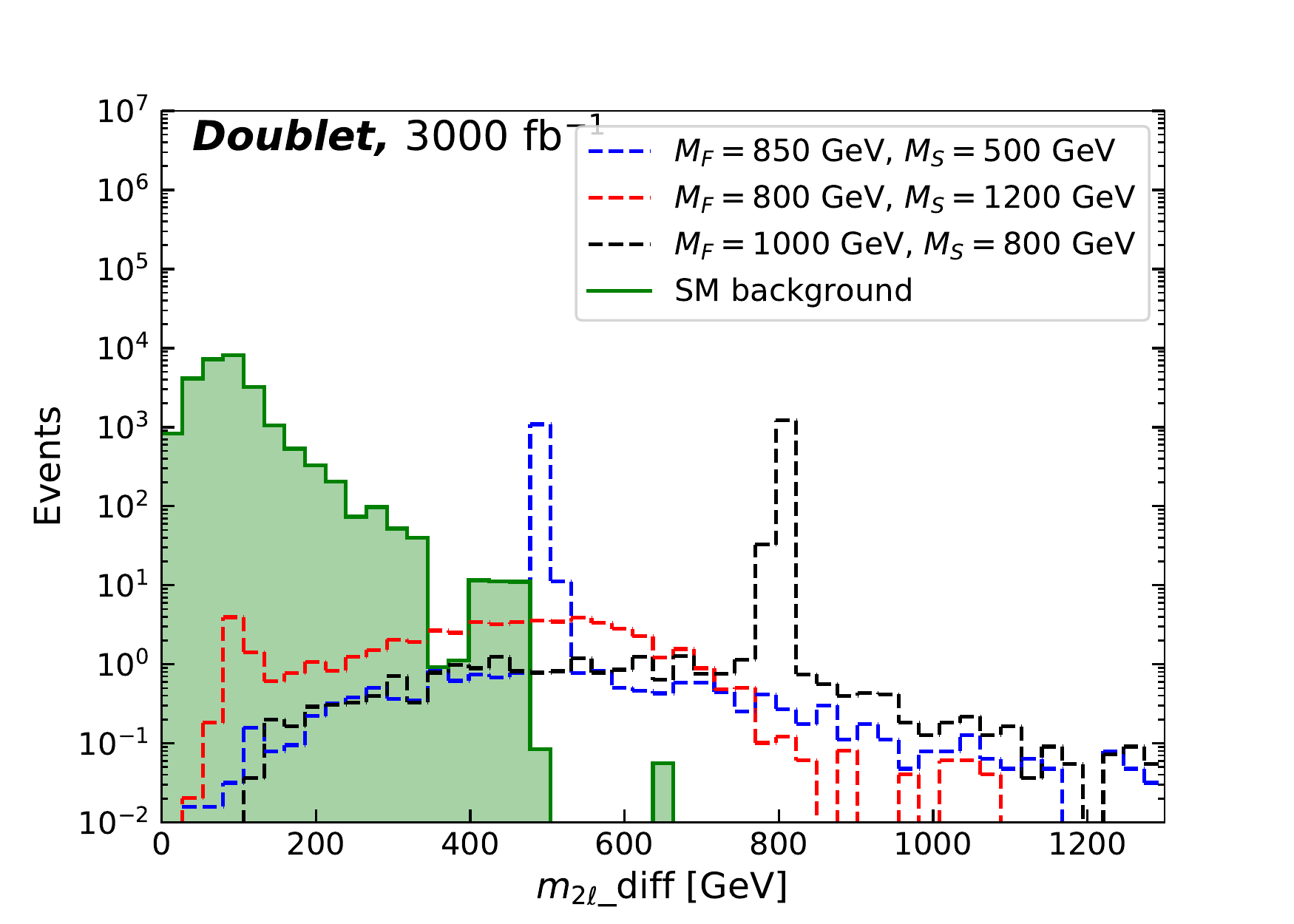}
	\includegraphics[width=0.49\textwidth]{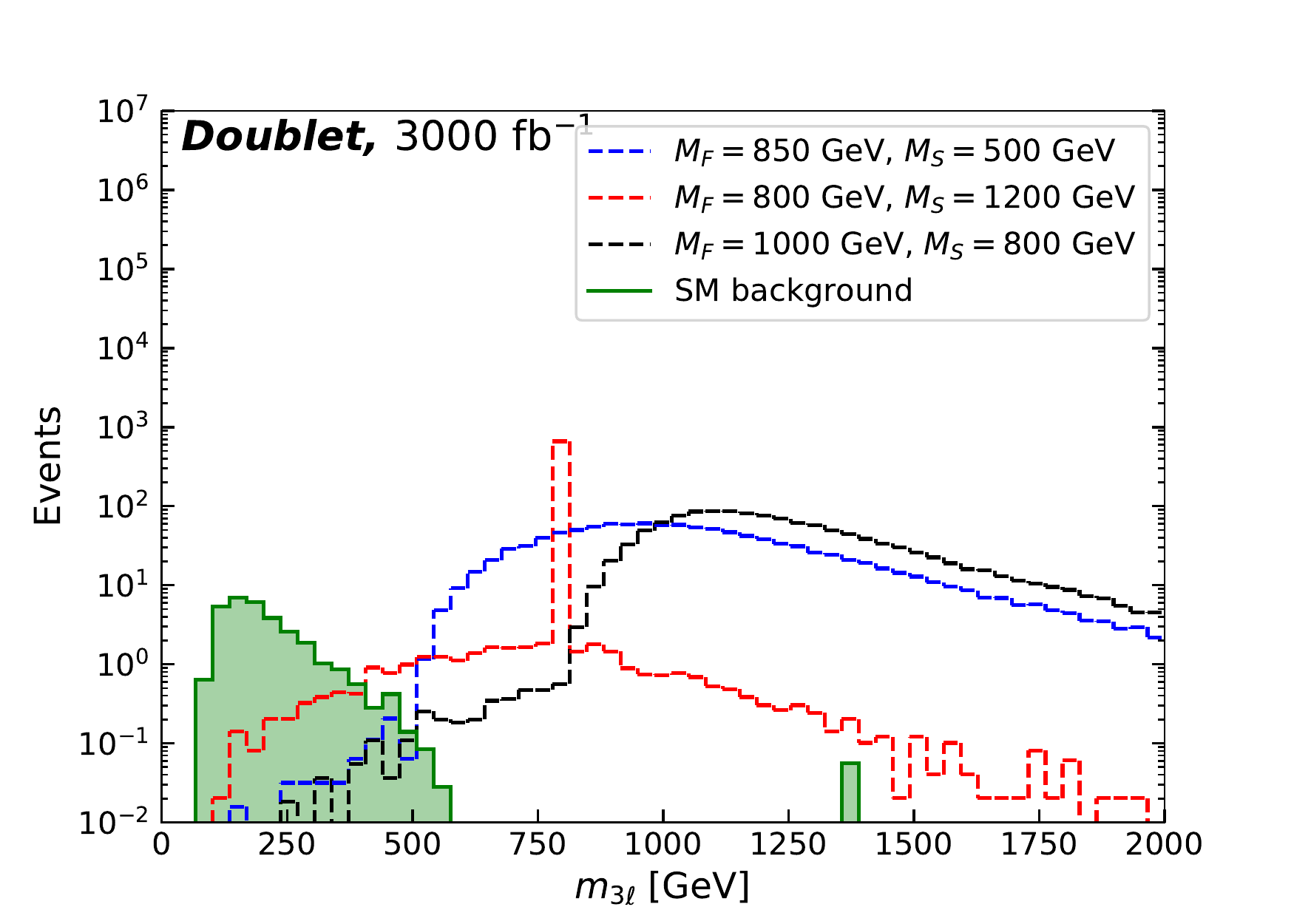}
	\includegraphics[width=0.49\textwidth]{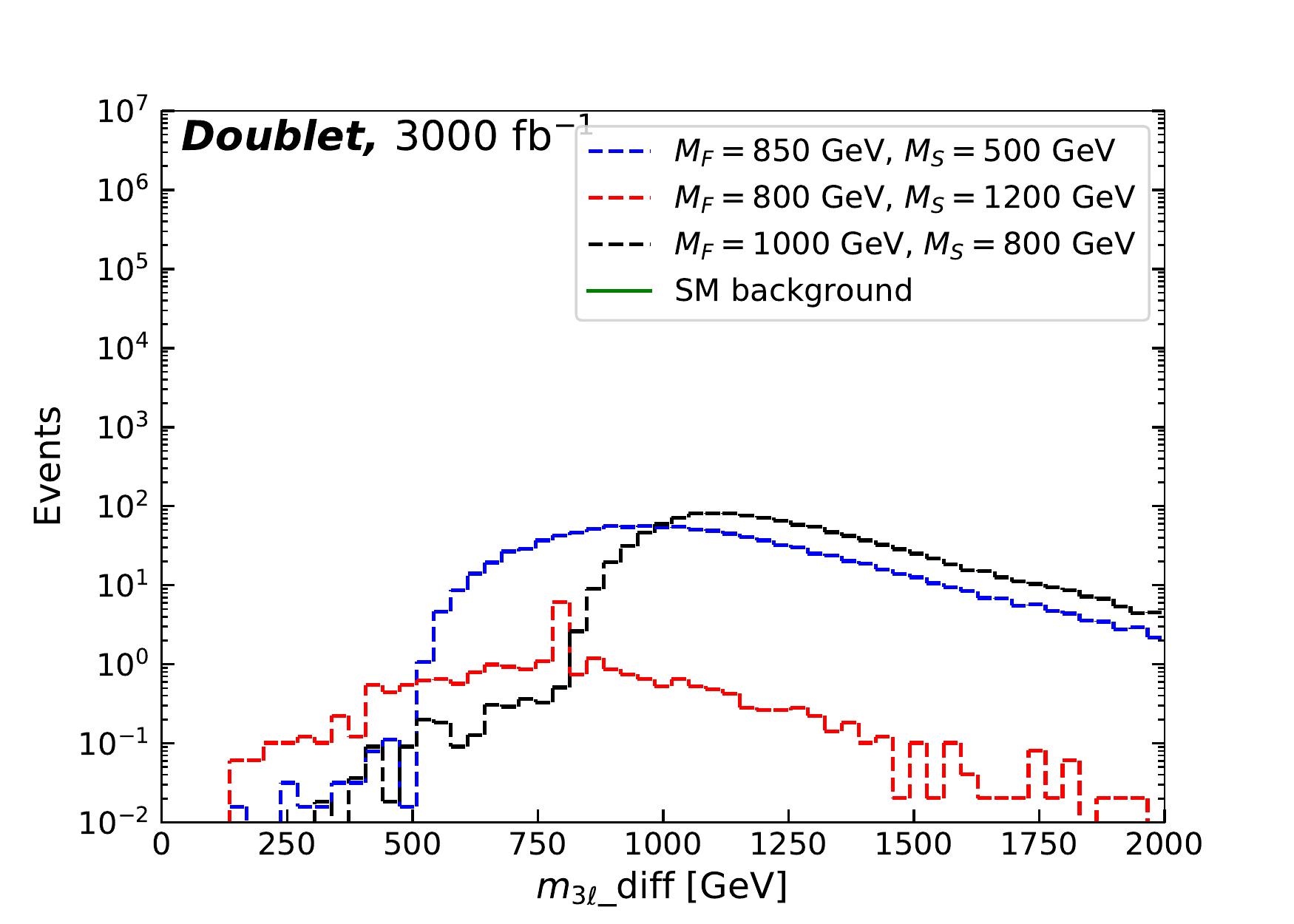}
	\caption{As in Fig.~\ref{fig:doublet-bench} but for  higher luminosity $3000$ fb${}^{-1}$ and $\sqrt{s}=14$~TeV.}
	\label{fig:doublet-bench-14tev}
\end{figure}

As shown  in Sec.~\ref{sec:run2}  the discovery of a BSM sector consisting of VLLs and new scalars with a non-trivial flavor structure remains a challenging task at 
Run 2.  
Here we study  the new observables $m_{2\ell}$,  $m_{2\ell}\_{\rm diff}$, $m_{3\ell}$ and $m_{3\ell}\_{\rm diff}$ for the benchmarks
scenarios of Sec.~\ref{sec:Null-test} at the HL-LHC, at  higher luminosity $3000~{\rm fb}^{-1}$, for  upgraded detectors and $\sqrt{s} = 14$~TeV \cite{ApollinariG.:2017ojx}. For the detector simulation we employ \textsc{DELPHES3} with the HL-LHC default card instead of the CMS default card. The distributions  before detector simulation are shown in Figs.~\ref{fig:singlet-bench-14tev} and \ref{fig:doublet-bench-14tev}  and after detector simulation in Figs.~\ref{fig:singlet-bench-14tev-delphes} and \ref{fig:doublet-bench-14tev-delphes}  for the singlet and doublet models, respectively. 

The HL-LHC set-up enhances event rates relative to Run 2, in both models and benchmarks, both  signal peaks and the SM background, according to $\sim 3000/150=20$. Despite the different detector settings, and the increased center of mass energy, the corresponding distributions from Run 2 and the HL-LHC are very similar. For example, 
the  singlet model  3000 fb$^{-1}$  plots in Fig.~\ref{fig:singlet-bench-14tev}  essentially look like scaled-up versions of the 150 fb$^{-1}$ ones shown in   Fig.~\ref{fig:singlet-bench}.
The scaling factors between  no  detector simulations and including them given in Tab.~\ref{tab:scale-factors}  remain also very similar  between the two LHC settings with larger scaling in the HL-LHC scenario due to improved detector settings, see
App.~\ref{app:delphes} for details.
For example, 
the  singlet model  3000 fb$^{-1}$  plots with detector simulation in Fig.~\ref{fig:singlet-bench-14tev-delphes}  essentially look like scaled-up versions of the 150 fb$^{-1}$ ones shown in   Fig.~\ref{fig:singlet-bench-delphes}, and similarly for the doublet model. At the HL-LHC
the new observables continue to feature great separation of BSM signals  from the SM background, just with (more) events.

In the $m_{2\ell}\_{\rm diff}$ spectra  the bins with $m_{2\ell}\_{\rm diff} \gtrsim 500$ GeV allow to search for  both on-shell and off-shell $S$-production. For the former, we find $\mathcal{O}(10^3)$ events ($\mathcal{O}(20)$ after detector simulation)  in the peaks of the $m_{2\ell}\_{\rm diff}$ distribution for both the singlet and doublet model.
The $ m_{3\ell}\_{\rm diff}$ distributions have  in both models  $\mathcal{O}(10-10^2)$ events per bin ($\mathcal{O}(1-10)$ events per bin after detector simulation) with the exception of the doublet benchmark with light VLLs  $M_F  \ll M_S$ (red curve in doublet model). Here, the  $ m_{3\ell}$ distribution turns out to be powerful
and produces up to $\mathcal{O}(10-10^2)$ events per bin after detector simulation.
The  $ m_{3\ell}$ distribution enhances also the peak in the singlet model with light VLLs and hierarchical spectrum $M_F=300$ GeV,  $M_S=800$ GeV (blue curves in singlet model) up to this level.

We conclude that at  the HL-LHC  the ($M_F$, $M_S$)-parameter space consistent with the $g-2$ anomalies can be probed and mass hierarchies extracted.

\section{Beyond $g-2$}\label{sec:beyondg2}

 \begin{figure}
	\centering
	\includegraphics[width=0.49\textwidth]{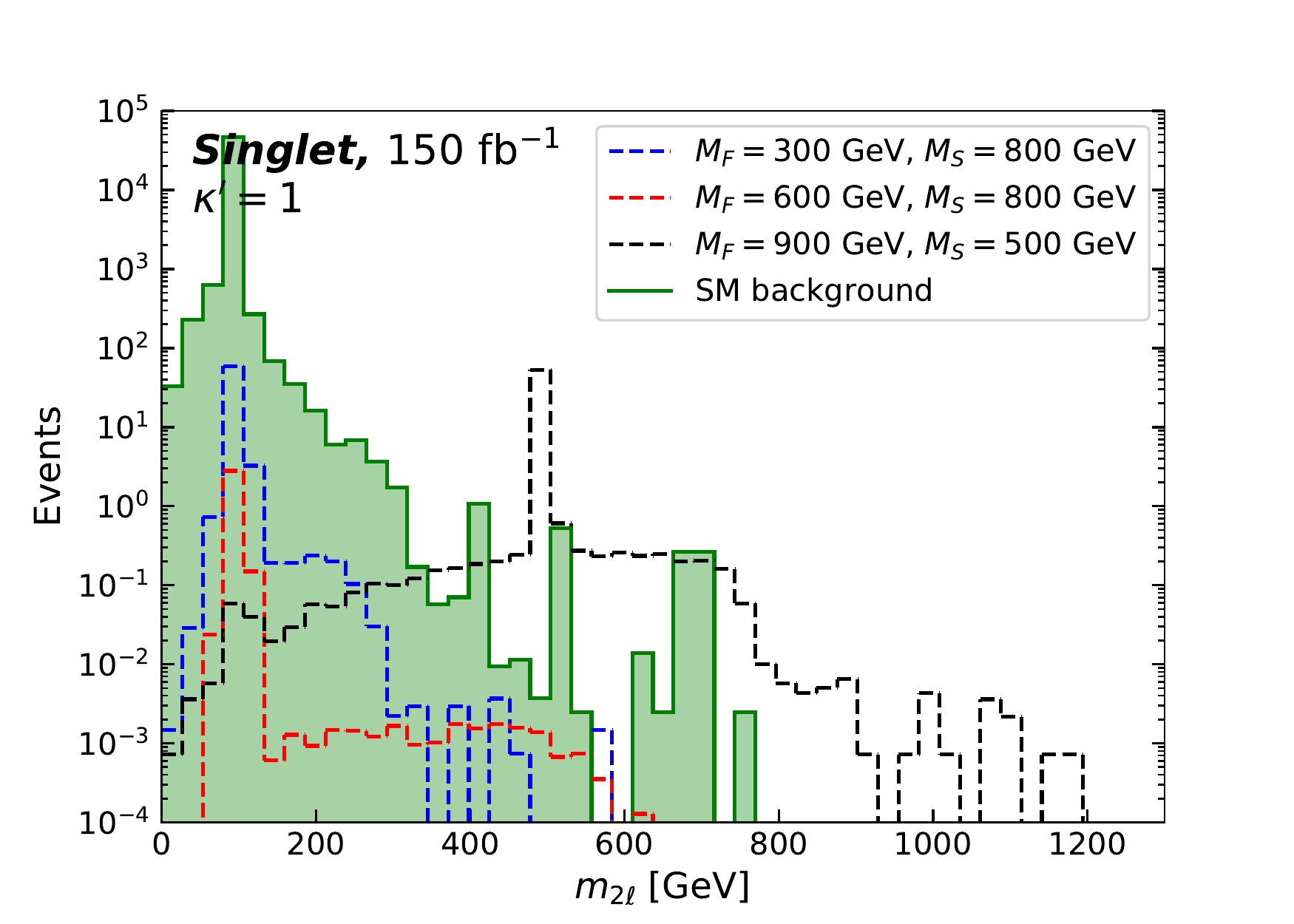}
	\includegraphics[width=0.49\textwidth]{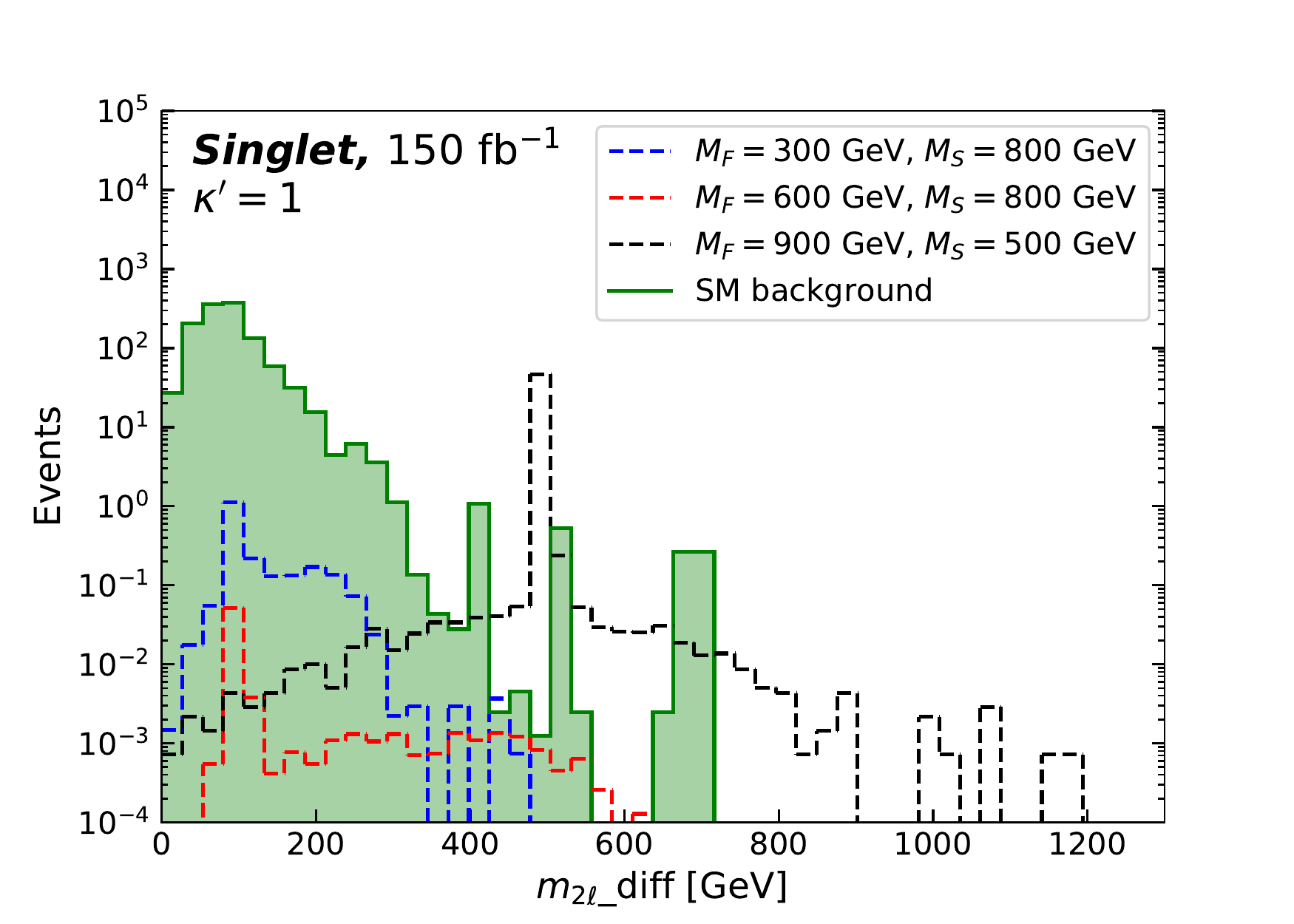}
	\includegraphics[width=0.49\textwidth]{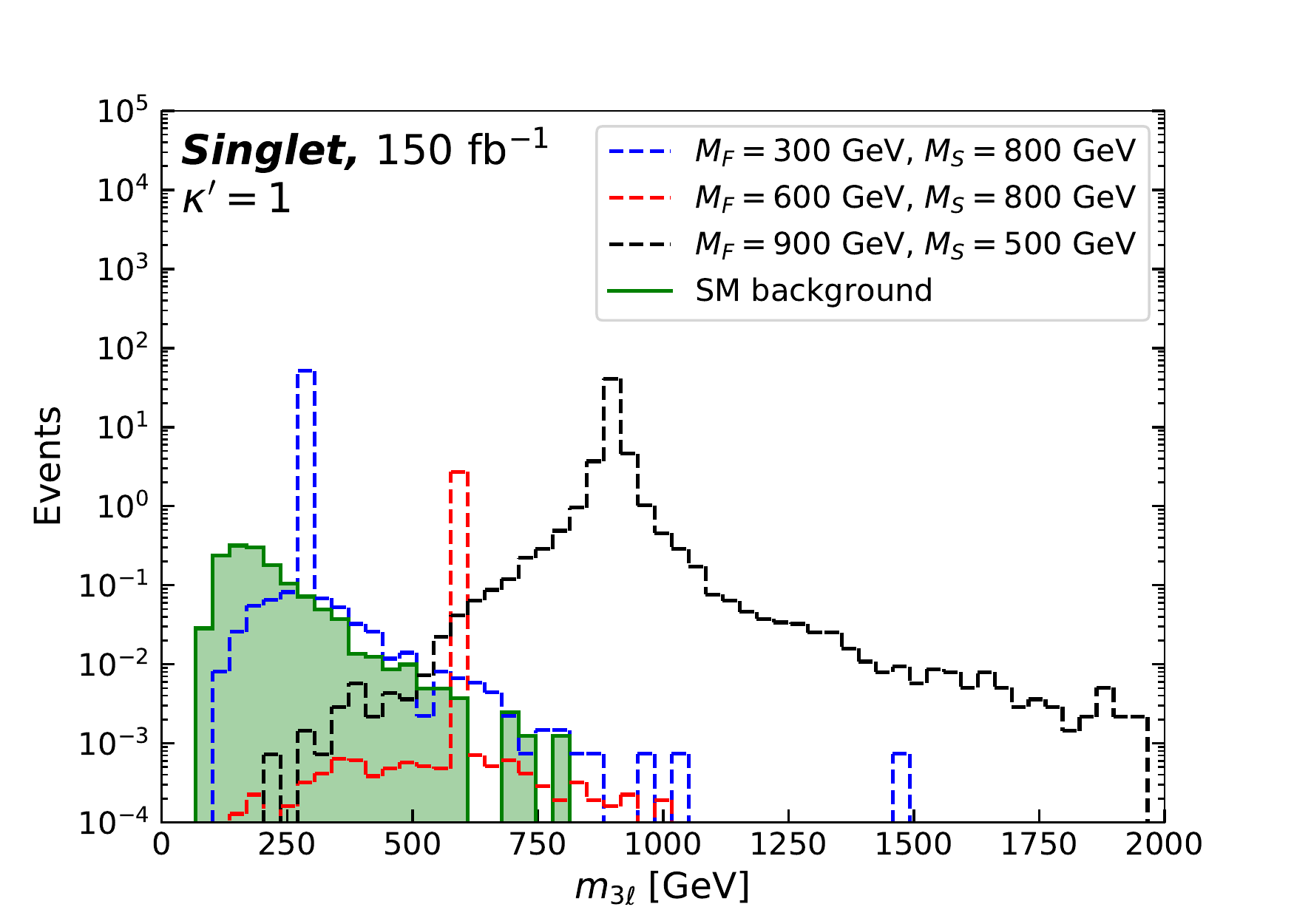}
	\includegraphics[width=0.49\textwidth]{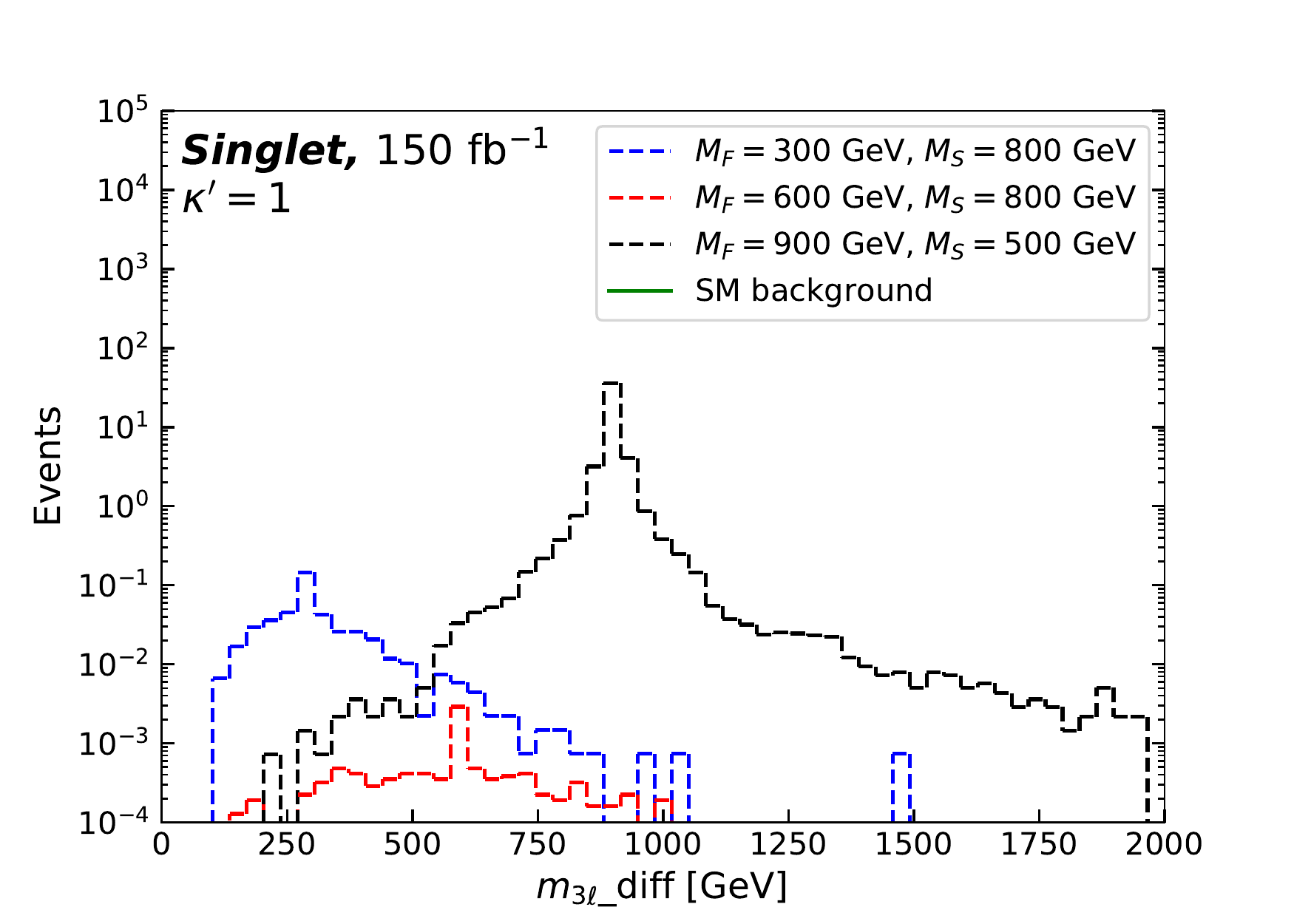}
	\caption{Di- and trilepton invariant mass distributions  $m_{2\ell}$,  $m_{2\ell}\_{\rm diff}$, $m_{3\ell}$, and $m_{3\ell}\_{\rm diff}$  for the singlet model with 
	$\kappa^\prime=1$, for the full Run 2 luminosity $150$ fb${}^{-1}$ and $\sqrt{s}=13$~TeV. 
	}
	\label{fig:singlet-bench-k1}
\end{figure}

 \begin{figure}
	\centering
	\includegraphics[width=0.49\textwidth]{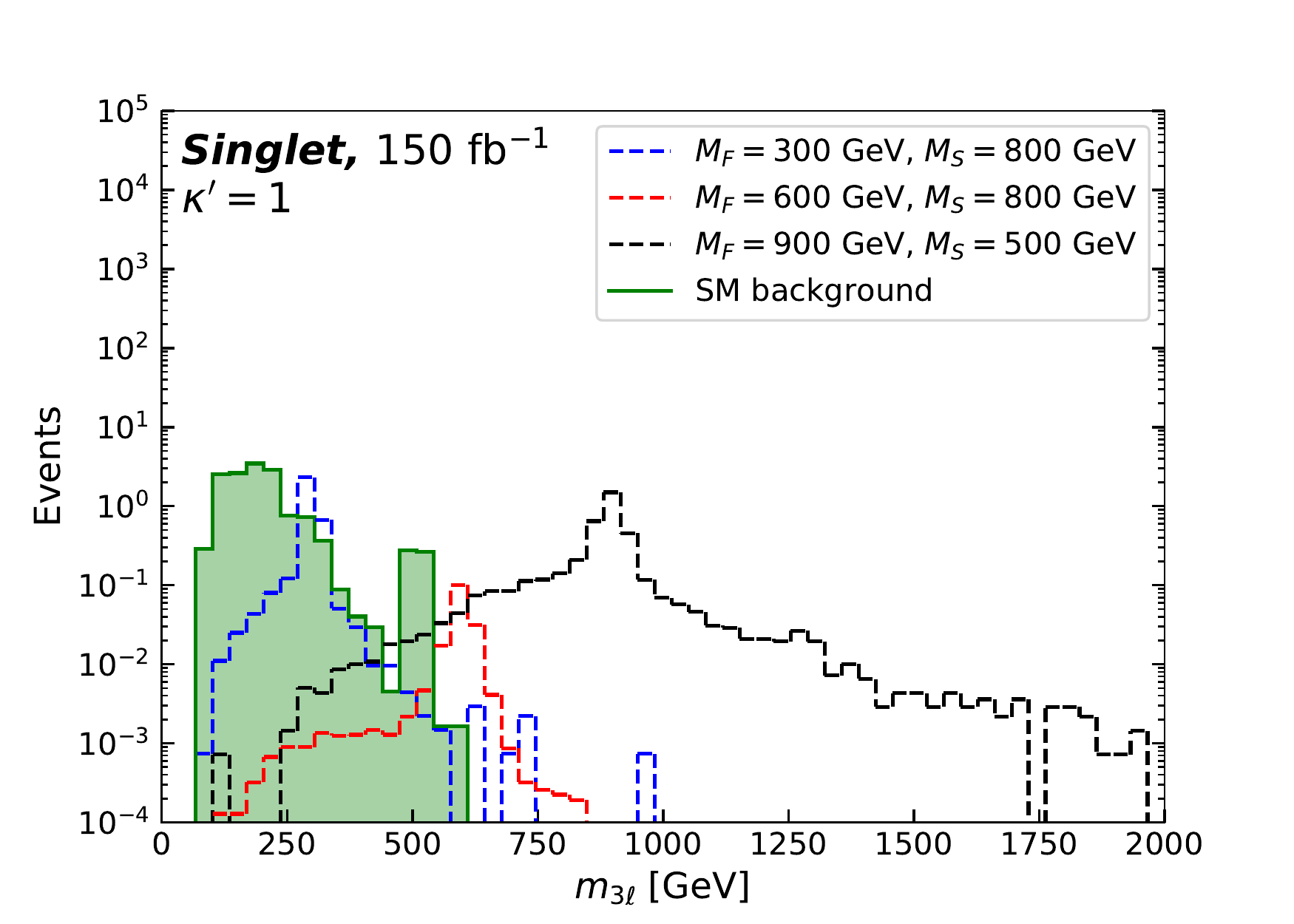}
	\includegraphics[width=0.49\textwidth]{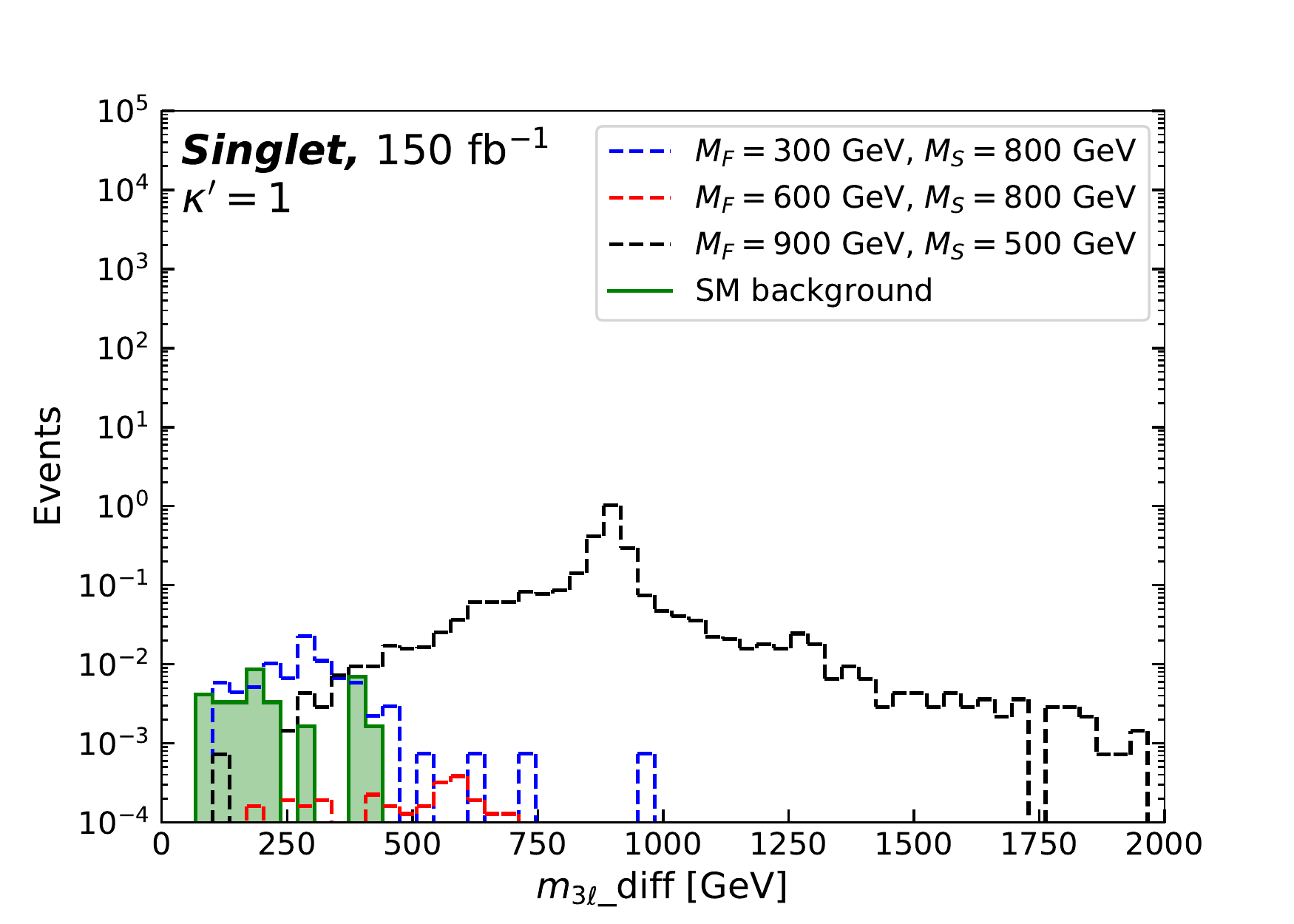}
	\caption{	
	Trilepton invariant mass distributions  $m_{3\ell}$ and $m_{3\ell}\_{\rm diff}$  for the singlet model with  $\kappa^\prime=1$ after detector simulation,
	 for the full Run 2 luminosity $150$ fb${}^{-1}$ and $\sqrt{s}=13$~TeV.}
	\label{fig:singlet-bench-k1-delphes}
\end{figure}

In the previous sections we have studied the VLL models in Eq.~\eqref{Lint-singlet} and Eq.~\eqref{Lint-doublet} in the parameter space where the coupling $\kappa^\prime$ is fixed by the BSM masses $M_F,M_S$ \eqref{eq:para}. In this section, we analyze the model space
beyond the $g-2$ constraint, entertaining the possibility of a shift in $\Delta a_\mu$ due to improved data and theory.

In general, the coupling $\kappa$ remains limited in magnitude from above  by $Z$ decays, inducing small but relevant effects in fermion mixing.
On the other hand, $\kappa^\prime$ is unconstrained  by electroweak data.  As it is already rather sizable in the benchmark (\ref{eq:para}), we investigate the implications of a reduced $\kappa^\prime$.
The latter implies a suppression of $\psi$ to $S$ plus lepton decays. Since these modes are the dominant ones for $M_F >M_S$, see Fig.~\ref{Fig:BR_psi},
the width of the $\psi$ in this region is proportional to $\kappa^{\prime 2}$.  We expect therefore narrower resonances in $M_F >M_S$
and a suppression of events in the region $M_F <M_S$.

One may wonder what happens if $\kappa^\prime$ vanishes. In this case the
models could still  produce lepton flavor violation-like signals for  $y\neq 0$, with $\psi\to S\ell$ happening  at order $y\theta$ and $S \to \ell \ell^{(\prime)}$ at order
$y \theta^2 v_h/M_F$ times the lepton Yukawa with the Higgs. Due to the $Z$-constraints on the mixing angle $\theta  < \mathcal{O}(10^{-2})$ the "diff"-observables would be strongly suppressed up to  some statistical noise.
This outcome holds also for other UV-safe models with flavorful VLLs in  representations of $SU(2)_L \times U(1)_Y$ which do not allow for a $\psi$-$S$-lepton
Yukawa  coupling ("$\kappa^\prime=0$") \cite{Hiller:2020fbu}. Furthermore, 
if $M_S$ were very heavy, in  all models with  mixed SM-BSM Yukawas  the "diff"-observables would be SM-like. Note that
the phenomenology of VLLs without any mixed SM-BSM Yukawas ("$\kappa=\kappa^\prime=0$") is markedly different and discussed for various exotic representations in \cite{Bond:2017wut}.

In the following we focus on the singlet model, since the allowed parameter space of BSM masses is larger, see Fig.~\ref{fig:parameter-space}.
For simplicity we use $\kappa^\prime=1$.
We find that the benchmark $M_F=800$~GeV, $M_S=500$~GeV and $\kappa^\prime=1$ is excluded by CMS data, even though the $g-2$ benchmark scenario with a larger coupling but the same BSM masses was found to fall within the allowed region (see Fig.~\ref{fig:L_T_delphes-benchmark}). Nevertheless, we observe that larger VLL masses  $M_F\sim900$~GeV are allowed for  $M_S\sim500$~GeV and $\kappa'=1$. Therefore, in this section we take $M_F=900$~GeV, $M_S=500$~GeV as one of our benchmarks.
We study as well the $\kappa'=1$ counterparts of the two remaining benchmarks considered in previous sections, which we find to be allowed.

The corresponding distributions of the new observables  at the $\sqrt{s}=13$ TeV LHC and the full Run 2 data set are shown in Fig.~\ref{fig:singlet-bench-k1}.
We observe that for $m_{2\ell}$ and $m_{2\ell}\_{\rm diff }$ (see Fig.~\ref{fig:singlet-bench-k1}, upper row), the patterns are very similar to the 
$g-2$ benchmarks shown   in Fig.~\ref{fig:singlet-bench}, with  resonance peaks when the scalars can be produced on-shell (black curves).
For $m_{3\ell}$ and $m_{3\ell}\_{\rm diff }$ (see Fig.~\ref{fig:singlet-bench-k1}, lower row), however, reducing  $\kappa^\prime$ leads to qualitatively different effects. 
As anticipated, these are mainly due to the VLLs' narrower widths, which can be seen in all benchmarks. $m_{3\ell}$ is the observable with more distinctive peaks regardless of the BSM mass hierarchy, with resonances  above the SM background and reaching at least $\mathcal{O}(10)$ events in the peak bin for all benchmarks. 
For $M_F>M_S$ (black curves)  $m_{3\ell}\_{\rm diff }$ is the optimal observable, since all SM background is suppressed and the number of events per bin barely decreases with respect to $m_{3\ell}$. For  $M_F<M_S$, the $m_{3\ell}\_{\rm diff }$  distributions are substantially depleted with respect to $m_{3\ell}$, and higher luminosities would be beneficial. 

In Fig.~\ref{fig:singlet-bench-k1-delphes}  we give  the  $m_{3\ell}$ and $m_{3\ell}\_{\rm diff }$ distributions after hadronization and detector simulation The $m_{2\ell}$ and $m_{2\ell}\_{\rm diff }$ spectra are very similar to the ones in Fig.~\ref{fig:singlet-bench-delphes} and therefore not shown. The $m_{3\ell}$ distributions show peaks for all three benchmarks, with $\mathcal{O}(1)$ events per bin  for both $M_F=300$~GeV (blue) and $M_F=900$~GeV (black). For the latter, the $m_{3\ell}\_{\rm diff }$ distribution allows for a null test, as the SM background is sufficiently suppressed while we find a peak with few events per bin in the $M_F=900$~GeV distribution. 

\section{Summary}
\label{sec:conclusion}

We  investigated  opportunities at the LHC and the HL-LHC  to search for flavorful vector-like leptons $\psi_i$ and new scalar singlets $S_{ij}$. 
Such BSM sector (\ref{Yukawa})  occurs in novel model building frameworks with favorable UV behavior  \cite{Litim:2014uca,Bond:2016dvk,Bond:2018oco}, and
particle physics phenomenology \cite{Bond:2017wut,Hiller:2019mou}.

We considered two explicit BSM models of this kind, featuring three generations of either $SU(2)_L$ singlet or doublet VLLs, which can also accommodate present data of the muon and electron $g-2$.
Key ingredients for flavor phenomenology are the mixed SM-BSM Yukawa couplings, the flavor matrix structure of the BSM scalars, the identification of lepton and VLL
flavor, and fermion mixing after electroweak symmetry breaking.
Although all BSM interactions \eqref{Lint-singlet} and \eqref{Lint-doublet} are  flavor-conserving, the decays of the VLLs through the $S_{ij}$ and their subsequent decay $S_{ij}\to \ell_i^+\ell_j^-$ lead to production of different-flavor lepton pairs, a signature we exploited to construct novel null tests of the SM:
The dilepton invariant masses $m_{2\ell}$ and $m_{2\ell}\_{\rm diff}$, which permit to look for scalar resonances, and the three-lepton invariant masses $m_{3\ell}$ and $m_{3\ell}\_{\rm diff}$, which are designed to reconstruct VLL masses, as described in Sec.~\ref{sec:Null-test}. The $\_{\rm diff}$ distributions are populated exclusively by invariant masses which contain at least two leptons of different flavor and opposite charge, which results in a strong suppression of the SM background and targets models with a non-trivial flavor structure affecting the charged lepton sector.
The background suppression is especially efficient for $m_{3\ell}\_{\rm diff}$, which makes this an excellent null test.

For our study we implemented the models into UFO models using \textsc{FeynRules}. Predictions for observables including dominant SM background processes at $pp$-colliders are computed with \MG together with \textsc{PYTHIA8} and \textsc{DELPHES3}.
We worked out constraints from a CMS search  in final states with at least four light leptons (electrons, or muons) \cite{Sirunyan:2019ofn}.
Results are summarized in Fig.~\ref{fig:parameter-space}, showing allowed regions (green and yellow circles)  of VLL and scalar masses while accommodating $g-2$ of the muon, (\ref{eq:para}). 
We find that, in general, regions around the $M_S = M_F$ line are excluded by data due to  the underlying enhancement of cross sections via on-shell $S$ production.
Lower limits for the VLL masses  are around 300 GeV in the singlet model and 800 GeV in the doublet model.
As such, our findings offer new constraints for the Planck safe models put forward in  \cite{Hiller:2019mou}.

Predictions for the new observables  $m_{2\ell}$,  $m_{2\ell}\_{\rm diff}$, $m_{3\ell}$, and $m_{3\ell}\_{\rm diff}$  after detector simulation are shown for several allowed benchmarks in Figs.~\ref{fig:singlet-bench-delphes} and \ref{fig:doublet-bench-delphes} for the singlet and doublet model, respectively, for the full Run 2 data set with 150 fb${}^{-1}$. 
The distributions exhibit  a highly discriminating power on the BSM mass hierarchy, but  suffer from marginal event rates and therefore would extremely benefit from higher  luminosity. 
At the HL-LHC, for $\sqrt{s}=14$ TeV and a luminosity of 3000 ${\rm fb}^{-1}$, we obtain $\mathcal{O}(10^2)$ events after detector simulation in some bins, see Figs.~\ref{fig:singlet-bench-14tev-delphes} and \ref{fig:doublet-bench-14tev-delphes} for the singlet and doublet model, respectively. 
Hence, these new, optimized observables are very promising for higher luminosity runs at the LHC, to discover and discern hierarchies in flavorful models with 
multi-lepton final states.

Studying more general versions of our models in Sec.~\ref{sec:beyondg2} we reduce $\kappa^\prime$, the  key Yukawa for filling the "diff"-distributions.
Results are shown in Fig.~\ref{fig:singlet-bench-k1-delphes}  for the invariant mass distributions after detector simulation. We again observe striking BSM signatures with
diagnosing power.
Let us also mention that the other colorless models with effectively $\kappa^\prime=0$ put forward as asymptotically safe extensions of the SM  \cite{Hiller:2020fbu}
are not contributing significantly to the "diff"-observables,  but could be probed  using $m_{2 \ell}$ and $m_{3 \ell}$ or  conventional VLL search strategies.

\section*{Acknowledgements}
We are grateful to Jonas Lindert for pointing out the importance of (multi-)jet contributions to the $ZZ$ background.
SB and CHF would like to thank Dennis Loose and B\"orn Wendland for helping with the  implementation of the models in MadGraph and related tools, and useful discussions. CHF is grateful to INFN, Sezione di Roma, for hospitality while this work was finalized.
DL is supported by the Science and Technology Research Council (STFC) under the Consolidated Grant ST/T00102X/1.
\appendix

\section{HL-LHC Distributions after detector simulation}\label{app:delphes}

 \begin{figure}
	\centering
	\includegraphics[width=0.49\textwidth]{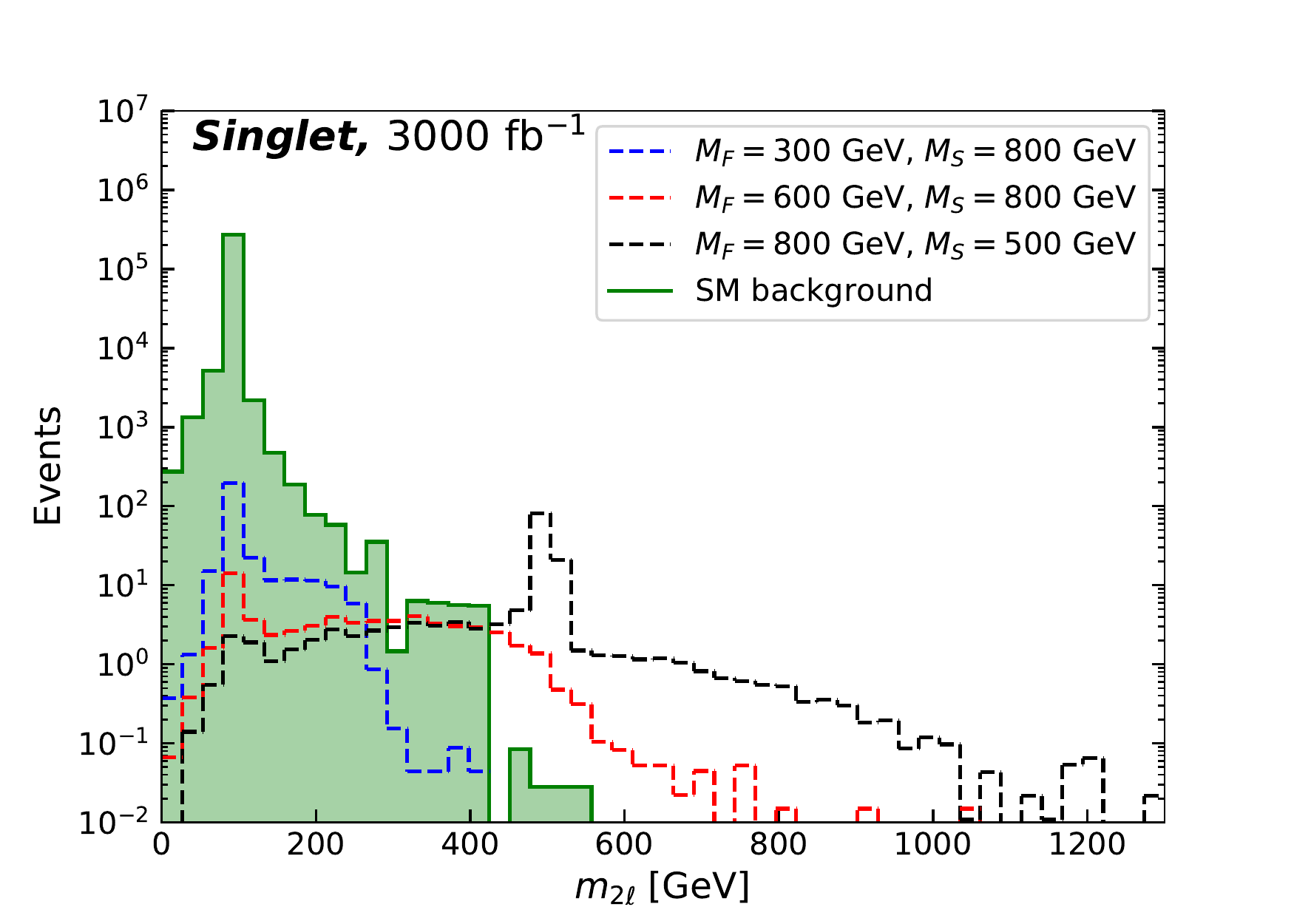}
	\includegraphics[width=0.49\textwidth]{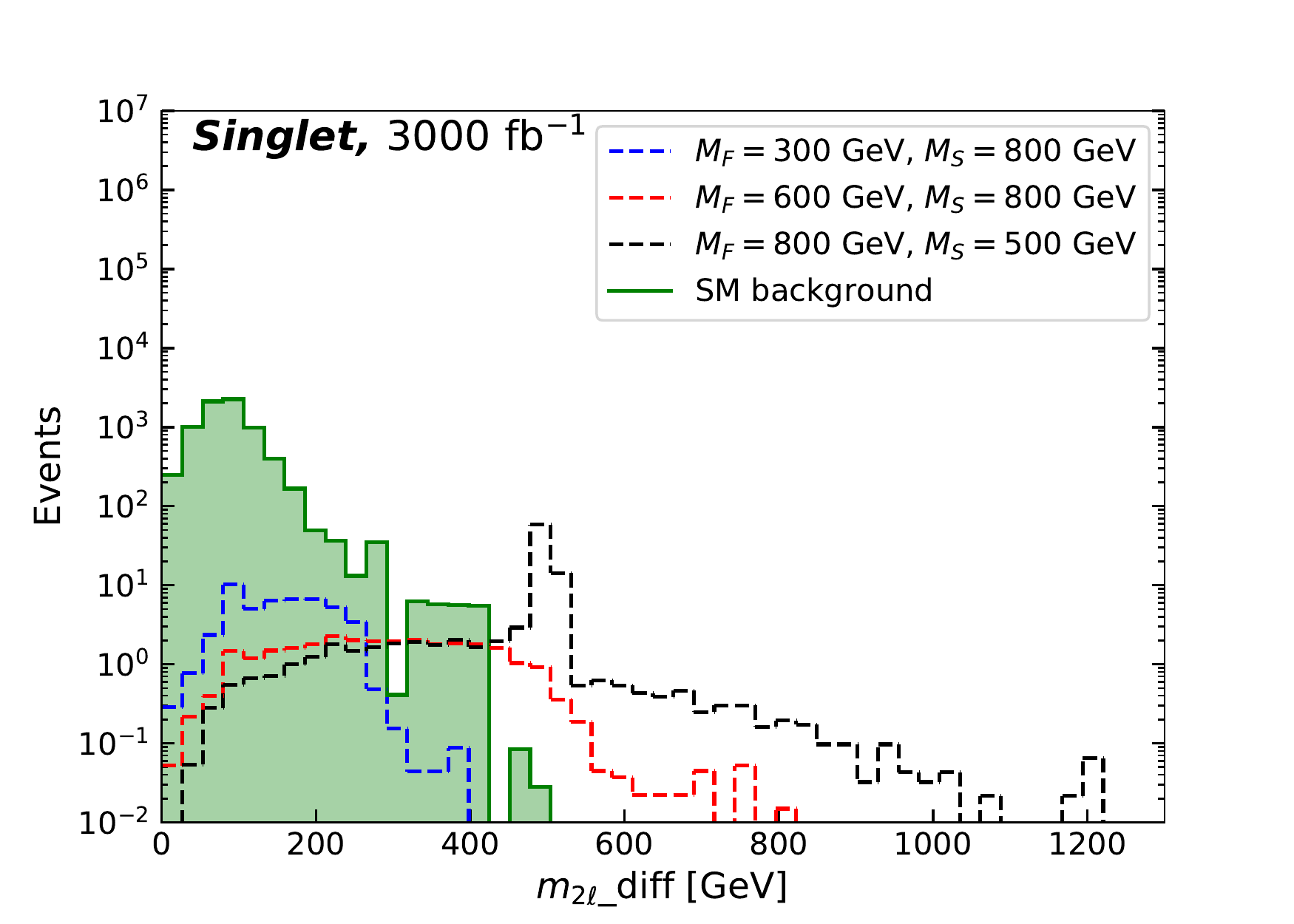}
	\includegraphics[width=0.49\textwidth]{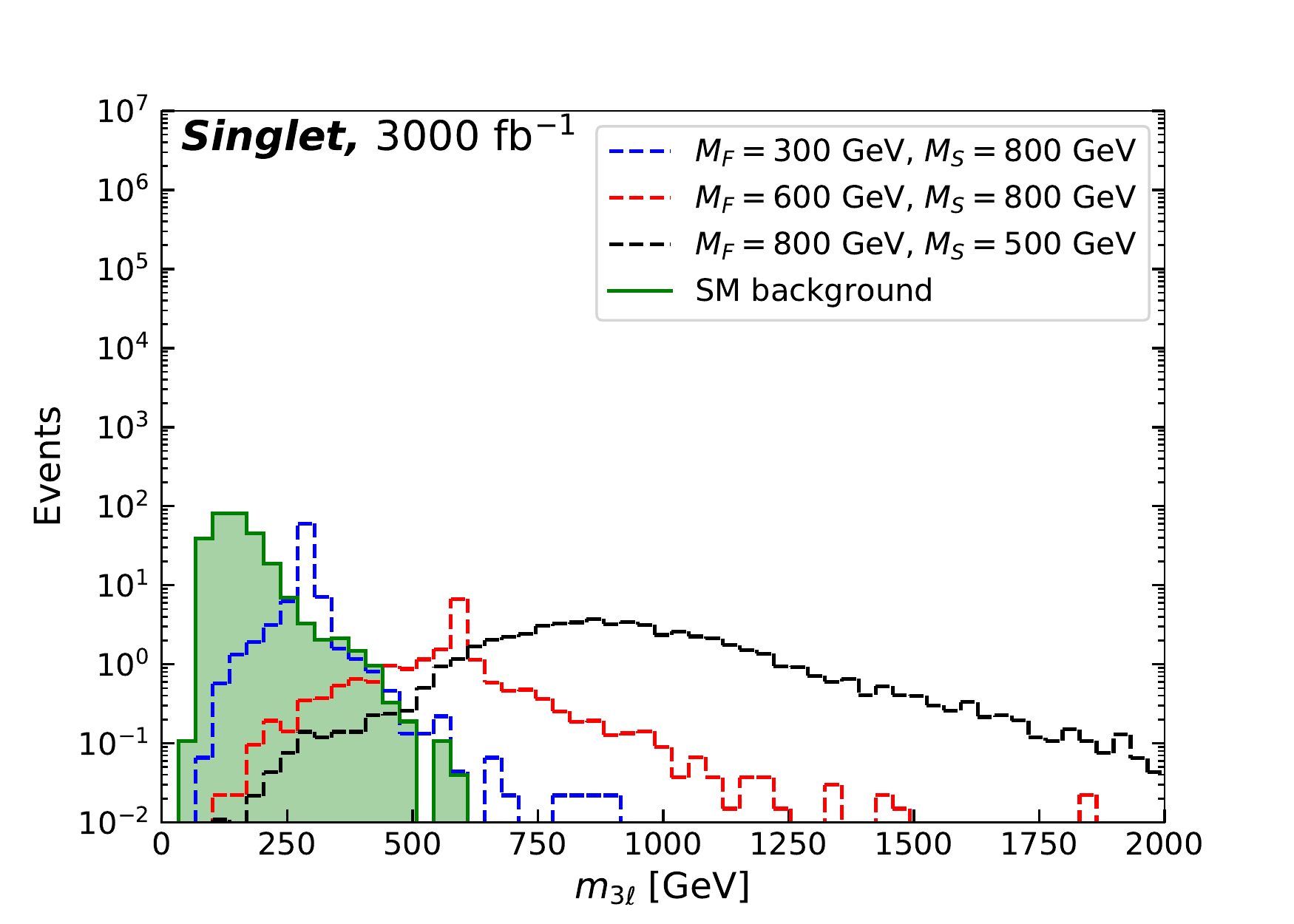}
	\includegraphics[width=0.49\textwidth]{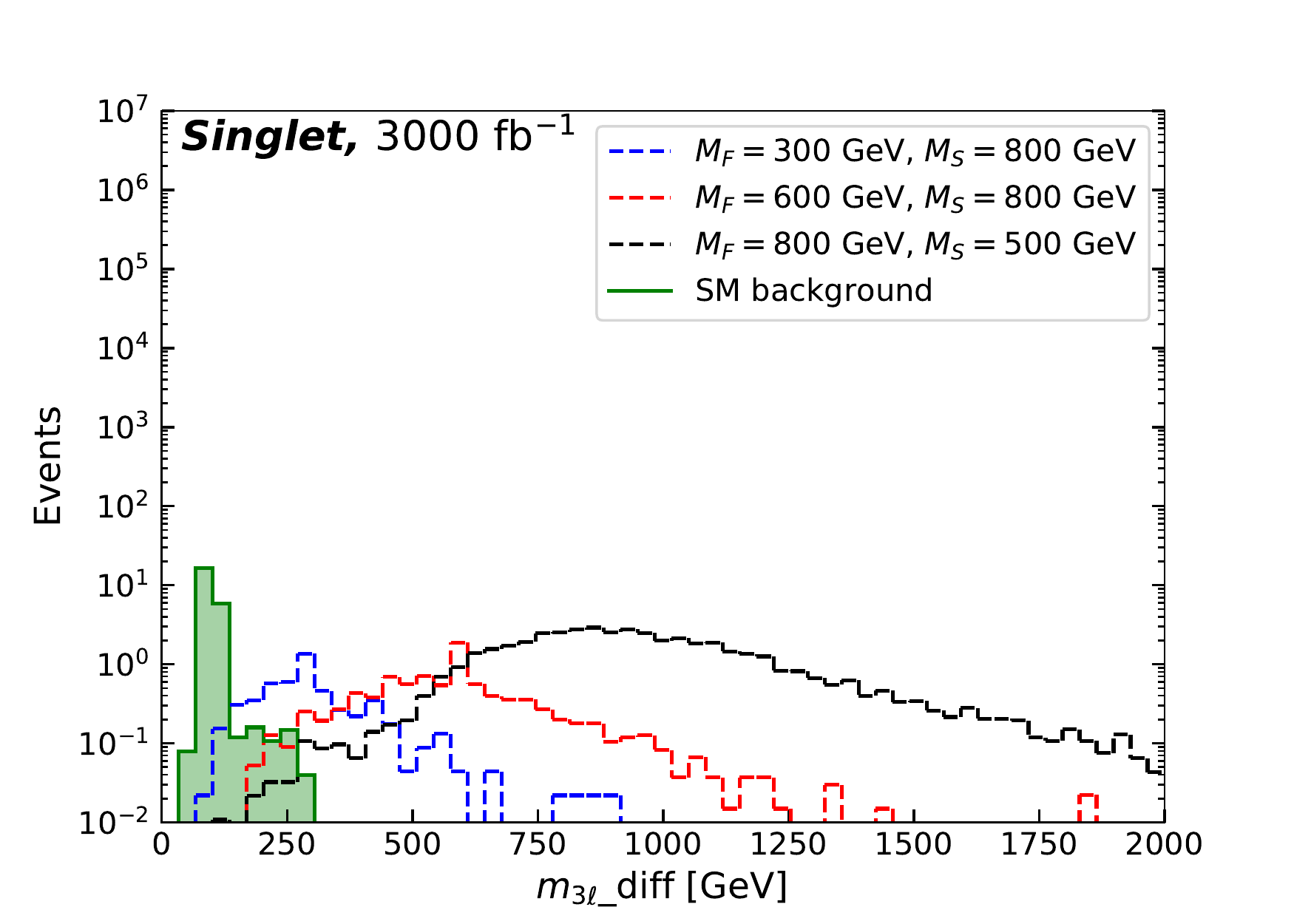}
	\caption{Di- and trilepton invariant mass distributions  $m_{2\ell}$,  $m_{2\ell}\_{\rm diff}$, $m_{3\ell}$, and $m_{3\ell}\_{\rm diff}$ after detector simulation for the singlet model. The observables are shown for different benchmarks of the VLLs and BSM scalar masses at a luminosity of $3000$ fb${}^{-1}$ and $\sqrt{s}=14$~TeV. The coupling $\kappa'$ is fixed according to Eq.~\eqref{Amu}.
	}
	\label{fig:singlet-bench-14tev-delphes}
\end{figure}

 \begin{figure}
	\centering
	
	\includegraphics[width=0.49\textwidth]{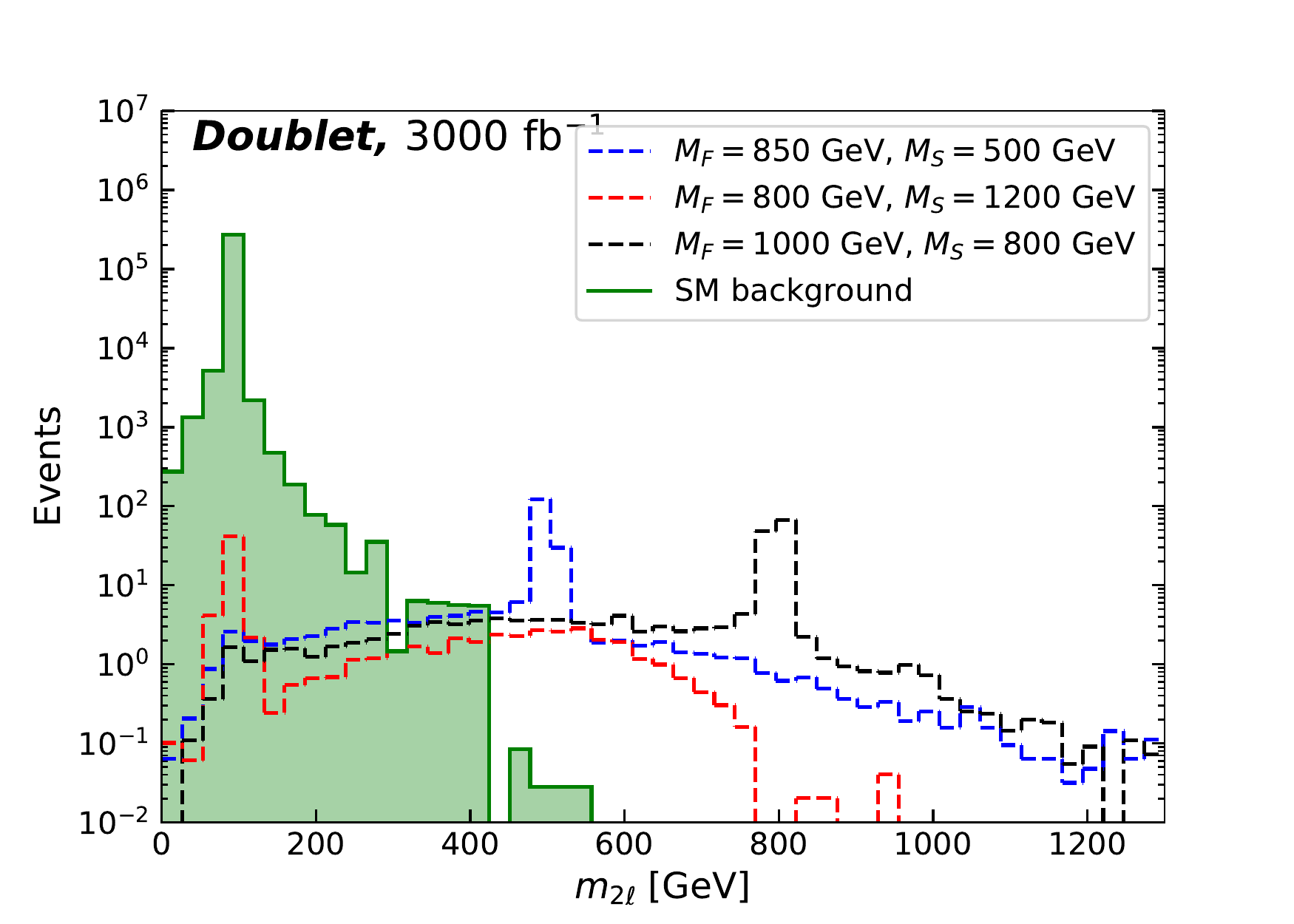}
	\includegraphics[width=0.49\textwidth]{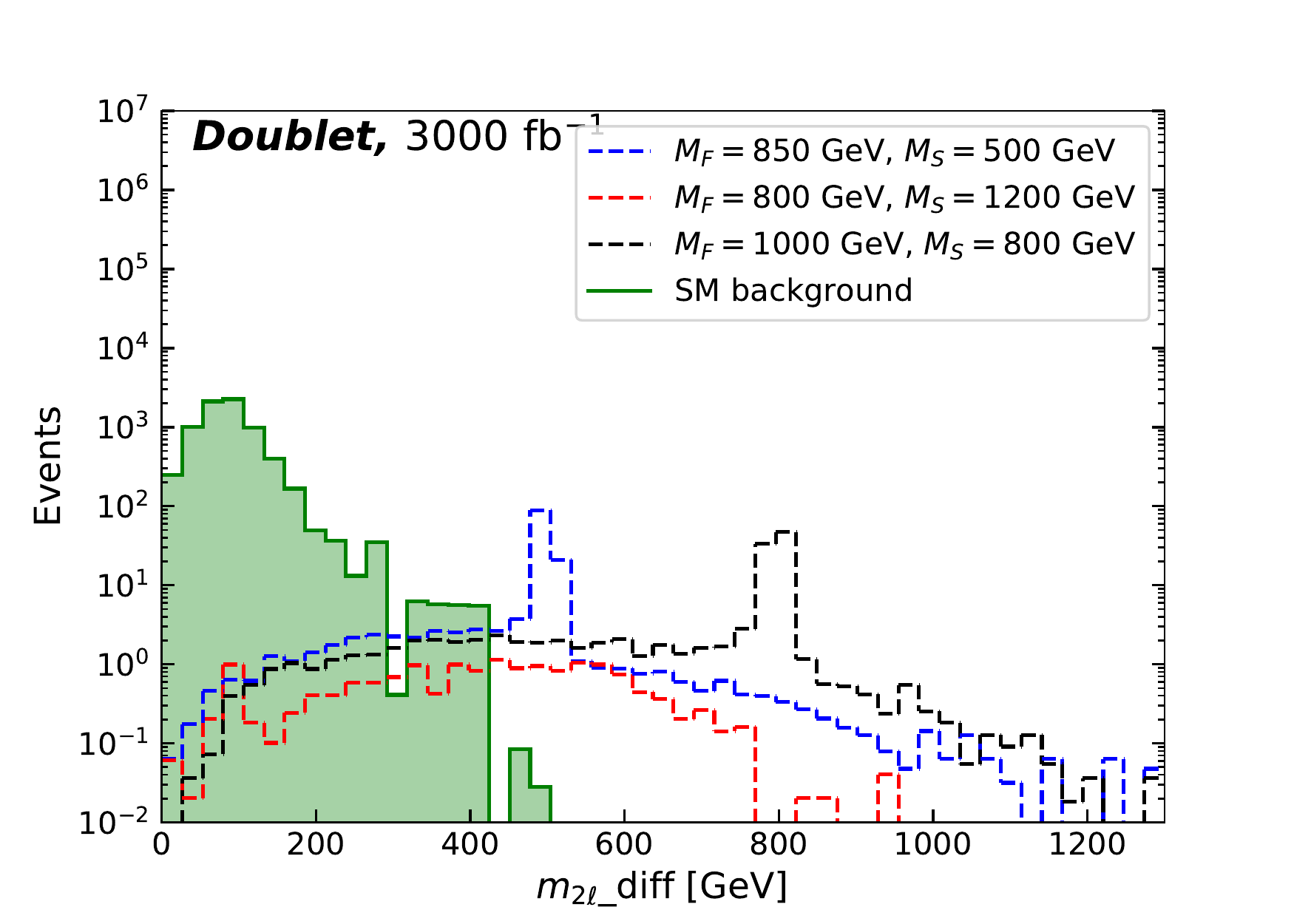}
	\includegraphics[width=0.49\textwidth]{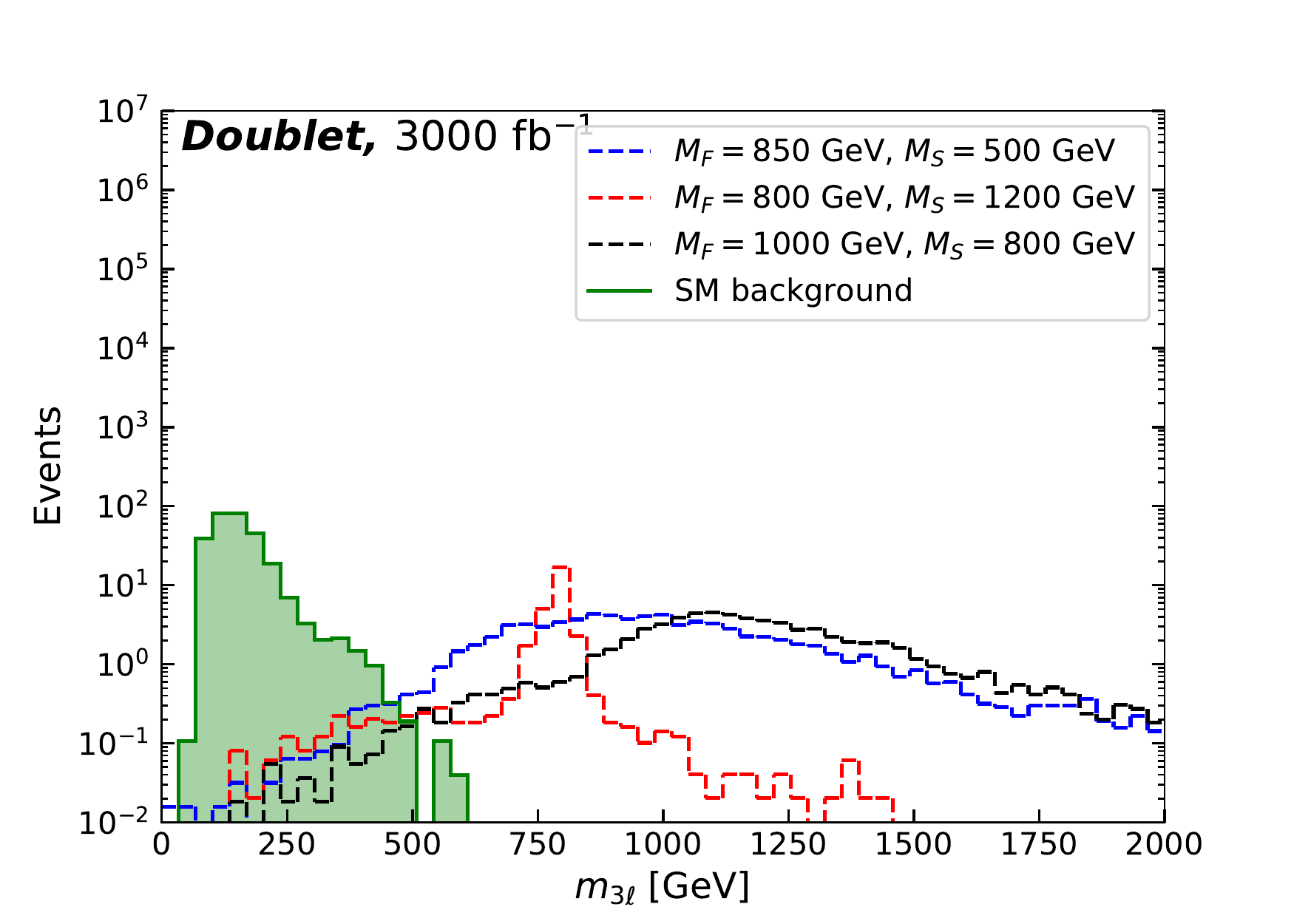}
	\includegraphics[width=0.49\textwidth]{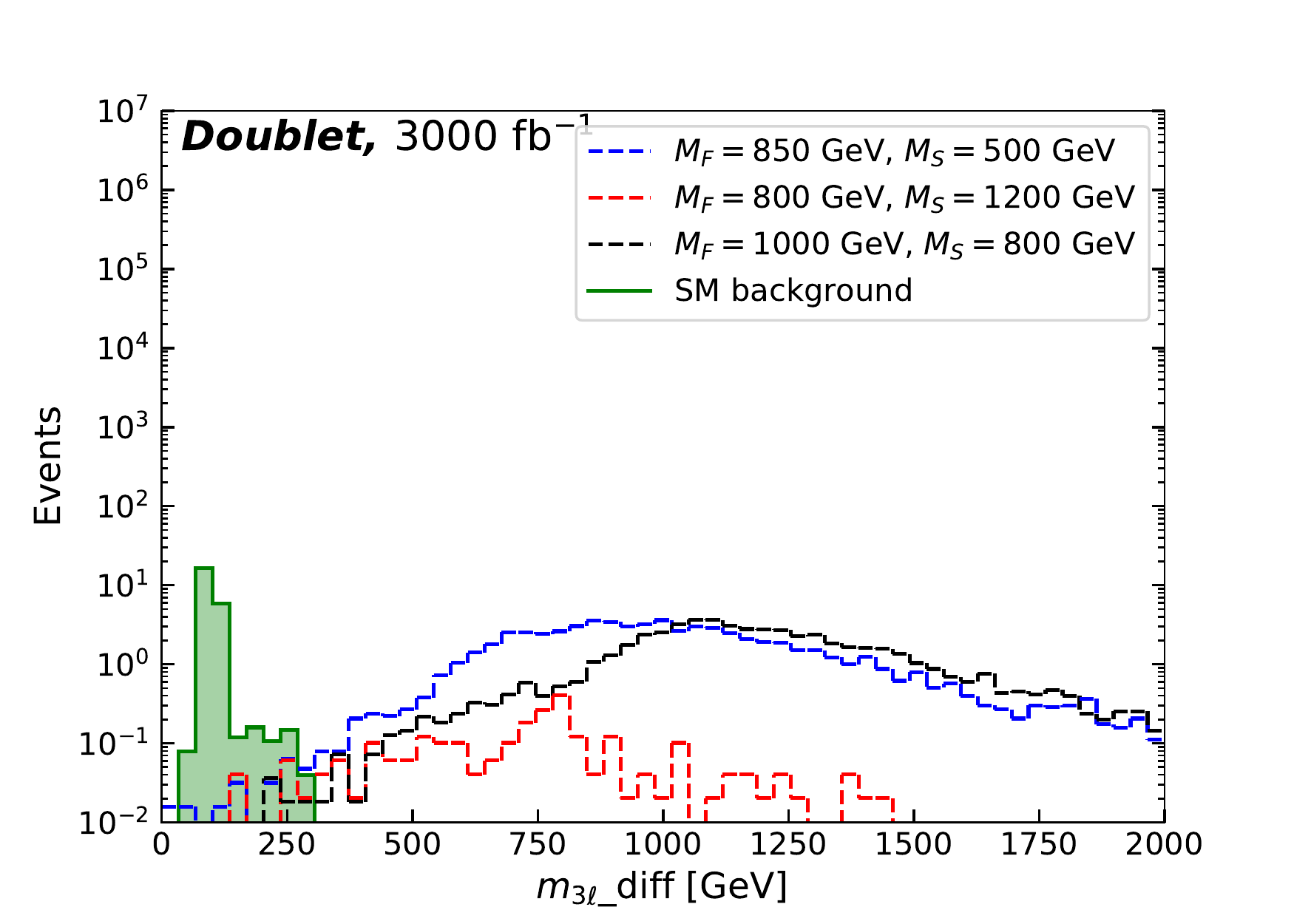}
	\caption{Di- and trilepton invariant mass distributions  $m_{2\ell}$,  $m_{2\ell}\_{\rm diff}$, $m_{3\ell}$, and $m_{3\ell}\_{\rm diff}$ after detector simulation  for the doublet model. The observables are shown for different benchmarks of VLLs and  BSM scalar masses  at a luminosity of $3000$ fb${}^{-1}$ and $\sqrt{s}=14$~TeV. The coupling $\kappa'$ is fixed according to Eq.~\eqref{Amu}.
	}
	\label{fig:doublet-bench-14tev-delphes}
\end{figure}

In Fig.~\ref{fig:singlet-bench-14tev-delphes} and Fig.~\ref{fig:doublet-bench-14tev-delphes} we show the $m_{2\ell}$,  $m_{2\ell}\_{\rm diff}$, $m_{3\ell}$, and $m_{3\ell}\_{\rm diff}$ distributions after detector simulation corresponding to Fig.~\ref{fig:singlet-bench-14tev} and Fig.~\ref{fig:doublet-bench-14tev} of Sec.~\ref{sec:Outlook}, respectively.
The detector simulation, implemented by showering the events with \textsc{PYTHIA8} and performing a fast detector simulation with \textsc{DELPHES3} with the HL-LHC default card for $3000~$fb$^{-1}$, depletes the distributions, similar to what happens in corresponding figures at 150~fb${}^{-1}$.
 In particular, we find that peaks around resonances are most affected, with baseline distributions suffering little changes. The depletion of peaks can be quantified through the scaling factors $f = N_{\rm peak, det}/N_{\rm peak}$, where $N_{\rm peak}$ ($N_{\rm peak, det}$) are the number of events in the most populated bin of a distribution before (after) detector simulation. Scaling factors for the benchmarks of the singlet and doublet model studied in Secs.~\ref{sec:Null-test} and \ref{sec:Outlook} are given in Tab.~\ref{tab:scale-factors}. In general we find that $f$ is in the $\mathcal{O}(10^{-1})$ - $\mathcal{O}(10^{-2})$ range.

This pattern is expected: As the detector energy resolution scales like $\Delta E \sim \sqrt{E}$, we expect the high-$m_{2(3)\ell}$ region to show larger differences due to finite resolution. Similarly, the peaks of distributions show the strongest effects, as the finite resolution results in general in a broadening of the peaks due to bin-to-bin migration of events.
The scaling factors in Tab.~\ref{tab:scale-factors} reflect exactly these effects, e.g. in the doublet model the $M_F=1000~$GeV and $M_S=800~$GeV scenario shows a stronger suppression of the peaks compared to the $M_F=850~$GeV and $M_S=500~$GeV benchmark. At the same time, scaling factors for the doublet $M_F=850~$GeV scenario are very similar to the singlet $M_F=800~$GeV benchmark.
In general, very sharp peaks of distributions are most affected, as a large number of events migrate out of the peak bin but only very few migrate into it from neighboring bins. 
Bins with very few events show as well significant scaling factors, since small changes in the event count due to bin-to-bin migration can have a significant impact. 
In general, we find that the improved resolution at the HL-LHC results in a smaller suppression of the peaks due to detector effects compared to Run 2 CMS. This leads to larger scaling factors in the case of the HL-LHC, as seen in Tab.~\ref{tab:scale-factors}.

\begin{table}[h]
	\centering
	\begin{tabular}{p{1.5cm}p{1.8cm}p{1.8cm}p{2.5cm}p{2.5cm}p{2.7cm}p{2.7cm}}\hline
	Model & $M_F$ (GeV) & $M_S$ (GeV) & $m_{2\ell}$ &  $m_{2\ell}\_{\rm diff}$ & $m_{3\ell}$ & $m_{3\ell}\_{\rm diff}$\\\hline 
		Singlet & 300 & 800 & 1/8* (1/7)* & 1/3* (1/3)* & 1/22 (1/20) & 1/19 (1/10) \\
		Singlet & 600 & 800 & 1/6* (1/5)* & 1/5* (1/6)* & 1/18 (1/15) & 1/16 (1/12) \\
		Singlet & 800 & 500 & 1/12 (1/10) & 1/17 (1/12) & 1/17** (1/14)** & 1/19** (1/16)** \\
    	Doublet & 850 & 500 & 1/14 (1/10) & 1/17 (1/16) & 1/16** (1/15)** & 1/20** (1/17)** \\
    	Doublet & 800 & 1200 & 1/16* (1/14)* & 1/6* (1/4)* & 1/60 (1/40) & 1/15 (1/15) \\
    	Doublet & 1000 & 800 & 1/36 (1/20) & 1/48 (1/26) & 1/27** (1/19)** & 1/34** (1/22)** \\ \hline
	\end{tabular}
	\caption{Scaling factors $f = N_{\rm peak, det}/N_{\rm peak}$ for the observables of Sec.~\ref{sec:Null-test} for different benchmarks, with $N_{\rm peak, det}$ $(N_{\rm peak})$ denoting the number of events at the peaks per bin  after (before) detector simulation for  $\sqrt{s} = 13$ TeV and a luminosity of 150~fb${}^{-1}$ and in  parentheses for $\sqrt{s} = 14$ TeV and 3000~fb${}^{-1}$. We marked with * (**) the cases where the peaks fall under SM background (resonances are broad). }
	\label{tab:scale-factors}
\end{table}

\bibliography{bhh-ASCol}

\end{document}